\documentclass[onecolumn,authoryear]{els-mrw} 

\usepackage{amsmath,amssymb,amsfonts,amsthm,makeidx,graphicx}
\usepackage{txfonts}
\usepackage{helvet}
\usepackage{xspace}
\usepackage{xcolor}

\newcommand{\de}{\ensuremath{\rm d}}

\newcommand{\Nexp}{\ensuremath{N_{\rm exp}}\xspace}

\newcommand{\Msun}{\ensuremath{\xspace\rm{M}_{\odot}}\xspace}
\newcommand{\hu}{\xspace \ensuremath{{\rm km \, s^{-1} \, Mpc^{-1}}\xspace}}

\newcommand{\Tobs}{\ensuremath{T_{\rm obs}}\xspace}

\def\eg{{\emph{e.g.~}}}
\newcommand{\apjl}{The Astrophysical Journal Letters}
\newcommand{\mnras}{MNRAS}

\newcommand{\apj}{The Astronomical Journal}
\newcommand{\prd}{Physical Review D}
\newcommand{\prl}{Physical Review Letters}

\newcommand{\jcap}{JCAP}
\newcommand{\aap}{Astronomy \& Astrophysics}

\newcommand{\nat}{Nature}

\begin{document}

\chapter{Gravitational Wave Cosmology}\label{chap1}

\author[1]{Antonella Palmese}%
\author[2]{Simone Mastrogiovanni}%

\address[1]{\orgname{Carnegie Mellon University}, \orgdiv{McWilliams Center for Cosmology and Astrophysics, Department of Physics}, \orgaddress{5000 Forbes Ave, Pittsburgh, PA 15213, USA}}
\address[2]{\orgname{INFN}, \orgdiv{Sezione di Roma}, \orgaddress{00185 Roma, Italy}}

\articletag{GW Cosmology}

\maketitle

%\begin{glossary}[Glossary]
%\term{kilonova} \\
%\term{jet afterglow} \\
%\end{glossary}

\begin{glossary}[Nomenclature]
\begin{tabular}{@{}lp{34pc}@{}}
BBH & Binary Black Hole\\
BNS & Binary Neutron Star\\
CBC & Compact Binary Coalescence\\
CDM & Cold Dark Matter\\
CE & Cosmic Explorer\\
CMB & Cosmic Microwave Background\\
EM & Electromagnetic\\
ET & Einstein Telescope\\
GW & Gravitational wave\\
GWTC & Gravitational Wave Transient Catalog\\
HBI & Hierarchical Bayesian Inference\\
KAGRA & Kamioka Gravitational wave detector\\
LIGO & Laser Interferometer Gravitational wave Observtory\\
LVK & LIGO/Virgo/KAGRA\\
PN & Post-Newtonian\\
XG & Next generation
\end{tabular}
\end{glossary}

\begin{abstract}[Abstract]
Since their first detection in 2015, gravitational wave observations have enabled a variety of studies, ranging from stellar evolution to fundamental physics. In this chapter, we focus on their use as ``standard sirens'', describing the different methodologies that can be adopted to measure cosmological parameters with compact object binaries from ground-based gravitational wave detectors. We cover the three main classes of standard siren measurements, showing how the expansion of the Universe can be constrained through Bayesian statistics both with gravitational wave observations alone and with the aid of electromagnetic emission from the electromagnetic counterpart of gravitational wave events and from galaxies. Finally, we summarize the existing measurements and prospects for future constraints on cosmological parameters.

\end{abstract}

\begin{BoxTypeA}[chap1:box1]{Key points}
\begin{itemize}
    \item Gravitational wave sources are ``standard sirens'' because their luminosity distance can be directly estimated from the amplitude of their signal. When combined with a redshift measurement, they enable measurements of the expansion of the Universe.
    \item The redshift information can be recovered through identification of the host galaxy of an electromagnetic counterpart (bright sirens), through the possible host galaxies contained within the gravitational wave localization region (dark siren galaxy catalog approach), or through measurements of the redshifted mass of compact object binaries (spectral sirens).
    \item The luminosity distance measurement of a standard siren, and therefore its resulting constraining power on the Hubble expansion, can be improved through the detection of higher order modes or precession in the gravitational wave signal, or with external measurements from the electromagnetic counterpart of a gravitational wave source.
    \item Bright, dark, and spectral sirens represent different classes of the same method. In particular, dark siren methods with galaxy catalogs and spectral sirens are ideally applied jointly to the data.
    \item Current measurements of the Hubble constant from standard sirens are not competitive with other, more mature cosmological probes, but have the potential to reach the $\sim 2\%$ precision during the fifth LIGO/Virgo/KAGRA observing run.
\end{itemize}
\end{BoxTypeA}

\section{Introduction}
The detection of gravitational waves (GWs) from compact binary coalescences (CBCs) such as binary black holes (BBHs), binary neutron stars (BNSs), or neutron star black holes (NSBHs) opened a new avenue to measure the cosmological expansion of the Universe. GW CBCs are the only astrophysical sources for which it is \emph{directly} possible to measure the distance at extragalactic scales well beyond the local group, and therefore they are referred to as ``standard sirens''.  Standard sirens offer an unprecedented opportunity for understanding how gravity works on cosmological scales and what the main actors driving the Universe's expansion are. 

The standard cosmological model describes many observations of our Universe such as its late-time accelerated expansion and the existence of a Cosmic Microwave Background (CMB). Nevertheless, the standard cosmological model suffers from issues that can potentially revolutionize our understanding of physics. On the observational side, there are discrepancies between independent measurements of the Hubble constant ($H_0$) between the value of $H_0=67.49 \pm 0.53 ~\hu$  estimated from the CMB (considered an early-time Universe probe) \citep{planckcolab} and $H_0 = 73.04 \pm 1.04 ~ \hu$ measured in the local Universe (through late-time Universe probes) through Cepheid-anchored Supernova Type Ia \citep{Riess2022}. This and other tensions can be signs of physics not described by the standard cosmological model, which is assumed to measure the local expansion of the Universe (i.e. $H_0$) from the CMB observations.
Measurements of the expansion rate of the Universe constitute a powerful probe to investigate the origin of these tensions. To directly measure the Universe expansion rate, we require cosmological sources for which it is possible to determine the distance and the recessional velocity (redshift). For sources observed through electromagnetic (EM) emission, such as galaxies or supernovae, it is typically possible to directly measure the redshift but not the distance. The distance is then obtained by building a Distance Ladder based on a range of distance indicators and the hypothesis that these sources are Standard Candles and their intrinsic luminosity is constant during all the cosmic epochs. 

Since their first detection in 2015 \citep{LIGOScientific:2016aoc}, GWs have provided us with a new tool for studying the Universe's expansion. GWs do not need a cosmological ladder and they directly provide the source distance.
Unfortunately, GWs without EM counterparts (``dark sirens'') do not directly provide the redshift of the source. Other complementary information or observations are required. In the last years, several new methods have been developed to exploit dark sirens for cosmology. The methods have in common one aspect; they need to anchor the GW source in redshift either using calibrations of the intrinsic binary mass, associations to other astronomical observations such as galaxy catalogs or Large-scale Structure (LSS) tracers. 

In this chapter, we provide a pedagogic introduction to the main methodologies currently in use for GW cosmology, their latest results and future prospects. As we will see later, GW cosmology is intimately related to GW population and astrophysical studies. For a dedicated review on that subject, we defer the reader to \citet{Callister:2024cdx}.
In Section~\ref{chap1:sec1} we review why GW sources are self-calibrating distance indicators and we introduce the statistical framework currently in use to infer cosmological expansion parameters from a set of GW detections. We also explain with some examples the three different methodologies used to provide a redshift to these sources. In Section~\ref{sec:curm} we  review the current measurements of the cosmological expansion parameters, focusing on $H_0$, with GW sources. Finally, in Section~\ref{sec:fut}, we  discuss future prospects for GW cosmology and draw our summary and conclusions in Section \ref{sec:summary}.

\section{Standard Sirens at cosmological scales}\label{chap1:sec1}

Compact binary coalescences at cosmological scales are the only known astrophysical sources for which it is possible to directly measure the luminosity distance $d_L$ with their GW emission. This peculiar property of CBCs has granted them the name of \textit{standard sirens} \citep{2005ApJ...629...15H}, as in contrast to standard candles, they are self-calibrating sources for which no cosmological ladder is needed to obtain their distance. The other ingredient needed for measuring the cosmic expansion is the source redshift $z$ to fit the distance-redshift relation. Within the standard cosmological model, this relation can be written as:
\begin{equation}
    d_L(z)=(1+z)\int_0^z\frac{{\rm d}z'}{H_0E(z')}\, ,
\end{equation}
where, the $H_0$ dependence is clear, and, for a flat $\Lambda$CDM scenario:
\begin{equation}
    E(z)\,=\sqrt{\Omega_{m}(1+z)^3+\Omega_\Lambda} ,
\end{equation}
with $\Omega_{m}$ and $\Omega_\Lambda=1-\Omega_{m}$ being the Universe matter and cosmological constant density respectively. For very low redshifts $z<<1$ one can Taylor expand the Hubble parameter $H(z)=H_0E(z)$ as:
\begin{equation}
    H(z)=H_0[1+(1+q_0)z+ (-q_0^2+j_0)z^2+\mathcal{O}(z^3)]\,\label{eq:Htaylor}
\end{equation}
where $q_0$ is the deceleration parameter and $j_0$ is the jerk. In this case, the assumptions about the background cosmological model are minimal.
Unfortunately, the GW emission does not provide any direct redshift estimate. Therefore, several methodologies have been proposed and used to assign a redshift to GW sources and employ them for cosmological studies.

This section briefly reviews why GW sources are standard sirens and what information can and cannot be extracted from the GW waveform. We then continue by introducing \textit{Hierachical Bayesian Inference}, the main statistical technique used to infer population and cosmological properties of a set of CBC detections prone to non-trivial selection biases. Finally, we explain with some examples how the redshift information can be included in this framework.

\subsection{Compact binaries at cosmological scales}\label{sec:cbc}

A compact binary coalescence is described by a set of parameters usually classified as \textit{intrinsic} or \textit{extrinsic}. Intrinsic parameters are the ones that describe features that are proper for each CBC. The intrinsic parameters are the two objects' masses, spins (with orientation), tidal deformability and orbital eccentricity. Instead, the extrinsic parameters are defined with respect to an observer, in our case at cosmological distances. The extrinsic parameters are the binary distance (or redshift), orbital inclination with respect to the line-of-sight (often referred to as $\iota$ or $\theta_{JN}$), the time of the merger at the detector $t_m$, the sky position (described using right ascension $\alpha$ and declination $\delta$) and the coalescence phase $\phi_c$ and polarization angle $\psi$.

The GW strain at the detector can be described as  
\begin{equation}
    h(t_m) = F_+(\alpha,\delta,\psi,t_m)h_+(t_m) + F_\times(\alpha,\delta,\psi,t_m)h_\times(t_m),
    \label{eq:waveformdetector}
\end{equation}
where $F_{+,\times}$ are the detectors response functions to the two GW polarizations $h_+$ and $h_\times$. These functions depend on the detector geometry, the time of arrival of the GW, the position of the source in the sky, and the polarization angle $\psi$. For ground-based GW detectors, the duration of CBC signals in the detector ranges from a few minutes to a few milliseconds. As such, we can consider $F_{+,\times}$ constant as their variation happens on time scales similar to the sidereal day ($\sim 86,400$~s).

The two polarizations of the GW can be composed of several modes, described in terms of spherical harmonics. The dominant mode contributing to the GW emission is the quadrupole mode corresponding to $l=2, m=2$ \citep{Maggiore:1900zz}. Other GW modes are strongly suppressed unless the binary displays a significant mass asymmetry. Moreover, the evolution of the GW waveform is typically described in terms of Post-Newtonian (PN) expansion. This consists of a description of the waveform expanded into the two components' velocities $\left[\frac{v^2}{c^2}\right]^{\rm PN}$ and in terms of orbital frequencies $f_{\rm orb}^{\frac{{\rm PN}-5}{3}}$. The GW waveform of binaries far from the merger, whose orbital frequency is smaller than their frequency corresponding to the last stable orbit, can be described using the 0 PN approximation. For the purpose of this introductory material, and to explain why CBCs are standard sirens, we assume that all the GW modes but the quadrupole one are negligible and the waveform can be described with the 0 PN approximation. We note that the property that GW sources are standard sirens is more general and can be proven even if the GW waveform has multiple modes and is described with higher terms in the PN order. For a more formal and general proof, see \citet{Maggiore:1900zz}.

Working in Fourier space and using the \textit{stationary phase approximation} \citep{Maggiore:2018sht}
assuming circular orbits, point masses and neglecting spins, gives, to lowest order in the PN expansion
\begin{eqnarray}
\tilde{h}_+(f)&=& A(f,{\cal{M}})\left[ \frac{1+\cos^2\iota}{2}\right] \exp{\left(i\Psi(f,{\cal{M}})\right)}
\label{eq:hO}\\
\tilde{h}_\times(f) &=& A(f,{\cal{M}}) [ \cos\iota]\exp{\left( \frac{\pi}{2} + i\Psi(f,{\cal{M}})\right)}
\label{eq:hOcross}
\end{eqnarray}
where 
\begin{equation}
    \Psi(f,{\cal{M}}) = 2\pi f t_{m} -\frac{\pi}{4} - \phi_c+ \frac{3}{128}\left(\frac{\pi G {\cal{M}}}{c^3}\right)^{-5/3} \frac{1}{f^{5/3}}
    \label{eq:phase}
\end{equation}
and
\begin{equation}
    A(f,{\cal{M}}) = \frac{1}{d} \frac{5}{24 \pi^{4/3}}
    \frac{(G {\cal{M}})^{5/6}}{c^{3/2}} \frac{1}{f^{7/6}}.
    \label{eq:amplitude}
\end{equation}
Above $f$ is the GW frequency, $d$ is the detector's physical distance from the source and
\begin{equation}
    {\cal{M}} \equiv \frac{(m_1m_2)^{3/5}}{(m_1+m_2)^{1/5}},
\end{equation}
is the \textit{chirp mass}, defined in terms of the binary components masses $m_1$ (for the primary component, the most massive object in the binary) and $m_2$ (for the secondary).
When a binary is at cosmological distances, we need to account for the fact that their frequency is redshifted by the expansion of the Universe. Using the subscripts $d$ for ``detector'', the frequency at the detector is
\begin{equation}
f_d=\frac{f}{1+z}.
\label{eq:redf}
\end{equation}
We can now try to express the GW waveform in terms of detector frequency. By noticing that both the time and frequency in the detector frame are redshifted, we can rewrite Eq.~\eqref{eq:phase} as 
\begin{equation}
\begin{split}
    \Psi(f_d,{\cal{M}}_d) &= 2\pi f_d(1+z) \frac{t_{m,d}}{1+z} -\frac{\pi}{4} - \phi_c+ \frac{3}{128}\left(\frac{\pi G {\cal{M}}}{c^3}\right)^{-5/3} \frac{1}{(1+z)^{5/3}f_d^{5/3}} \\
    &= 2\pi f_d t_{m,d} -\frac{\pi}{4} - \phi_c+ \frac{3}{128}\left(\frac{\pi G {\cal{M}}_d}{c^3}\right)^{-5/3} \frac{1}{f_d^{5/3}},
\end{split}\label{eq:phase2}
\end{equation}
where we have defined the chirp mass at the \textit{detector} $\mathcal{M}_d=(1+z) \mathcal{M}$. Eq.~\eqref{eq:phase2} implies that for an observer at a cosmological distance, the phase of the GW waveform has the same analytical expression that one would have at the source, with the difference that one should define a mass that is redshifted. 
We can also make the same substitutions into Eq.~\eqref{eq:amplitude} for the amplitude of the GW, obtaining
\begin{equation}
    A(f_d,{\cal{M}}_d) = \frac{1}{d(1+z)} \frac{5}{24 \pi^{4/3}}
    \frac{(G {\cal{M}}_d)^{5/6}}{c^{3/2}} \frac{1}{f_d^{7/6}}.
\end{equation}
We notice that for an observer \textit{today} ($z=0$), the physical distance $d$ is equal to the comoving distance and $d(1+z)$ is therefore the luminosity distance of the source $d_L$. It follows that the amplitude of the GW can be rewritten as
\begin{equation}
    A(f_d,{\cal{M}}_d) = \frac{1}{d_L} \frac{5}{24 \pi^{4/3}}
    \frac{(G {\cal{M}}_d)^{5/6}}{c^{3/2}} \frac{1}{f_d^{7/6}}.
    \label{eq:amplitude2}
\end{equation}
Eq. \eqref{eq:amplitude2} tells us that the amplitude of a GW observed at a cosmological distance has the same scaling as it would have in the source frame, with the difference that the physical distance should be replaced by the luminosity distance and the mass by the redshifted chirp mass. Note that this is not valid for mergers involving a neutron star, in which the tidal deformability can play a role in breaking the degeneracy with redshift \citep{Messenger:2011gi, DelPozzo:2015bna,Chatterjee:2021xrm}.

To summarize the discussion above, Eq.~\eqref{eq:phase2} and Eq.~\eqref{eq:amplitude2} imply that the GW waveform for an observer at a cosmological distance has the same analytical form (in terms of binary parameters) as the one seen by an observer at the source frame. However, for an observer at a cosmological distance, one should replace the distance with the luminosity distance and the chirp mass with the redshifted chirp mass.
A crucial consequence of this is that an observer at a cosmological distance is only able to measure the redshifted chirp mass (from the GW phase) and the luminosity distance (from the amplitude). An observer at cosmological scales is not able to directly measure the source mass, as this is completely degenerate with redshift. In other words, no redshift information can be extracted from the GW waveform \emph{directly}. 

Another interesting aspect of the GW waveform is that the amplitude factor $A$, and hence the luminosity distance, is strongly degenerate with the determination of $\cos \iota$ (see Eqs.~\ref{eq:hO}-\ref{eq:hOcross}). In other words, from a GW waveform point of view, a close-by binary with its orbital plane edge-on with respect to the observer would be similar to a GW waveform for a farther binary face-on with respect to the observer. However, note that this is not a complete degeneracy and that it can be greatly alleviated by including in the GW waveform the effect of spin-induced precession \citep{Graff:2015bba,2018PhRvL.121b1303V}, higher-order GW modes \citep{LIGOScientific:2020zkf} and possibly GW signals from the post-merger phase of BNS \citep{CalderonBustillo:2020kcg}. 
\begin{figure}[htp]
%	\hspace*{-0.7cm}
\centering
	\includegraphics[width=0.4\textwidth]{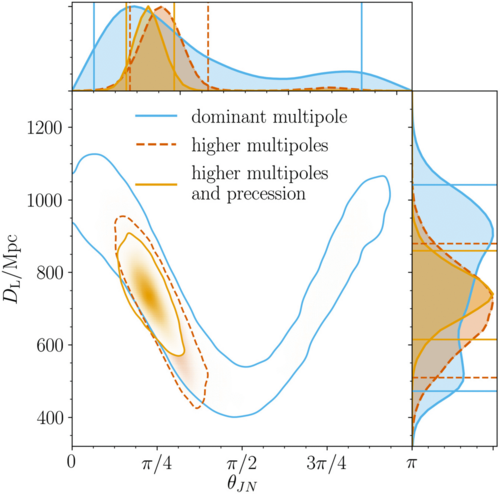}
    \caption{Inferred luminosity distance (vertical axis) and orbital viewing angle (horizontal axis) from GW190412. The different colors indicate posteriors from waveform models that include, or do not, precession and higher order modes of on the GW waveform. Figure reproduced from \citet{LIGOScientific:2020zkf}.}
    \label{fig:GW190412}
\end{figure}
As an example, Figure~\ref{fig:GW190412} shows the distance and $\cos \iota$ determination for GW190412, a GW event for which it was possible to observe the presence of higher-order modes \citep{LIGOScientific:2020zkf}. When the data are analyzed with waveform models including higher-order modes, the degeneracy between luminosity distance and $\cos \iota$ is alleviated. Determining precisely the luminosity distance is crucial for cosmology, as the precision on the cosmological parameters that we are able to obtain strongly depends on the precision with which we are able to infer the luminosity distance for the source.

To conclude our discussion, measuring the cosmic expansion requires sources for which luminosity distance and redshift are observed. Since for GWs, the redshift is \textit{completely degenerate} with the source mass, GW sources can not be used to measure the universe expansion unless additional observations or hypotheses to obtain a redshift are made. In the next sections, we review the current possibilities for assigning a redshift to GW sources.

\subsection{Hierarchical Bayesian inference}
\label{sec:hbi}

Before diving into the details of the methods used for GW cosmology, we need to introduce the basics of \textit{Hierarchical Bayesian Inference} (HBI), the main statistical tool used to infer cosmological and population properties from current GW observations. HBI is a statistical tool developed to describe heterogeneous populations of sources observed in incomplete and noisy datasets. An in-depth discussion and derivation of the HBI likelihood is provided by \citet{2022hgwa.bookE..45V}.

With the term heterogeneous populations, we indicate that the population might not be trivial to describe, for instance, the mass distribution of binary black holes can strongly deviate from a simple power law. With the term incomplete data set, we indicate that the observed population can be different from the astrophysical population, as there is a selection bias that depends on \eg the binary masses. Finally, with the term noisy data set, we indicate that the properties of the binary, such as detector masses and luminosity distance are not precisely measured.
HBI describes the detection of a population of CBCs with a ``modified'' Poissonian likelihood \citep{Mandel_2019}. Let us introduce a population of GW sources dependent on a set of parameters ${\bf \Lambda},$ that describe the production rate of binaries in terms of {the GW} binary parameters, $\theta$. In other words, we are modeling a differential number of sources
\begin{equation}
    \frac{\de N}{\de t \de \theta}({\bf \Lambda}) = \frac{\de N}{\de t}p_{\rm pop}(\theta|{\bf \Lambda}),
\end{equation}
that can be also described in terms of a population probability $p_{\rm pop}(\theta|{\bf \Lambda})$. To give a practical example, $\theta$ could be the detector masses of the binary $m_d$, $p_{\rm pop}(m_d|{\bf \Lambda})$ a simple power law distribution, and ${\bf \Lambda}$ the set of parameters governing the shape of the power law. The likelihood of obtaining $N_{\rm obs}$ observations, each described by some parameters $\theta$, in a data collection $\{x\}$, for non overlapping signals, for a given observing time $T_{\rm obs}$ from a population of events with a constant rate and in presence of selection biases is given by
\begin{equation}\label{eq:fund}
\begin{split}
    \mathcal{L}(\{x\}|{\bf \Lambda}) &\propto e^{-N_{\rm exp}({\bf \Lambda})} \prod_{i=1}^{N_{\rm obs}} \int  \mathcal{L}(x_i|\theta) \frac{\de N}{\de t \de \theta}({\bf \Lambda}) \de t \de \theta \\
    & \propto  e^{-N_{\rm exp}({\bf \Lambda})} \prod_{i=1}^{N_{\rm obs}} T_{\rm obs} \int \mathcal{L}(x_i|\theta) \frac{\de N}{\de t \de\theta}({\bf \Lambda}) \de \theta.
\end{split}
\end{equation}
Eq.~\eqref{eq:fund} is referred to as a ``hierarchical likelihood''. In Eq.~\eqref{eq:fund}, $\mathcal{L}(x_i|\theta)$ is the GW likelihood for a single event $i$, a statistical function that quantifies how precisely we are able to measure the binary parameters $\theta$ from the GW data, while \Nexp is the expected number of GW detections in a given observing time \Tobs. The likelihood in Eq.~\eqref{eq:fund} represents an inhomogeneous Poisson process with selection biases; indeed, by closely inspecting its analytical form it strongly resembles the likelihood of a standard Poisson process.
Another central quantity of the hierarchical likelihood is the ``expected number of GW detections'' $N_{\rm exp}({\bf \Lambda})$, which is related to the selection bias and can be evaluated as: 
\begin{equation}
    N_{\rm exp}({\bf \Lambda}) = T_{\rm obs} \int p_{\rm det}(\theta) \frac{dN}{\de t \de \theta} \de \theta, 
    \label{eq:nexp}
\end{equation}
where $p_{\rm det}(\theta)$ is a detection probability  
\begin{equation}
    p_{\rm det}(\theta) = \int_{x \in \rm{detectable}} \mathcal{L}(x_i|\theta) \de x.  
\end{equation}
Typically, the detection probability is not known analytically, unless some simplifying assumptions are made. We refer to \citet{Hitchhiker} for an introductory example in the context of GW cosmology. The current approach to evaluate selection biases is to use Monte Carlo simulations of injected and detected events \citep{Tiwari:2017ndi,2024A&A...682A.167M}, often referred to as ``injections''. The injections are used to evaluate the {signal detectable} volume that can be explored in the parameter space and correct for selection biases. 
The hierarchical likelihood in Eq.~\eqref{eq:fund} can be written into a more compact form. We can analytically marginalize over \Nexp by using a ``scale-free'' prior $\pi(N_{\rm exp}) \propto 1/N_{\rm exp}$. With this choice, we can simplify the form of the likelihood as
\begin{equation}
    \mathcal{L}(x|{\bf \Lambda}) \propto  \prod_{i=1}^{N_{\rm obs}} \frac{ \int \mathcal{L}(x_i|\theta) p_{\rm pop}(\theta|{\bf \Lambda}) \de \theta}{ \int  p_{\rm det}(\theta) p_{\rm pop}(\theta|{\bf \Lambda}) \de \theta}.
    \label{eq:fund_scalefree}
\end{equation}
The scale-free version of the hierarchical likelihood allows us to describe the distribution of CBCs in terms of population distributions. For the purposes of this introductory material, we use the scale-free version of the hierarchical likelihood in order to simplify our discussions for GW cosmology.

Let us introduce how cosmological information from the GW is encoded in the likelihood. We recall that $\theta$ are the binary parameters that we measure from the GW detectors. According to what is discussed in Section~\ref{sec:cbc}, the relevant binary parameters measured for cosmology are the luminosity distance $d_L$, the inclination angle $\iota$ (as it is strongly degenerate with $d_L$), the two detector masses $\mathbf{m}_d=(m_{1,d},m_{2,d})$ as they are related to the source frame masses with a redshift factor $\mathbf{m}_d=\mathbf{m}(1+z)$ and the sky position $\mathbf{\Omega}=(\alpha,\delta)$. The only two binary parameters that are degenerate with redshift are the luminosity distance and the detector masses, the other parameters do not carry any relevant redshift information but are important for GW cosmology for the motivations we explore in the following sections. Note that, in principle, one should consider all the measured binary parameters (including spins). However, in order to simplify our discussion and notation, here we simply write equations in terms of these four parameters which are the most crucial for cosmology. To measure cosmological parameters, we need to rewrite Eq.~\eqref{eq:fund_scalefree} in terms of redshift. This means that we need to perform a transformation from the detector frame variables $(d_L,\mathbf{m}_d,\cos\iota)$ to the source frame variables $(z,\mathbf{m},\cos\iota)$. This can be easily done by assuming a cosmological model and a set of cosmological parameters, here considering only $H_0$ to keep the notation simple. When we perform this transformation, the likelihood becomes
\begin{equation}
    \mathcal{L}(x|{\bf \Lambda}) \propto  \prod_{i=1}^{N_{\rm obs}} \frac{ \int \mathcal{L}(x_i|z,\mathbf{m},\cos\iota,\mathbf{\Omega},H_0) p_{\rm pop}(z,\mathbf{m},\cos\iota,\mathbf{\Omega}|{\bf \Lambda})\, \de z \, \de \mathbf{m} \, \de \cos\iota \, \de \mathbf{\Omega}}{ \int  p_{\rm det}(z,\mathbf{m},\cos\iota,\mathbf{\Omega},H_0) p_{\rm pop}(z,\mathbf{m},\cos\iota,\mathbf{\Omega}|{\bf \Lambda})\, \de z \, \de \mathbf{m} \, \de\cos\iota \, \de \mathbf{\Omega}}.
 \label{eq:hiecosmo}
\end{equation}

\subsubsection{Obtaining redshift information from the source mass of the binaries}\label{sec:galaxymethod}

One possibility of obtaining redshift information from GWs alone is to make assumptions about the source mass distribution of CBCs \citep{Taylor2012,Farr_2019,2021PhRvD.104f2009M,Ezquiaga:2020tns,spectral_sirens,Mali:2024wpq}. The idea is the following: since the source mass is related to the detector mass with a redshift factor, and since we measure detector mass, by making some assumptions on the source mass distribution we can obtain implicit redshift information. This can be done by explicitly modeling the population distribution of CBCs in the source frame, namely
\begin{equation}
    p_{\rm pop}(z,\mathbf{m},\cos\iota, \mathbf{\Omega}|{\bf \Lambda})=p_{\rm pop}(z|{\bf \Lambda}) p_{\rm pop}(\mathbf{m}|{\bf \Lambda}) p_{\rm pop}(\cos\iota|{\bf \Lambda}) p_{\rm pop}(\mathbf{\Omega}|{\bf \Lambda}).
    \label{eq:ratespecsiren}
\end{equation}
Note that in writing Eq.~\eqref{eq:ratespecsiren}, we have already made an important simplified assumption that the source mass of CBCs is not dependent on redshift. This may not be the case as the mass spectrum of CBCs could evolve in redshift due to several astrophysical processes, see \citet{2021hgwa.bookE..16M} for a review. The only distributions that in this case bring additional redshift information are the distribution of CBCs in redshift and source masses, namely $p_{\rm pop}(z|{\bf \Lambda}), p_{\rm pop}(\mathbf{m}|{\bf \Lambda})$. The distributions over inclination angle and sky positions are typically considered to be isotropic and not dependent on any population parameter ${\bf \Lambda}$.

\begin{figure}
    \centering
    \includegraphics[width=0.45\linewidth]{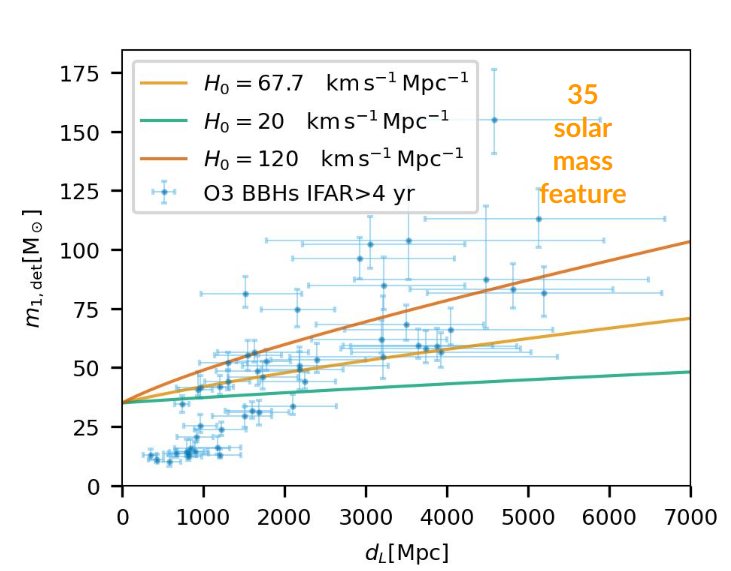}
    \includegraphics[width=0.45\linewidth]{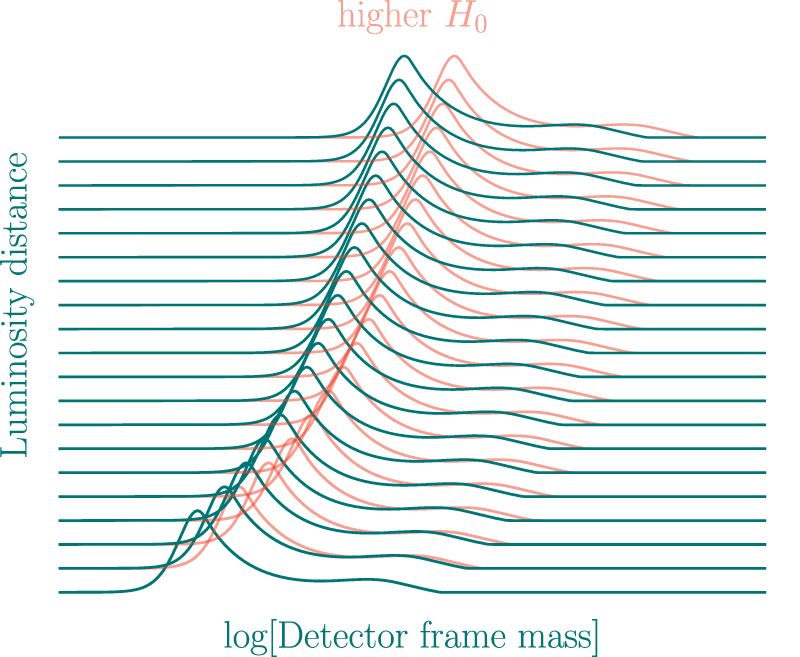}
    \caption{\textit{Left:} distribution of GW events reported in GWTC-3 \citep{GWTC_3} with an inverse False Alarm Rate higher than 4 years in terms of their estimated luminosity distance (horizontal axis) and primary detector frame mass (vertical axis). The panel also reports how a source frame mass of 35 \Msun would be recast onto the luminosity distance versus the detector mass plane for different values of $H_0$. \textit{Right:} Evolution of a source mass spectrum described by a \textsc{Power Law + Peak} model \citep{2018ApJ...856..173T} in the detector mass (horizontal axis) luminosity distance (vertical axis) plane for different values of $H_0$ (see colours). The figure is reproduced from \citet{Chen:2024gdn}.}
    \label{fig:specex}
\end{figure}

To understand why the source frame mass spectrum can be informative on the redshift of CBCs and cosmology, let us reference to Figure~\ref{fig:specex}. The left panel shows the distribution of BBHs reported in GWTC-3 \citep{GWTC_3} with an inverse False Alarm Rate higher than 4 years in terms of their estimated luminosity distance and primary detector frame mass. If we assume that the BBHs are only produced with 35 \Msun, namely $p_{\rm pop}(\mathbf{m}|{\bf \Lambda})=\delta(\mathbf{m}-35 \Msun)$, then we have a mass scale that we can exploit for cosmology. The $35~ \Msun$ mass scale is recast in terms of $d_L,m_d$ according to the cosmological model, in our case according to an $H_0$ value. From the left panel of Figure~\ref{fig:specex}, we can see that a flat $\Lambda$CDM model with $H_0=67.7 \, \hu$ seems to follow the overdensity of BBHs observed between 30-50 \Msun in detector frame. We warrant that this is only an example to understand how the source mass distribution can be informative on the redshift of GW events, robust scientific results should be obtained with full HBI.

Naturally, we also do not expect the distribution of CBCs to be a Dirac delta distribution. That is why, in current GW cosmological studies, flexible phenomenological models for the mass distribution are adopted \citep{LIGOScientific:2018jsj,LIGOScientific:2020kqk, KAGRA:2021duu}. The right panel of Figure~\ref{fig:specex} shows how a \textsc{Power Law + Peak} model \citep{2018ApJ...856..173T}, would evolve in the detector masses versus luminosity distance plane for different cosmological models. This different evolution for the two cosmological models is the motivation for which we are able to infer cosmological parameters using mass information. While considering a mass spectrum that is more complicated than a delta function renders the analysis less simple, it is clear that only relying on one mass feature or scale may not be feasible given that, as mentioned above, these may evolve over redshift due to astrophysical processes, such as the cosmic metallicity evolution over redshift \citep{2021hgwa.bookE..16M}. One concern about the spectral siren method is that the population model assumed, in particular the mass distribution and its redshift evolution, needs to be accurate \citep{Mastrogiovanni:2021wsd,Mukherjee:2021rtw}. 
If astrophysical processes move the mass scale over redshift, then these processes become degenerate with the cosmology. On the other hand, if multiple features are present, as they are found to be, in the mass spectrum, it is unlikely that they will all move in the same manner due to the underlying astrophysics. This motivates the use of the entire mass spectrum and multiple bumps and features to constrain the Universe expansion, as well as the name \emph{spectral sirens} \citep{spectral_sirens}.  The magnitude of the systematic bias induced by a possible redshift evolution of the mass spectrum is currently under debate. Recent studies based on the simulation of populations of BBHs with a redshift-dependent mass spectrum \citep{Pierra:2023deu, Agarwal:2024hld}, shows that this can be a potential source of systematic bias for the future. For this reason, non-parametric models reconstructing the masses and redshift (luminosity distance) distribution can be employed \citep{MaganaHernandez:2024uty,2024arXiv241023541N,2025ApJ...978..153F}.

\subsubsection{Obtaining redshift information from galaxy surveys}

Another possibility for obtaining redshift information for GW events is to employ galaxy surveys. This idea was originally presented in \citet{schutz}, the first article to present the idea of standard sirens, and it consists of using the GW sky localization, in terms of luminosity distance and sky position, to identify possible host galaxies from a galaxy survey. Electromagnetic observational surveys typically directly estimate the redshift of the galaxies using spectroscopic or photometric techniques. Therefore, using a given cosmological model, one can convert the redshift of the galaxies to luminosity distance and try to match them with the localized GW source.

This type of technique can be implemented into the HBI scheme by defining a population distribution conditioned on the observation of the galaxy catalogue CAT, namely
\begin{equation}
    p_{\rm pop}(z,\mathbf{m},\cos\iota, \mathbf{\Omega}|{\bf \Lambda},{\rm CAT})=p_{\rm pop}(z,\mathbf{\Omega}|{\rm CAT},{\bf \Lambda}) p_{\rm pop}(\mathbf{m}|{\bf \Lambda}) p_{\rm pop}(\cos\iota|{\bf \Lambda}).
    \label{eq:ratecatalog}
\end{equation}
Note that, differently from Eq.~\eqref{eq:ratespecsiren}, here we can not separate the population distribution in terms of sky position and redshift as for the galaxy catalogue this is dictated by the galaxies observed in the survey. Moreover, let us note that the distribution of CBCs $p_{\rm pop}(z,\mathbf{\Omega}|{\rm CAT},{\bf \Lambda})$ does not have to match the distribution of galaxies in the catalogues $p_{\rm gal}(z,\mathbf{\Omega}|{\rm CAT})$ as GW sources can be preferentially hosted by a particular type of galaxy, \eg luminous galaxies. In other words, if we are able to assign to each galaxy a GW host probability $p_{\rm host}(\zeta|{\bf \Lambda})$, where $\zeta$ is a collection of galaxy parameters to describe the GW host probability, then
\begin{equation}
p_{\rm pop}(z,\mathbf{\Omega}|{\rm CAT},{\bf \Lambda}) \propto \int \delta(z-z_g) p_{\rm gal}(z_g,\mathbf{\Omega},\zeta|{\rm CAT}) p_{\rm host}(\zeta,z_g|{\bf \Lambda}) \de \zeta \de z_g.
\end{equation}
In the above equation, $z_g$ is the redshift distribution of the galaxies and the $\delta$ function is added to model that if no galaxy is present, GW events can not be present.

Another important aspect for this type of analysis is that galaxy surveys are typically not complete as they are ``flux-limited'', meaning that EM surveys can only detect galaxies brighter than a given apparent magnitude threshold. Therefore, when constructing the expected distribution of the CBC sources, one should account for the missing galaxies. This can be done by defining the overall distribution of CBC sources as 
\begin{equation}
p_{\rm pop}(z,\mathbf{\Omega}|{\rm CAT},{\bf \Lambda})= f(z,\mathbf{\Omega}) p_{\rm pop}(z,\mathbf{\Omega}|{\rm CAT},{\bf \Lambda}) +[1-f(z,\mathbf{\Omega})] p_{\rm pop}(z|{\bf \Lambda}) p_{\rm pop}(\mathbf{\Omega}|{\bf \Lambda}),
\label{eq:comp}
\end{equation}
where $f(z,\mathbf{\Omega})$ is the fraction of galaxies we expect to be in the catalog at redshift $z$ and sky position $\mathbf{\Omega}$. The completeness function is dependent on the sky position as there could be flux limitations due to the field-of-view of actual electromagnetic instruments or particular sky locations. As an example, most of the extragalactic surveys in the optical band are strongly incomplete over the direction of the Galactic plane. Let us note that, in Eq.~\eqref{eq:comp}, the second term on the right-hand side is the same term encoded in Eq.~\eqref{eq:ratespecsiren}.
The completeness function can be computed as
\begin{equation}
f(z,\mathbf{\Omega}) = \frac{\int^\infty_{M_{\rm thr}(z,\mathbf{\Omega})} {\rm Sch}(M,z)p_{\rm host}(M,z|{\bf \Lambda}) \de M}{\int^\infty_{M_{\rm faint}} {\rm Sch}(M,z)p_{\rm host}(M,z|{\bf \Lambda}) \de M},
\end{equation}
where ``Sch'' denotes the Schechter function, namely the distribution of galaxies in absolute magnitudes, $M_{\rm faint}$ its faint end boundary and $M_{\rm thr}(z,\Omega)$ the apparent magnitude threshold for the survey. 
Before looking into some practical examples of this method, let us make an important remark. The source mass spectrum always enters in Eq.~\eqref{eq:ratecatalog} and it competes with the distribution of galaxies in the localization area to identify the possible redshift of the GW source. If the galaxy catalog is strongly incomplete, or the localization of GW sources is large and includes thousands of galaxies, the choice of the mass spectrum will still have a crucial impact on cosmological inference. That is why it is important even in this case to marginalize over several mass spectrum models.

\begin{figure}[htp!]
\centering   
\includegraphics[width=1.0\textwidth]{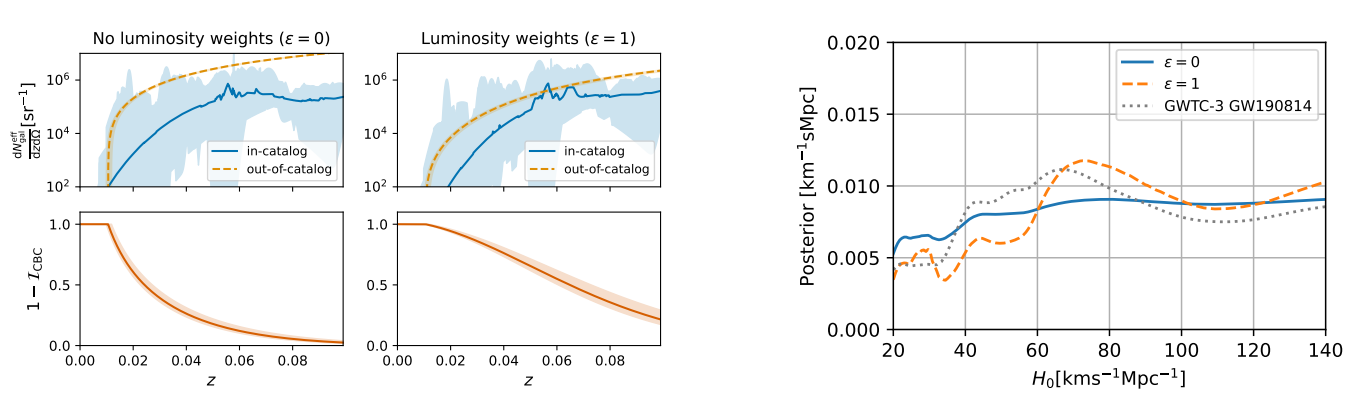}
    \caption{\emph{Left:} Top panels, the number density of galaxies that can be CBC hosts as a function of redshift, compared with the completeness correction. The two panels are generated for two GW host probability models proportional to $L^\epsilon$. Bottom panels, completeness fraction of the galaxy catalog used in \citet{Mastrogiovanni2023} as a function of redshift.  \emph{Right:} Hubble constant posteriors generated for GW190814 \citep{LIGOScientific:2020zkf} using the two galaxy number densities in the left panel. Figures from \citet{Mastrogiovanni2023}.}
    \label{fig:galexa}
\end{figure}

The left panel in Figure~\ref{fig:galexa} shows how the distribution of galaxies in a galaxy catalog can be informative on redshift. The top plots of the panel show the differential number of galaxies that can host GW signals, assuming a host probability proportional to the luminosity of the galaxy $L^\epsilon$. The differential number of CBC hosts is reported for the sky localization area of GW190814 \citep{LIGOScientific:2020zkf}, one of the best localized GW sources. We can observe that there is an overdensity of galaxies around redshift 0.04-0.06, which translates to a $H_0$ posterior (right panel) with a local mode in the $H_0$ tension region. 
Another relevant aspect is that the possible redshift of GW sources is not entirely given by the galaxies in the catalog, but it is corrected with a completeness, or ``out-of-catalog'', correction. The fraction governing the mixture of the ``in-catalog'' and ``out-of-catalog'' components is the estimated completeness fraction in Eq.~\eqref{eq:comp}. As we can see from the bottom panels of Figure~\ref{fig:galexa}, the completeness rapidly deteriorates as a function of redshift and at redshifts 0.04-0.06, the redshift localization is entirely dominated by the completeness correction.
A final interesting aspect is that, if the GW host probability prefers luminous galaxies as hosts for GW events, then the galaxy catalog is more complete from a GW perspective. This is a consequence of the fact that more luminous galaxies are detectable to higher redshifts.

\subsubsection{Bright Standard Sirens}

A final possibility to obtain a redshift estimate of the GW source is with observations complementary to the GW data, \eg a direct observation of the GW EM counterpart. Electromagnetic counterparts may often allow us to identify the host galaxy, thus the redshift and sky position of the GW event. Moreover, using astrophysical prescriptions for the EM counterpart, one can also obtain an estimation of $\cos \iota$.
In other words, we have a complementary data set $y$ and an additional likelihood $\mathcal{L}_{\rm EM}(y|z, \cos \iota)$ that can measure redshift and inclination angle. 
The likelihood for this type of event can be modified as follows
\begin{equation}
    \mathcal{L}(x|{\bf \Lambda}) \propto  \prod_{i=1}^{N_{\rm obs}} \frac{ \int \mathcal{L}(x_i|z,\mathbf{m},\cos\iota,\mathbf{\Omega},H_0)  \mathcal{L}_{\rm EM}(y_i|z,\cos\iota,\mathbf{\Omega})  p_{\rm pop}(z,\mathbf{m},\cos\iota,\mathbf{\Omega}|{\bf \Lambda})\, \de z \, \de \mathbf{m} \, \de \cos\iota \, \de \mathbf{\Omega}}{ \int  p_{\rm det}^{\rm GW+EM}(z,\mathbf{m},\cos\iota,\mathbf{\Omega},H_0) p_{\rm pop}(z,\mathbf{m},\cos\iota,\mathbf{\Omega}|{\bf \Lambda})\, \de z \, \de \mathbf{m} \, \de\cos\iota \, \de \mathbf{\Omega}}.
 \label{eq:hiecosmo_bright}
\end{equation}
Note that in the above equation, the GW detection probability has to be replaced with the probability of measuring a GW \textit{and} its EM counterpart. 

It is now instructive to make some simplified assumptions to better understand the hierarchical likelihood in Eq.~\eqref{eq:hiecosmo_bright}. Let us assume that with the EM counterpart, we are not able to obtain any information on $\cos \iota$ while we are perfectly able to infer the redshift of the source $z_s$ and the sky position $\mathbf{\Omega}_s$. Then the EM likelihood would be $\mathcal{L}_{\rm EM}(y_i|z,\cos\iota,\mathbf{\Omega})\propto \delta(\mathbf{\Omega}-\mathbf{\Omega}_s) \delta(z-z_s)$. With this assumption, we can integrate analytically over $z,\cos\iota$ and $\mathbf{\Omega}$ the numerator of Eq.~\eqref{eq:hiecosmo_bright}. Moreover, as the redshift is perfectly known, the source frame mass distribution will not introduce any reference\footnote{This is valid in the limit that the GW likelihood can be separated into two independent terms, one dependent on the mass and the other on the luminosity distance, see \citet{2021PhRvD.104f2009M} for more details.}. Before rewriting Eq.~\eqref{eq:hiecosmo_bright}, let us also make the assumption that the most limiting selection for detection is given by GW detectors so that we can approximate $p_{\rm det}^{\rm GW+EM}(\cdot) \approx p_{\rm det}^{\rm GW}(\cdot)$. The GW detection horizon is mostly a function of luminosity distance (for a narrow range of masses), as such the explorable redshift volume, namely the denominator of Eq.~\eqref{eq:hiecosmo_bright}, scales as $H_0^3$. The selection bias basically accounts for the missing sources close to the luminosity distance detection threshold.
The hierarchical likelihood can be simplified as
Eq.~\eqref{eq:hiecosmo_bright} can be rewritten as
\begin{equation}
    \mathcal{L}(x|{\bf \Lambda}) \propto  \prod_{i=1}^{N_{\rm obs}} \frac{ \mathcal{L}(x_i|d_L(z_s,H_0))}{H_0^3}.
    \label{eq:hiesimp}
\end{equation}
and if we do not even have detection thresholds, the likelihood reduces to the standard multiplication of independent likelihoods for $N_{\rm obs}$ independent observations.
\begin{equation}
    \mathcal{L}(x|{\bf \Lambda}) \propto  \prod_{i=1}^{N_{\rm obs}} \mathcal{L}(x_i|d_L(z_s,H_0)).
\end{equation}

To summarize, for GW sources that are only detectable at low redshifts, the simplified likelihood in Eq.~\eqref{eq:hiesimp} is a good proxy to estimate the posterior of $H_0$, unless peculiar velocities should be accounted for (see Section \ref{sec:curm}).

We note that even bright standard sirens can potentially be prone to systematics on $H_0$. Possible sources of systematics for bright sirens are related to the determination of the binary inclination angle \citep{Chen:2020dyt,Salvarese:2024jpq,Mancarella:2024qle}, mismodeling of the EM detection probability \citep{Chen:2023dgw} and calibration uncertainties for the GW strain data \citep{Huang:2022rdg}.

\subsection{Cross-correlation techniques}

Cross-correlation of spatial distributions of GW sources with other Large Scale Structure (LSS) tracers, such as galaxies, is another technique that was proposed for GW cosmology \citep{Namikawa:2015prh,Oguri:2016dgk,Namikawa:2016edr}. This method has been further developed in \citet{Zhang:2018nea,Scelfo:2020jyw,Bera:2020jhx,Libanore:2020fim,Mukherjee:2020hyn,diaz22,Ferri:2024amc,Ghosh:2023ksl,Zazzera:2024agl} in the context of cross-correlations between GWs and galaxies and in \citet{Scelfo:2021fqe} in the context of cross-correlation between GWs and neutral hydrogen intensity maps. Here we do not go into the details of the method, but we provide a general explanation for this basics based on the angular power spectra.
 
The idea behind this methodology is that the spatial distribution of GW sources, and in particular their anisotropies, should be correlated with the anisotropies of other LSS tracers. 
The spatial distribution of LSS tracers is typically studied in terms of density fluctuations $\delta^{\rm X}(\mathbf{\Omega},z)$, where ${\rm X}$ indicates the species of the LSS tracer, e.g. GW sources, galaxies, or neutral hydrogen densities. These density fluctuations are linked to the matter density fluctuations $\delta_m$ with a set of bias parameters $b_{\rm X}(z)$. From the density fluctuations, it is possible to study anisotropies by computing the auto-angular power spectrum $\mathcal{C}^{\rm X,X}_l(z)$ and the cross-angular power spectrum $\mathcal{C}^{\rm X,Y}_l(z)$ between the different species. In this approach, $l$ indicates the angular scale on which the anisotropies correlate and the power spectrum is allowed to evolve over redshift. The potential constraint on cosmological parameters, such as $H_0$, is given by the fact that GW sources are detected in luminosity distance space while galaxies and neutral hydrogen intensity maps are detected in redshift space. By varying the cosmological model it is possible to calculate $\mathcal{C}^{\rm GW,X}_l(z)$ and $\mathcal{C}^{\rm GW,GW}_l(z)$ and compare them with their expected values. This method can also be applied considering the correlation function and the 3D power spectra, see \citet{Mukherjee:2020hyn} for more details.

There are challenges when applying this technique. From a detection point of view, in order to be able to measure the cross-correlation signal at the relevant angular scales of the anisotropies, we need the GW sources to be well-localized. Well-localized GW sources are currently limited by the detectors' sensitivity and their duty cycles. From a modeling point of view, in order to calculate the expected cross-correlation signal $\mathcal{C}^{\rm GW,X}_l(z)$, we need to model the redshift distribution of the sources, their matter bias parameter, magnification and evolution biases. The addition of these ingredients can result in the definition of nuisance parameters on which the analysis needs to marginalize upon. 
From a modelling point of view, non-isotropic detector sensitivities of the GW detectors and assumptions about the population of BBHs are also aspects to consider.
The former can introduce artificial anisotropies for the GWs at the angular scales relevant to the GW detectors' antenna patterns, but it can be corrected and taken into account \citep{diaz22,Afroz:2024joi,Ferri:2024amc}. 
The latter can either impact the GW bias parameter or introduce a systematic bias on the redshift, in particular due to the mass spectrum choices. Regarding the bias parameter, several studies have calculated the GW bias parameter varying population assumptions on GW host probability, time-delay and rate modelling in redshift
\citep{Scelfo:2020jyw,Libanore:2020fim,Libanore:2021jqv,Peron:2023zae,Dehghani:2024wsh,Zazzera:2024agl}, arguing that is possible to marginalize upon it. Regarding the mass spectrum, to our knowledge, there are no systematic studies on how this could impact the GW luminosity distance--redshift conversion for the cross-correlation analysis. However, \citet{Mukherjee:2022afz} argues that with current data these effects are negligible.

\section{Current measurements}
\label{sec:curm}

\subsection{Bright standard sirens}

The detection of the first confirmed EM counterpart associated with a GW event occurred on August 17, 2017. The event, GW170817 \citep{ligobns}, was the first GW detection from a binary neutron star merger, and it was accompanied by a variety of EM counterparts observed by the astronomical community across the EM spectrum \citep{MMApaper}. First, the short Gamma Ray Burst GRB 170817A \citep{LIGOScientific:2017zic} was detected within $\sim 1.7$ s of the GW event, confirming the nature of at least a fraction of short GRBs as relativistic jets launched from neutron star mergers and enabling tests of the speed of gravity. Following the GRB, a kilonova \citep{Li:1998bw,Metzger_2010}, an optical/infrared transient powered by the radioactive decay of heavy $r$-process nuclei synthesized in the ejecta, was identified by several teams \citep{Arcavi_2017,Coulter_2017,Lipunov,Soaressantos_2017,Tanvir_2017,Valenti_2017}. The kilonova identification first allowed for the localization of the GW source to sub-arcsecond precision, and subsequent association to a specific host galaxy, NGC 4993 \citep[e.g.][]{Blanchard,Palmese_2017}, whose redshift could be used for the first standard siren analysis \citep{firststandardsiren}. Once the kilonova faded and the field became observable again, it was possible to identify a rising afterglow component \citep[e.g.][]{Margutti_2018} from the jet interactions with the circumstellar material.

\begin{figure}[htp!]
\centering   \includegraphics[width=0.6\textwidth]{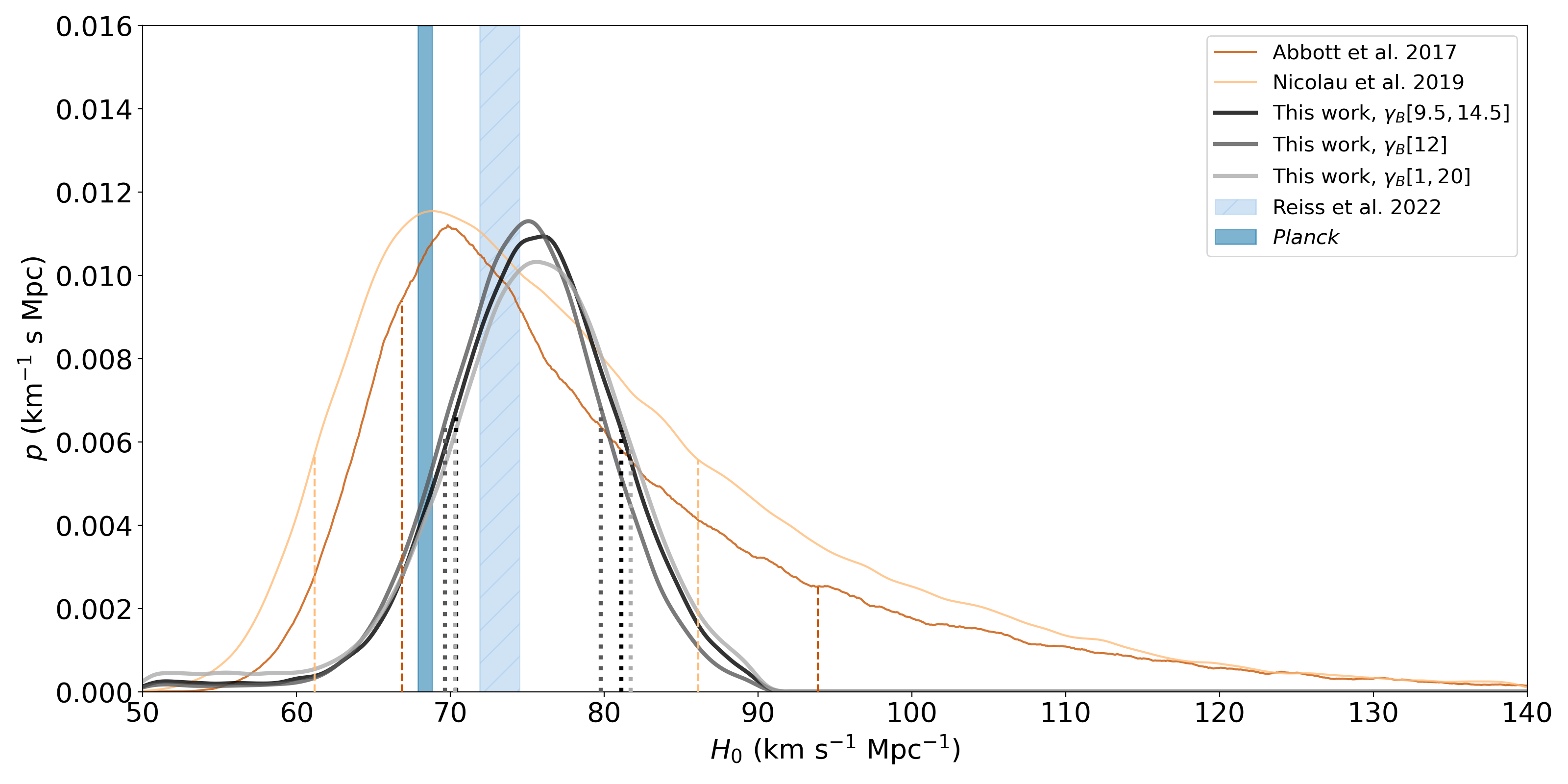}\includegraphics[width=0.4\textwidth]{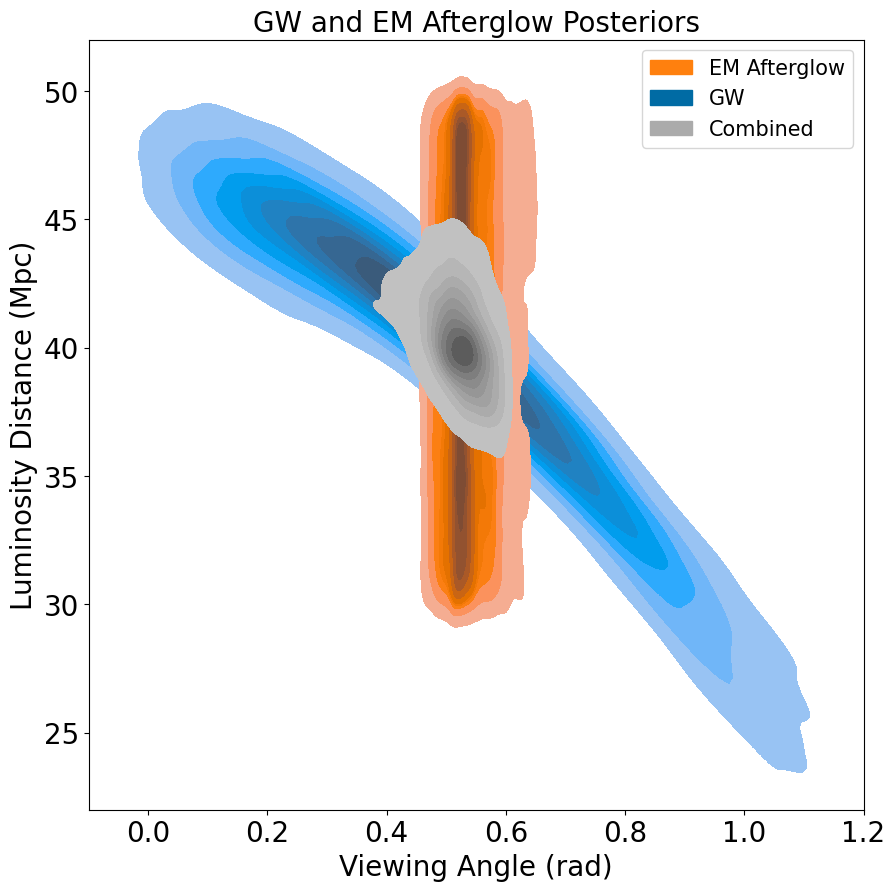}
    \caption{\emph{Left:} Hubble constant posterior distributions from various right siren measurements using GW170817, including the first standard siren analysis \citep{firststandardsiren} in dark orange, a revised measurement taking into account uncertainties in the peculiar velocity estimation method \citep{Nicolaou_2020} in light orange, and an analysis that takes into account EM observations of the afterglow \citep{Palmese_2024} in black. For each posterior, the corresponding vertical lines show the 68\% CI. SH0ES \citep{Riess2022} and \emph{Planck} 1$\sigma$ constraints are shown by the vertical bands.  \emph{Right:} 10-90\% CI contours of the luminosity distance-viewing angle posterior from GW observations of GW170817 (blue), from EM observations of the afterglow (orange), and from the combination of the two (grey). Figures from \citet{Palmese_2024}.}
    \label{fig:H0bright}
\end{figure}

The observed redshift of the GW host galaxy cannot be directly plugged into Eq. \eqref{eq:hiecosmo_bright}, since it is affected by motion that is not related to cosmological expansion alone. In other words, the observed redshift $z_{\rm obs}$ of a galaxy is typically given by:
\begin{equation}
    (1+z_{\rm obs}) = (1+z_{\rm cos})(1+z_{\rm pec})(1+z_{\rm Sun})
\end{equation}
where $z_{\rm cos}$ is the cosmological redshift of interest, $z_{\rm pec}$ arises from the peculiar motion of the galaxy due to local dynamics, and $z_{\rm Sun}$ is the redshift due to the motion of the Sun with respect to the CMB. Although $z_{\rm Sun}$ can be computed, $z_{\rm pec}$ can be more challenging to estimate, but it is necessary to subtract out the peculiar motion component and estimate the Hubble constant from a standard siren measurement. Typically a peculiar velocity can be estimated from a peculiar velocity survey including galaxies with known distances (so that the peculiar motion can be estimated for a given Hubble expansion at that distance) or through velocity field reconstruction.
The first standard siren measurement \citep{firststandardsiren} assumed a recessional velocity of $3327 \pm 72 ~\rm{km/s}$ and a peculiar velocity of $310\pm 150~\rm{km/s}$, finding $H_0 = 70^{+12}_{-8}~ \hu$ and a posterior shown in dark orange in the left panel of Figure \ref{fig:H0bright}. This measurement relies on velocity estimates of neighboring galaxies around NGC 4993 from a peculiar velocity survey, weighted following a Gaussian smoothing kernel centered on the GW170817 host. However, the choice of the size of the smoothing kernel is somewhat arbitrary, so ideally this choice would be marginalized over to take into account this source of systematic uncertainty \citep{Nicolaou_2020}. When this is considered in the standard siren analysis, the uncertainty increases and \citet{Nicolaou_2020} finds $H_0 =  68.6^{+14.0}_{-8.5}~ \hu$, with a posterior shown in light orange in the left panel of Figure \ref{fig:H0bright}.

It is worth noting that NGC 4993 has been identified as part of a group of galaxies, but most of the galaxies used in the smoothing kernel approach do not belong to this group and are therefore possibly unrelated to the peculiar motion of NGC 4993. If one tries to use the peculiar velocity and redshift of the galaxy group to perform a standard siren analysis, then a question arises as to which galaxies in existing peculiar velocity surveys are part of the group. \citet{Howlett_2020} use Bayesian Model Averaging to marginalize over the different group choices, also finding, as expected, an increased Hubble constant uncertainty, $H_0 =  66.8^{+13.4}_{-9.2}~ \hu$. Finally, one could choose to perform a reconstruction of the velocity field using galaxy redshifts in the region within a forward model framework based on N-body simulations to infer an improved proper motion as in \citet{Mukherjee2020}. Overall, the uncertainty on the peculiar velocity is to be considered a major source of uncertainty on the final $H_0$ estimation from GW170817.
%%Maybe add naive error estimation

Another major source of uncertainty for current standard siren measurements arises from the degeneracy between the distance and inclination angle in a GW waveform amplitude highlighted in Section \ref{sec:cbc}. The degeneracy can be broken either by GW observations of higher-order modes or precession, or by EM observations of the counterpart. GW170817 offered the opportunity to consider both cases. First, precession effects in the waveform allow for improved luminosity distance constraints when the neutron stars dimensionless spin magnitude $\chi$ prior is allowed to reach values as high as $0.89$ \citep{LIGOScientific:2018hze}. The low spin prior ($\chi<0.05$) on the other hand, chosen for consistency with the fastest pulsars in binaries that will merge within a Hubble time observed in our Galaxy, does not favor the presence of precessing effects in the GW signal, resulting in broader constraints on inclination angle, luminosity distance, and therefore, on $H_0$ ($H_0 = 70^{+13}_{-7}~ \hu$ using the high-spin priors, versus  $H_0 = 70^{+19}_{-8}~ \hu$ with low-spin priors). It is worth noting that in view of subsequent GW detections, priors motivated by Galactic neutron stars from EM observations may not be fully justified, as GW observations may probe different BNS populations than those observed in EM waves in the Milky Way \citep{Abbott_2020_GW190425}. 

Typically, the jet prompt or afterglow EM emission, or the kilonova, depends on the viewing angle $\theta_v$ of the observer with respect to the jet axis, rather than on the inclination angle with respect to the binary angular momentum $\iota$ defined in Section \ref{sec:cbc}. Assuming that the jet axis is aligned with the binary angular momentum for the case of prompt and afterglow jet emission, the GW observations allow us to discern between clock-wise versus counter-clockwise rotation (thus $\iota$ can range between 0 and 180 deg) while EM  emission would not be sensitive to the direction of the binary rotation (thus $\theta_v$ ranges between 0 and 90 deg). This would similarly apply to the kilonova emission, although in that case no assumption about the jet direction is needed. Therefore the relation between viewing and inclination angle is:
\begin{equation}
    \theta_v = \min(\iota, \pi - \iota). 
\end{equation}
 Taking into account this relation, the degeneracy between the viewing angle and the luminosity distance from GW observations alone can be seen in the 2D posterior of GW170817 in Figure \ref{fig:H0bright}. Breaking this degeneracy to obtain an improved distance constraint would also result in an improved $H_0$ constraint. For off-axis jets, such as that associated with GW170817, the jet afterglow, arising from broadband synchrotron emission due to a relativistic outflow that interacts with the circumstellar medium, peak time and light curve width are expected to depend on the jet geometry (especially on the jet opening angle) and the viewing angle. It is therefore possible to assume or marginalize over possible jet structures to obtain a viewing angle constraint from afterglow observations. Moreover, since this event was at a luminosity distance of only 40 Mpc, it was possible to observe the superluminal motion of the jet with radio very long baseline interferometry data \citep{mooley}, and derive viewing angle constraints based on it to obtain $H_0=70.3^{+5.3}_{-5.0} ~\hu$ \citep{Hotokezaka}. Using the first observations of the afterglow light curve it was also possible to place improved $H_0$ constraints \citep{Guidorzi:2017ogy}. When using the jet opening angle constraints from the jet motion \citep{Mooley_2022} as a prior and effectively marginalizing over a range of jet structures while fitting a comprehensive afterglow lightcurve from X-rays to radio over 3.5 years of observations, \citet{Palmese_2024} find $H_0=75.46^{+5.34}_{-5.39}$ km s$^{-1}$ Mpc$^{-1}$ (black posterior in the left-hand side panel of Figure \ref{fig:H0bright}), leading to a $\sim 7\%$ $H_0$ precision measurement. Note that the aforementioned analyses utilizing jet afterglow and superluminal motion observations make different assumptions about the peculiar velocity, leading to different precision levels even for similar viewing angle constraints. Inferring the viewing angle from the jet afterglow can also cause biases \citep{Gianfragna}, due, for example, to mismodeling of the jet structure.

%Could add cartoon of binary, afterglow, and kilonova structure, with viewing and inc angles

The relation between jet prompt and afterglow emission viewing angle and the GW inclination angle actually only holds if the jet is launched in the same direction as the binary angular momentum, which is typically assumed, but not necessarily true. The offset between the jet and the angular momentum directions, if it exists, may cancel out after combining a number of events if it is randomly oriented, but \citet{muller2024} explore the effect of a possible offset on the $H_0$ estimation.

Another method of estimating the viewing angle from EM observations concerns the kilonova. While mostly isotropic in its emission, a kilonova is expected to comprise of different components following different geometries, such as, in the case of a binary neutron star merger, a blue polar component from squeezed dynamical ejecta, a more isotropic, planar, redder dynamical ejecta component, as well as a more spherical wind ejecta.  Therefore, it is possible to constrain the viewing angle from kilonova emission and derive an improved standard siren measurement \citep{2020ApJ...888...67D}. However, depending on the geometry of the different kilonova components, extremely different constraints on the viewing angle can be obtained \citep{2021MNRAS.502.3057H}, thus careful multimessenger standard siren analyzes should take into account this potential source of systematics when using kilonovae to constrain the viewing angle.

It is worth noting that waveform modeling systematics are not currently considered a major source of systematics for current generation GW detectors \citep{kunert}.

%%%EM Selection effects here or in future prospects?

\subsection{Dark standard sirens - galaxy catalog approach}

\begin{figure}
    \centering
    \includegraphics[width=0.5\linewidth]{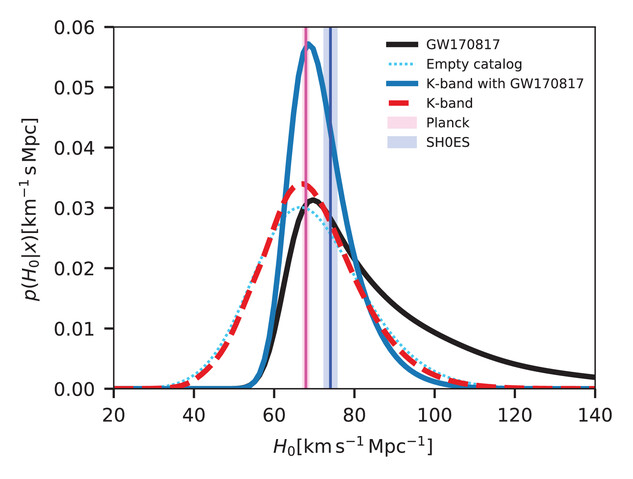}\includegraphics[width=0.5\linewidth]{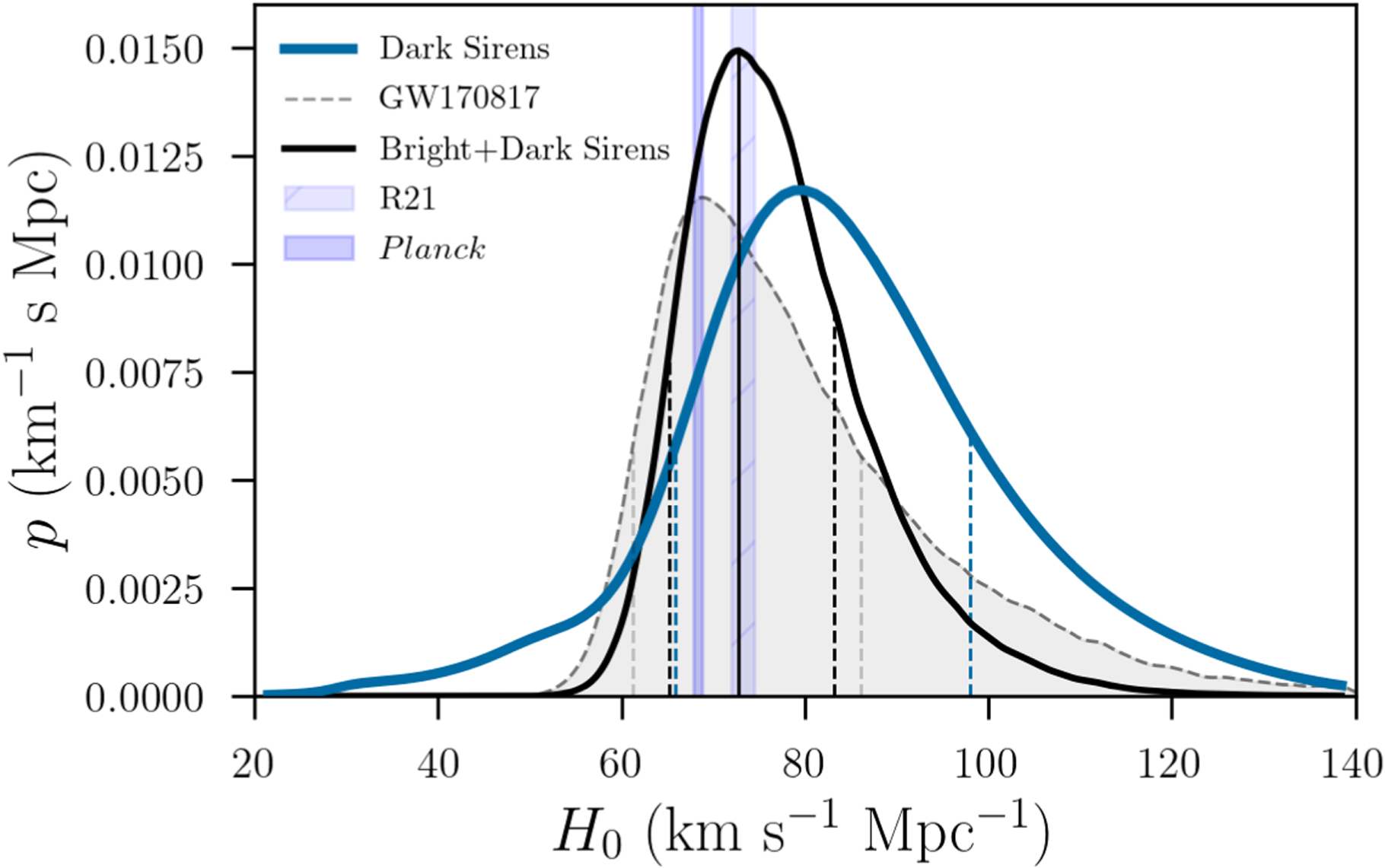}
    \caption{\emph{Left:} Hubble constant posteriors from the GWTC-3 dark siren analysis of \citet{GWTC3_expansion}. The dashed posterior is recovered using GLADE galaxies weighted by $K-$band luminosity and 46 dark sirens, and combined with the GW170817 bright siren measurement (in solid black), gives the solid blue posterior. The dotted line shows the posterior from the dark siren analysis when no galaxy information is used. From \citet{Mastrogiovanni:2024mqc}. \emph{Right:} Hubble constant posteriors from the GWTC-3 dark siren analysis of \citet{darksiren_DESI}, using 8 well-localized events and galaxy information from DES and the DESI Legacy Survey. From \citet{darksiren_DESI}.}
    \label{fig:darksirenresults}
\end{figure}

Owing to the overwhelmingly larger number of GW events with no EM counterpart compared to the one source with a confirmed multimessenger association, a wide range of dark siren analyses have been performed. The first 3-detector high confidence BBH detection led to a relatively small sky localization of $\sim 60$ deg$^2$ for GW170814 \citep{LIGOScientific:2017ycc}, fully covered by the Dark Energy Survey (DES; \citealt{DESoverview}) Year 3 galaxy catalog. The combination of the GW posterior with photometric redshifts from DES allowed the first dark siren from a BBH \citep{DES:2019}. Further analyses including events from the first two LIGO/Virgo observing runs (O1 and O2), and later O3, make use of the GLADE \citep{2018MNRAS.479.2374D,Dalya_2022} galaxy catalog for other GW events \citep{LIGOScientific:2019zcs,GWTC3_expansion}. The O3 analysis from 46 dark sirens yields $H_0=67^{+13}_{-12} \ \rm{km \ s^{-1} \ Mpc^{-1}}$ when galaxies are weighted using their $K-$band luminosity, which combined with the GW170817 bright siren analysis in \citet{firststandardsiren} gives $H_0=68^{+8}_{-6} \ \rm{km \ s^{-1} \ Mpc^{-1}}$ \citep{GWTC3_expansion}, a $\sim 40\%$ improvement in precision compared to the O2 results. The resulting $H_0$ posteriors for these two constraints are shown in the left panel of Figure \ref{fig:darksirenresults}. In addition, the same figure shows the posterior that one would derive from the dark siren measurement alone with an empty galaxy catalog. It is clear that the K-band weighted posterior is only slightly more informative than the case where no galaxy redshift prior is used, implying that most of the dark siren constraining power in this case is brought by the mass distribution assumed, i.e. this is effectively a spectral siren measurement with an assumed mass distribution following a \textsc{Power Law + Peak} model with fixed parameters. \citet{2021PhRvD.104f2009M,GWTC3_expansion} find indeed a significant impact of mass and rate distribution assumptions on Hubble constant measurements from dark sirens when the galaxy catalog is not complete at the relevant redshifts for the GW events. In other words, if the redshift prior from the galaxy catalog is not informative, the redshift information is coming from the mass spectrum. As expected from the spectral siren method, features in the mass distribution are found to correlate with $H_0$ \citep{GWTC3_expansion}, and especially the location of one of the most prominent features of the mass distribution - the $\sim30~M_\odot$ Gaussian bump, i.e. different assumptions on the Gaussian bump mean mass will significantly shift and may potentially bias the $H_0$ constraint. These findings highlight the need to jointly fit the mass distribution and cosmological parameters \citep{Mastrogiovanni2023,2023JCAP...12..023G}, especially when galaxy catalogs are incomplete at the redshifts of interest, and show how dark siren analyses with galaxy catalogs and spectral sirens are different faces of the same coin.

Another avenue to overcome mass distribution-cosmology degeneracies in dark siren analyses with galaxy catalogs consists of focusing only on well-localized events with extensive galaxy survey coverage to the necessary depth to provide constraining redshift information for the GW events. This is the case for the analyses using DES \citep{palmese20_sts}, the DESI Legacy Survey \citep{darksiren_DESI,Bom24_darksiren}, the DECam Local Volume Exploration (DELVE) Survey \citep{Alfradique24}, and more recently, the DESI spectroscopic survey \citep{ballard}. Unlike the GLADE catalog, containing the brightest galaxies out to $\sim 130$ Mpc and mostly designed to aid in multimessenger follow-up campaigns, the aforementioned state-of-the-art photometric surveys contain bright galaxies out to $z\sim 1$, thus covering the redshifts of interest for the dark sirens in questions (all at $>200$ Mpc), although over a smaller footprint ($\sim 16,000$ sq. deg for the DESI Legacy Survey versus full sky for GLADE). The analysis of \citet{darksiren_DESI} uses 8 O3 dark sirens and Legacy Survey Imaging, and finds $ H_0=79.8^{+19.1}_{12.8} ~\hu$, and $H_0=72.77^{+11.0}_{-7.55}~\hu$ when combining with the GW170817 bright siren posterior of \citet{Nicolaou_2020}. Dark sirens improve bright siren $H_0$ precision by about $\sim 30\%$, showing how dark sirens can significantly contribute to standard siren measurements.

On the galaxy catalog side, two important potential sources of systematics have been taken into account: galaxy weighting and photometric redshifts. In section \ref{sec:galaxymethod} we have introduced a probability $p_{\rm host}$ that weights each galaxy in the dark siren formalism based on its probability of hosting a GW event given its properties. Naively, one may expect that a galaxy with larger stellar mass contains more stars, including black holes and neutron stars, than a lower stellar mass galaxy, and should therefore be assigned a higher $p_{\rm host}$ value. For this reason, several works have explored stellar mass (or rest frame $K$-band luminosity, which is expected to correlate with stellar mass to first order) weighting. On the other hand, if time delays between star formation and binary merger are typically short, star formation rate (traced to first order by UV or $B$-band rest frame wavelengths) may also be a good tracer of GW host probability. To complicate the picture, isolated binary formation predicts a metallicity dependence on BH mass. At the current level of sensitivity of dark siren analyses, $K-$band or $B-$band luminosity weighting does not significantly affect the constraints reported above from \citet{GWTC3_expansion}, mostly because the GLADE catalog is incomplete at the distances of the GW events. In addition, the luminosity weighting of more complete catalogs currently does not have a significant impact on the $H_0$ posterior compared to the uniform weighting \citep{darksiren_DESI}. In the future, however, as the dark siren measurements precision improve, it will become crucial to estimate the impact of galaxy weighting \citep{Perna2024,hanselman2024} given realistic expectations from the major CBC formation channels.

Finally, when photometric redshifts (photo-$z$'s) are used in place of spectroscopic redshifts, it is important to account for any systematics that may arise from inaccurate redshift estimation \citep{palmese20_sts}. To account for any biases in photo-$z$'s, one may use a control sample with known redshifts to estimate the expected bias at different redshifts, to then fold such bias estimation in the $H_0$ posterior and marginalize over it. Another factor to consider is that typically individual galaxies' redshift posteriors are not Gaussian, and in galaxy surveys they are often estimated to ensure that the overall redshift distribution, rather than the single galaxies' estimates, are accurate. Similarly here, in a dark siren analysis at the current level of localization volume precision where hundreds and more often thousands of galaxies are marginalized over, it is not crucial that the single galaxies' redshift is extremely precise or accurate, but rather that the overall redshift distribution along a line of sight is.

Additional dark siren analyses have been run on existing GW events, overall finding consistent results while also constraining modified gravity models \citep{Finke_2021}.

\subsection{Spectral sirens}

\begin{figure}
    \centering
    \includegraphics[width=0.7\linewidth]{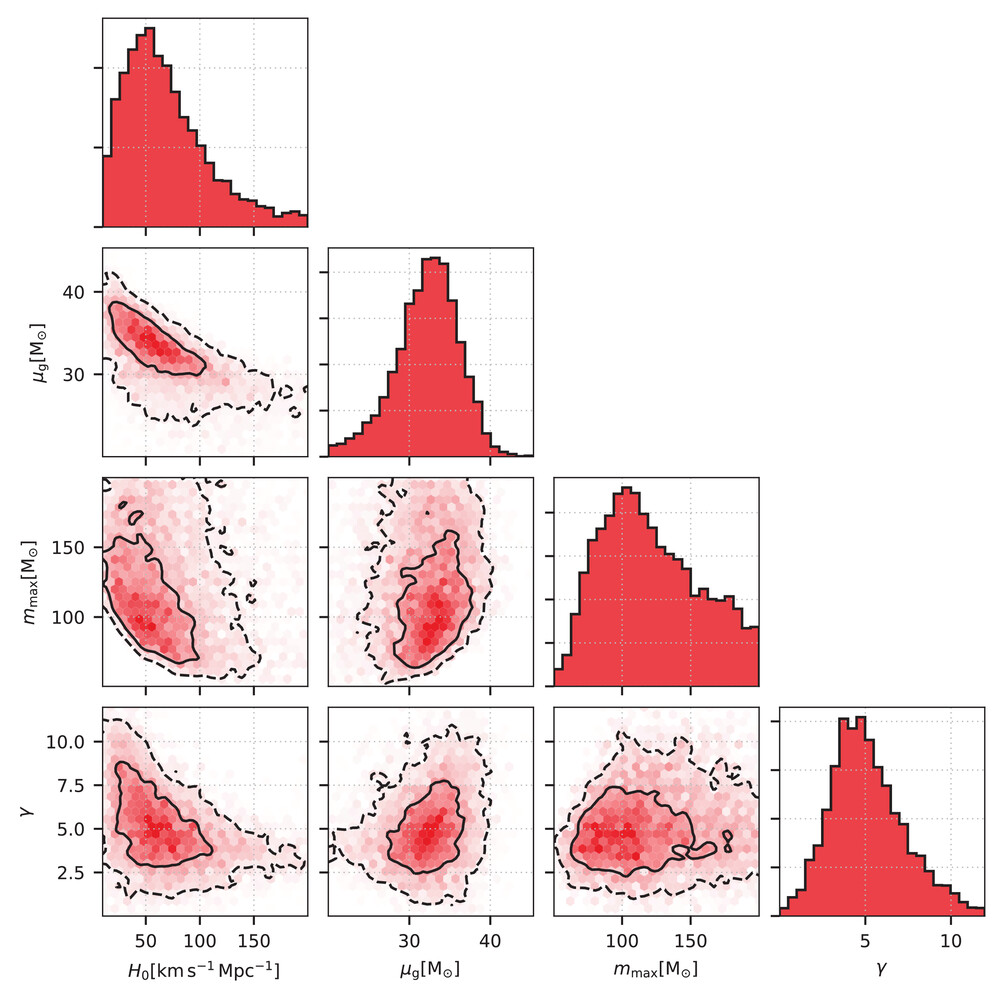}
    \caption{Spectral siren constraints on population parameters assuming a \textsc{Power Law + Peak} model and Hubble constant from \citet{GWTC3_expansion}. The population parameters included here are the mean $\mu_g$ of the Gaussian peak in the mass distribution, the BH maximum mass $m_{\rm max}$ and the rate evolution parameter $\gamma$. Figure from \citet{Mastrogiovanni:2024mqc}.}
    \label{fig:spectral_GWTC-3}
\end{figure}

As emphasized in the previous sections, the mass distribution of BBHs, and especially sharp features, has a strong dependence on cosmological parameters, and can therefore be used to probe the Hubble expansion. This was first achieved in \citet{GWTC3_expansion}, where 42 BBH mergers from GWTC-3 were used to infer simultaneously the population properties and the Hubble parameter, finding $H_0= 50^{+37}_{-30}~ \hu$ assuming the \textsc{Power Law + Peak} model. When combined with the bright siren analysis of GW170817, the constraint becomes $H_0=68^{+12}_{-8}~ \hu$.  Figure \ref{fig:spectral_GWTC-3} shows the posteriors resulting from this analysis, where $\mu_g$ represents the mean Gaussian peak in the mass distribution, $m_{\rm max}$ is the maximum allowed mass of a BH and $\gamma$ is a parameter related to the rate evolution of the CBC mergers $R$ as a function of redshift as in $R(z)\propto(1+z)^\gamma$. It is clear that the model fits for a Gaussian bump with mean around $\sim 35~M_\odot$, which sets a source mass scale that allows to fit for $H_0$. The dependence between these two parameters is clear from the figure given the degeneracy in their 2D contour plot. The maximum mass also shows degeneracy with $H_0$, but its constraints are currently broad. The current level of statistics, distance reach, and precision of GWTC-3 sources does not allow us to establish meaningful constraints on the dark energy equation of state or the Universe matter density \citep{GWTC3_expansion}. However, an analysis following a similar method can place constraints on the modification of gravity \citep{Ezquiaga:2021ayr,Mancarella_2022,Leyde:2022fsc,MaganaHernandez:2021zyc}.

Recent analyses have also reviseted the inference of $H_0$ by considering a possible evolution in redshift of the source mass spectrum. \citet{Karathanasis:2022rtr} used a parametrized approach based on the Pair Instability Supernova process, finding no evidence for an evolution of the mass spectrum and consistent $H_0$ values with the non-evolving redshift case. \citet{MaganaHernandez:2024uty} considered a non-parametric binned Gaussian Processes to model the mass distribution finds a broad $H_0$ constraint, which combined with the GW170817 bright siren measurements yields $H_0=73.0^{+13.3}_{-7.7}~ \hu$.

\subsection{Cross-correlation}

The cross-correlation method has been applied to the 8 well-localized (sky localization of 30 deg$^2$ at 68.3\% credible interval) GW sources from GWTC-3 and the photometric galaxy surveys 2MPZ and WISE-SuperCOSMOS \citep{Mukherjee:2022afz}. %The selected GW sources were GW170818, GW190412, GW190814, GW190701\_203306, GW190720\_000836, GW200129\_0065458, GW200224\_222234, and GW200311\_115853. 
The maximum angular scale set for the analysis was driven by the localization of GW sources and is set to $l \leq 30$. 
Due to the shot noise, the authors found that the cross-correlation signal between galaxies and GW sources was not measurable and that the posterior on $H_0$ was not significantly constraining. However, the posteriors obtained contain some mode in the explored $H_0$ range that when combined with GW170817 as a bright siren, results in $H_0=75^{+11}_{-6} \hu$.
The authors explored different choices for the tomographic redshift bins ($\delta z=0.05, 0.1$), multipole bins ($\delta l=5,15$) and maximum redshift $z_{\rm max}=0.5, 2$ finding that their dark siren posterior is sensitive to these choices. %The results are not significantly impacted when taking into account GW170817 and its EM counterpart. 

\section{Future prospects}
\label{sec:fut}

In less than 10 years, gravitational wave astronomy has gone from zero to hundreds of CBC detections. While the current level of precision on $H_0$ from standard siren measurements is not enough to weigh in the Hubble tension, the coming years are expected to bring an exponential growth to the number of GW sources, enabling precision cosmological measurements. Specifically, a competitive constraint from GW cosmology that could confirm consistency with either early- or late-time Universe probes needs to reach a $\sim 2\%$ precision \citep{chen17}.

\subsection{Bright standard sirens}

Forecasts show that given the current level of sensitivity of the GW detectors, about 50 bright standard sirens are needed to discern between the discrepant Hubble constant measurements \citep{2019PhRvL.122f1105F}. However, as of 2024, GW170817 is the only GW event with a confirmed EM counterpart. This is due at least in part to the BNS merger rate being on the low-end of what was previously estimated. Fifty BNS multimessenger detections are still in reach for the fifth LVK observing run (O5) \citep{kunnumkai2024detecting}, expected to start in 2027, and multimessenger detections of NSBHs (which would constitute excellent bright sirens; \citealt{2018PhRvL.121b1303V,Colombo:2023une}) may also provide a significant number of standard sirens in the near future \citep{Kunnumkai_2024_NSBH}. Note that if only the loudest events are considered, only $\sim 5$ bright sirens are sufficient to reach the precision level of the SH0ES measurements \citep{Kiendrebeogo}, however, its precision will also be dependent on peculiar velocity corrections \citep{Nimonkar:2023pyt}. It is worth to consider that these measurements occur in the local Universe, as BNS mergers and most of the EM-bright NSBH mergers are only detected out to a few hundred Mpc with current generation GW detectors. As such, standard siren measurements with these sources are mostly only sensitive to the Hubble constant rather than to other cosmological parameters that enter the Hubble parameter, so that $H_0$ can be measured in a cosmological model-agnostic manner through Eq. \eqref{eq:Htaylor}, providing an ideal probe to understand the Hubble tension independently of both early and late time Universe probes assumptions and systematics (recall that the CMB $H_0$ estimate assumes a specific cosmological model).

Some tentative associations of BBH mergers with flaring activity in Active Galactic Nuclei (AGN) have also been claimed \citep{graham20,graham23,Cabrera}. Motivated by the fact that AGN accretion disks offer a promising avenue to produce high-mass BBHs through gas accretion and hierarchical mergers \citep[e.g.][]{McKernan12}, EM emission from these events has been explored as the presence of the disk gas allows for shocks of the Hill's sphere around the kicked remnant black hole and accretion onto it \citep[e.g.][]{BartosRapid,mckernanRam}, as well as jet shock breakout, shock cooling \citep[e.g.][]{tagawa23}, and afterglow in case a relativistic jet is formed. If confirmed, BBH mergers in AGN disks offer a new avenue to measure the Hubble constant through bright standard sirens \citep{Alves}, as well as other cosmological parameters due to the larger distance reach of BBH detections compared to BNSs. It is possible that due to the intrinsically variable nature of AGNs and the large GW sky localizations, confident associations between individual GW BBH detections and AGN flares will be challenging with current generation detectors. Thus, the uncertainty of the association has to be folded into the standard siren analysis to avoid biasing the cosmology results \citep{Palmese_AGN}. Even with this additional source of uncertainty, standard siren measurements with AGN flares can produce competitive constraints $H_0$ in the near future \citep{Bom24}, provided that a significant fraction of BBHs actually occurs in AGN disks.

In the next decade, next generation (XG) ground based GW detectors such as Cosmic Explorer (CE) and Einstein Telescope (ET) are expected to reveal close to the entire population of CBCs in the Universe. A GW detector network made of one ET and at least one CE will reach a sub-percent precision on $H_0$ with hundreds of bright sirens \citep{Chen:2024gdn}. 
Even considering the Transient High Energy Sources and Early Universe Surveyor (THESEUS) \citep{2021ExA....52..183A} along with only one ET detector a $0.40~\hu$ precision on $H_0$ is expected  in five years (i.e. $\sim 166$ bright sirens at $z< 4.3$), assuming a $\Lambda$CDM model with variable curvature \citep{Califano:2022cmo}. For what concerns cosmologies allowing for different dark energy models, it may be possible to reach a 1\% precision on $H_0$ with ten years of observations \citep{Califano:2022syd} assuming a similar configuration. It is also worth noting that bright sirens with XG will reach a percent level precision in distance in the local Universe, thus enabling precision measurements of the peculiar velocity field and $\sigma_8$, the root mean square of the amplitude of matter perturbations over 8 $h^{-1}$ Mpc scale, as well as on popular modified gravity and dark energy equation of states models \citep{Mastrogiovanni:2020gua,Leyde:2022fsc,Cozzumbo:2024vxw,Afroz:2024lou}. 

\subsection{Dark standard sirens}

The promise of dark sirens with galaxy catalogs mostly relies on the localization precision of the GW detectors and on the availability of spectroscopic galaxy catalogs complete to the redshifts of interest. A percent level precision on $H_0$ may be achieved with dark sirens in O5 assuming a complete spectroscopic galaxy catalog and the most constraining 100 BBH mergers \citep{Borghi:2023opd}. 

It is important to note that the loudest events with smallest localization volumes are the ones that are expected to provide the largest constraining power to a dark siren analysis with galaxy catalogs. Indeed, the comoving volume encompassed by an event, given a prior on cosmological parameters, will scale with the number of galaxies we need to marginalize over. When a dark siren is so well localized that only one or few galaxies are contained within the localization volume, we call them ``golden dark sirens''.  Such events act effectively as (or close to) bright sirens, and are expected to enable a few percent precision on $H_0$ \citep{Borhanian}. A 2\% precision may be reached with current generation ground based detectors with golden dark sirens, although this may not be possible given that the expected number of golden dark siren for such network is only ${\cal O}(1)$ \citep{Chen:2024gdn}. For what concerns XG detectors, a sub-percent precision on $H_0$ will be possible for dark sirens with each of BNS, NSBH, and BBH sources if the network is composed of one ET and at least one CE detector. In the future, it will also be possible to use cross-correlation techniques for GW events combined with galaxy catalogs to probe the Hubble expansion (e.g. \citealt{diaz22}) modifications of gravity and dark energy equation of state \citep{Balaudo:2022znx,Mukherjee:2020mha}. At last, constraints can also be placed using BNSs as spectral sirens with XG \citep{Taylor2012}, as the number of BNS detections will grow significantly compared to current generation GW detectors, and the majority of them will be at such a distance and inclination that detecting the EM counterpart will be challenging \citep[e.g.][]{kaur2024}.

\section{Summary and outlook}\label{sec:summary}

Gravitational wave standard sirens offer a new method to probe the expansion of the Universe, in a way that is for the most part independent of other cosmological probes. As such, they offer a unique perspective on the Hubble tension. Current standard siren results presented can achieve a $\gtrsim7\%$ precision on $H_0$, and are consistent with both SH0ES and CMB $H_0$ measurements. This level of precision is expected given that they are a relatively new probe (the first standard siren measurement is only seven years old at the time of writing) and the number of detections is limited. While the number of standard sirens grows over time with new observing runs by the GW detectors, the cosmology community is exploring different avenues to improve upon available measurements and identify all possible sources of systematics. Multimessenger observations of bright standard sirens will be crucial to enable precision measurements of the Hubble constant, while GW localization precision and spectroscopic galaxy catalog availability are central for dark siren measurements with the galaxy catalog approach. Accurate measurements of the population parameters of CBCs is also essential both for spectral siren analyses and for dark siren measurements with galaxy priors.

The end of the 2020s and 2030s will be exciting times for gravitational wave cosmology. A few percent precision on the Hubble constant may be possible with standard sirens during the fifth LVK observing run, expected to start at the end of the 2020s, finally weighing into the Hubble tension business. With XG observatories we will enter the era of precision GW cosmology, and measurements of the dark energy equation of state and amplitude of fluctuations will also be achieved.

\begin{ack}[Acknowledgments]

Figures reproduced under CC BY license \url{https://creativecommons.org/licenses/by/4.0/}. AP is supported by NSF Grant No. 2308193. S.M. is supported by ERC Starting Grant No. 101163912–GravitySirens

\end{ack}

\bibliographystyle{Harvard}
\bibliography{reference}

\begin{thebibliography*}{144}
\providecommand{\bibtype}[1]{}
\providecommand{\natexlab}[1]{#1}
{\catcode`\|=0\catcode`\#=12\catcode`\@=11\catcode`\\=12
|immediate|write|@auxout{\expandafter\ifx\csname natexlab\endcsname\relax\gdef\natexlab#1{#1}\fi}}
\renewcommand{\url}[1]{{\tt #1}}
\providecommand{\urlprefix}{URL }
\expandafter\ifx\csname urlstyle\endcsname\relax
  \providecommand{\doi}[1]{doi:\discretionary{}{}{}#1}\else
  \providecommand{\doi}{doi:\discretionary{}{}{}\begingroup \urlstyle{rm}\Url}\fi
\providecommand{\bibinfo}[2]{#2}
\providecommand{\eprint}[2][]{\url{#2}}

\bibtype{Article}%
\bibitem[Abbott et al.(2016)]{LIGOScientific:2016aoc}
\bibinfo{author}{Abbott BP} and  et al. (\bibinfo{collaboration}{LIGO Scientific, Virgo}) (\bibinfo{year}{2016}).
\bibinfo{title}{{Observation of Gravitational Waves from a Binary Black Hole Merger}}.
\bibinfo{journal}{{\em Phys. Rev. Lett.}} \bibinfo{volume}{116} (\bibinfo{number}{6}): \bibinfo{pages}{061102}. \bibinfo{doi}{\doi{10.1103/PhysRevLett.116.061102}}.
\eprint{1602.03837}.

\bibtype{Article}%
\bibitem[{Abbott} et al.(2017)]{firststandardsiren}
\bibinfo{author}{{Abbott} BP}, \bibinfo{author}{{Abbott} R}, \bibinfo{author}{{Abbott} TD}, \bibinfo{author}{{Acernese} F}, \bibinfo{author}{{Ackley} K}, \bibinfo{author}{{Adams} C}, \bibinfo{author}{{Adams} T}, \bibinfo{author}{{Addesso} P}, \bibinfo{author}{{Adhikari} RX}, \bibinfo{author}{{Adya} VB} and  \bibinfo{author}{et~al.} (\bibinfo{year}{2017}), \bibinfo{month}{Nov.}
\bibinfo{title}{{A gravitational-wave standard siren measurement of the Hubble constant}}.
\bibinfo{journal}{{\em \nat}} \bibinfo{volume}{551}: \bibinfo{pages}{85--88}. \bibinfo{doi}{\doi{10.1038/nature24471}}.
\eprint{1710.05835}.

\bibtype{Article}%
\bibitem[Abbott et al.(2017{\natexlab{a}})]{ligobns}
\bibinfo{author}{Abbott BP}, \bibinfo{author}{Abbott R}, \bibinfo{author}{Abbott TD}, \bibinfo{author}{Acernese F}, \bibinfo{author}{Ackley K} and  et al. (\bibinfo{collaboration}{LIGO Scientific Collaboration and Virgo Collaboration}) (\bibinfo{year}{2017}{\natexlab{a}}), \bibinfo{month}{Oct}.
\bibinfo{title}{Gw170817: Observation of gravitational waves from a binary neutron star inspiral}.
\bibinfo{journal}{{\em Phys. Rev. Lett.}} \bibinfo{volume}{119}: \bibinfo{pages}{161101}. \bibinfo{doi}{\doi{10.1103/PhysRevLett.119.161101}}.
\bibinfo{url}{\url{https://link.aps.org/doi/10.1103/PhysRevLett.119.161101}}.

\bibtype{Article}%
\bibitem[Abbott et al.(2017{\natexlab{b}})]{LIGOScientific:2017zic}
\bibinfo{author}{Abbott BP} and  et al. (\bibinfo{collaboration}{LIGO Scientific, Virgo, Fermi-GBM, INTEGRAL}) (\bibinfo{year}{2017}{\natexlab{b}}).
\bibinfo{title}{{Gravitational Waves and Gamma-rays from a Binary Neutron Star Merger: GW170817 and GRB 170817A}}.
\bibinfo{journal}{{\em Astrophys. J. Lett.}} \bibinfo{volume}{848} (\bibinfo{number}{2}): \bibinfo{pages}{L13}. \bibinfo{doi}{\doi{10.3847/2041-8213/aa920c}}.
\eprint{1710.05834}.

\bibtype{Article}%
\bibitem[Abbott et al.(2017{\natexlab{c}})]{LIGOScientific:2017ycc}
\bibinfo{author}{Abbott BP} and  et al. (\bibinfo{collaboration}{LIGO Scientific, Virgo}) (\bibinfo{year}{2017}{\natexlab{c}}).
\bibinfo{title}{{GW170814: A Three-Detector Observation of Gravitational Waves from a Binary Black Hole Coalescence}}.
\bibinfo{journal}{{\em Phys. Rev. Lett.}} \bibinfo{volume}{119} (\bibinfo{number}{14}): \bibinfo{pages}{141101}. \bibinfo{doi}{\doi{10.1103/PhysRevLett.119.141101}}.
\eprint{1709.09660}.

\bibtype{Article}%
\bibitem[Abbott et al.(2019{\natexlab{a}})]{LIGOScientific:2018jsj}
\bibinfo{author}{Abbott BP} and  et al. (\bibinfo{collaboration}{LIGO Scientific, Virgo}) (\bibinfo{year}{2019}{\natexlab{a}}).
\bibinfo{title}{{Binary Black Hole Population Properties Inferred from the First and Second Observing Runs of Advanced LIGO and Advanced Virgo}}.
\bibinfo{journal}{{\em Astrophys. J. Lett.}} \bibinfo{volume}{882} (\bibinfo{number}{2}): \bibinfo{pages}{L24}. \bibinfo{doi}{\doi{10.3847/2041-8213/ab3800}}.
\eprint{1811.12940}.

\bibtype{Article}%
\bibitem[Abbott et al.(2019{\natexlab{b}})]{LIGOScientific:2018hze}
\bibinfo{author}{Abbott BP} and  et al. (\bibinfo{collaboration}{LIGO Scientific, Virgo}) (\bibinfo{year}{2019}{\natexlab{b}}).
\bibinfo{title}{{Properties of the binary neutron star merger GW170817}}.
\bibinfo{journal}{{\em Phys. Rev. X}} \bibinfo{volume}{9} (\bibinfo{number}{1}): \bibinfo{pages}{011001}. \bibinfo{doi}{\doi{10.1103/PhysRevX.9.011001}}.
\eprint{1805.11579}.

\bibtype{Article}%
\bibitem[Abbott et al.(2020{\natexlab{a}})]{Abbott_2020_GW190425}
\bibinfo{author}{Abbott BP} and  et al. (\bibinfo{year}{2020}{\natexlab{a}}), \bibinfo{month}{Mar.}
\bibinfo{title}{Gw190425: Observation of a compact binary coalescence with total mass 3.4 m}.
\bibinfo{journal}{{\em The Astrophysical Journal Letters}} \bibinfo{volume}{892} (\bibinfo{number}{1}): \bibinfo{pages}{L3}.
ISSN \bibinfo{issn}{2041-8213}. \bibinfo{doi}{\doi{10.3847/2041-8213/ab75f5}}.
\bibinfo{url}{\url{http://dx.doi.org/10.3847/2041-8213/ab75f5}}.

\bibtype{Article}%
\bibitem[Abbott et al.(2020{\natexlab{b}})]{LIGOScientific:2020zkf}
\bibinfo{author}{Abbott R} and  et al. (\bibinfo{collaboration}{LIGO Scientific, Virgo}) (\bibinfo{year}{2020}{\natexlab{b}}).
\bibinfo{title}{{GW190814: Gravitational Waves from the Coalescence of a 23 Solar Mass Black Hole with a 2.6 Solar Mass Compact Object}}.
\bibinfo{journal}{{\em Astrophys. J. Lett.}} \bibinfo{volume}{896} (\bibinfo{number}{2}): \bibinfo{pages}{L44}. \bibinfo{doi}{\doi{10.3847/2041-8213/ab960f}}.
\eprint{2006.12611}.

\bibtype{Article}%
\bibitem[Abbott et al.(2021{\natexlab{a}})]{LIGOScientific:2019zcs}
\bibinfo{author}{Abbott BP} and  et al. (\bibinfo{collaboration}{LIGO Scientific, Virgo, VIRGO}) (\bibinfo{year}{2021}{\natexlab{a}}).
\bibinfo{title}{{A Gravitational-wave Measurement of the Hubble Constant Following the Second Observing Run of Advanced LIGO and Virgo}}.
\bibinfo{journal}{{\em Astrophys. J.}} \bibinfo{volume}{909} (\bibinfo{number}{2}): \bibinfo{pages}{218}. \bibinfo{doi}{\doi{10.3847/1538-4357/abdcb7}}.
\eprint{1908.06060}.

\bibtype{Article}%
\bibitem[Abbott et al.(2021{\natexlab{b}})]{LIGOScientific:2020kqk}
\bibinfo{author}{Abbott R} and  et al. (\bibinfo{collaboration}{LIGO Scientific, Virgo}) (\bibinfo{year}{2021}{\natexlab{b}}).
\bibinfo{title}{{Population Properties of Compact Objects from the Second LIGO-Virgo Gravitational-Wave Transient Catalog}}.
\bibinfo{journal}{{\em Astrophys. J. Lett.}} \bibinfo{volume}{913} (\bibinfo{number}{1}): \bibinfo{pages}{L7}. \bibinfo{doi}{\doi{10.3847/2041-8213/abe949}}.
\eprint{2010.14533}.

\bibtype{Article}%
\bibitem[Abbott et al.(2023{\natexlab{a}})]{GWTC_3}
\bibinfo{author}{Abbott R}, \bibinfo{author}{Abbott TD}, \bibinfo{author}{Acernese F}, \bibinfo{author}{Ackley K}, \bibinfo{author}{Adams C}, \bibinfo{author}{Adhikari N}, \bibinfo{author}{Adhikari RX}, \bibinfo{author}{Adya VB}, \bibinfo{author}{Affeldt C}, \bibinfo{author}{Agarwal D}, \bibinfo{author}{Agathos M}, \bibinfo{author}{Agatsuma K}, \bibinfo{author}{Aggarwal N}, \bibinfo{author}{Aguiar OD}, \bibinfo{author}{Aiello L}, \bibinfo{author}{Ain A}, \bibinfo{author}{Ajith P}, \bibinfo{author}{Akcay S}, \bibinfo{author}{Akutsu T}, \bibinfo{author}{Albanesi S}, \bibinfo{author}{Allocca A}, \bibinfo{author}{Altin PA}, \bibinfo{author}{Amato A}, \bibinfo{author}{Anand C}, \bibinfo{author}{Anand S}, \bibinfo{author}{Ananyeva A}, \bibinfo{author}{Anderson SB} and  \bibinfo{author}{Others} (\bibinfo{collaboration}{LIGO Scientific Collaboration, Virgo Collaboration, and KAGRA Collaboration}) (\bibinfo{year}{2023}{\natexlab{a}}), \bibinfo{month}{Dec}.
\bibinfo{title}{Gwtc-3: Compact binary coalescences observed by ligo and virgo during the second part of the third observing run}.
\bibinfo{journal}{{\em Phys. Rev. X}} \bibinfo{volume}{13}: \bibinfo{pages}{041039}. \bibinfo{doi}{\doi{10.1103/PhysRevX.13.041039}}.
\bibinfo{url}{\url{https://link.aps.org/doi/10.1103/PhysRevX.13.041039}}.

\bibtype{Article}%
\bibitem[Abbott et al.(2023{\natexlab{b}})]{GWTC3_expansion}
\bibinfo{author}{Abbott R} and  et al. (\bibinfo{collaboration}{LIGO Scientific, Virgo, KAGRA}) (\bibinfo{year}{2023}{\natexlab{b}}).
\bibinfo{title}{{Constraints on the Cosmic Expansion History from GWTC\textendash{}3}}.
\bibinfo{journal}{{\em Astrophys. J.}} \bibinfo{volume}{949} (\bibinfo{number}{2}): \bibinfo{pages}{76}. \bibinfo{doi}{\doi{10.3847/1538-4357/ac74bb}}.
\eprint{2111.03604}.

\bibtype{Article}%
\bibitem[Abbott et al.(2023{\natexlab{c}})]{KAGRA:2021duu}
\bibinfo{author}{Abbott R} and  et al. (\bibinfo{collaboration}{KAGRA, VIRGO, LIGO Scientific}) (\bibinfo{year}{2023}{\natexlab{c}}).
\bibinfo{title}{{Population of Merging Compact Binaries Inferred Using Gravitational Waves through GWTC-3}}.
\bibinfo{journal}{{\em Phys. Rev. X}} \bibinfo{volume}{13} (\bibinfo{number}{1}): \bibinfo{pages}{011048}. \bibinfo{doi}{\doi{10.1103/PhysRevX.13.011048}}.
\eprint{2111.03634}.

\bibtype{Article}%
\bibitem[Afroz and Mukherjee(2024{\natexlab{a}})]{Afroz:2024lou}
\bibinfo{author}{Afroz S} and  \bibinfo{author}{Mukherjee S} (\bibinfo{year}{2024}{\natexlab{a}}), \bibinfo{month}{12}.
\bibinfo{title}{{Multi-Messenger Cosmology: A Route to Accurate Inference of Dark Energy Beyond CPL Parametrization from XG Detectors}} \eprint{2412.12285}.

\bibtype{Article}%
\bibitem[Afroz and Mukherjee(2024{\natexlab{b}})]{Afroz:2024joi}
\bibinfo{author}{Afroz S} and  \bibinfo{author}{Mukherjee S} (\bibinfo{year}{2024}{\natexlab{b}}).
\bibinfo{title}{{Prospect of precision cosmology and testing general relativity using binary black holes \textendash{} galaxies cross-correlation}}.
\bibinfo{journal}{{\em Mon. Not. Roy. Astron. Soc.}} \bibinfo{volume}{534} (\bibinfo{number}{2}): \bibinfo{pages}{1283--1298}. \bibinfo{doi}{\doi{10.1093/mnras/stae2139}}.
\eprint{2407.09262}.

\bibtype{Article}%
\bibitem[Agarwal et al.(2024)]{Agarwal:2024hld}
\bibinfo{author}{Agarwal A} and  et al. (\bibinfo{year}{2024}), \bibinfo{month}{12}.
\bibinfo{title}{{Blinded Mock Data Challenge for Gravitational-Wave Cosmology-I: Assessing the Robustness of Methods Using Binary Black Holes Mass Spectrum}} \eprint{2412.14244}.

\bibtype{Article}%
\bibitem[{Alfradique} et al.(2024)]{Alfradique24}
\bibinfo{author}{{Alfradique} V}, \bibinfo{author}{{Bom} CR}, \bibinfo{author}{{Palmese} A}, \bibinfo{author}{{Teixeira} G}, \bibinfo{author}{{Santana-Silva} L}, \bibinfo{author}{{Drlica-Wagner} A}, \bibinfo{author}{{Riley} AH}, \bibinfo{author}{{Mart{\'\i}nez-V{\'a}zquez} CE}, \bibinfo{author}{{Sand} DJ}, \bibinfo{author}{{Stringfellow} GS}, \bibinfo{author}{{Medina} GE}, \bibinfo{author}{{Carballo-Bello} JA}, \bibinfo{author}{{Choi} Y}, \bibinfo{author}{{Esteves} J}, \bibinfo{author}{{Limberg} G}, \bibinfo{author}{{Mutlu-Pakdil} B}, \bibinfo{author}{{No{\"e}l} NED}, \bibinfo{author}{{Pace} AB}, \bibinfo{author}{{Sakowska} JD} and  \bibinfo{author}{{Wu} JF} (\bibinfo{year}{2024}), \bibinfo{month}{Feb.}
\bibinfo{title}{{A dark siren measurement of the Hubble constant using gravitational wave events from the first three LIGO/Virgo observing runs and DELVE}}.
\bibinfo{journal}{{\em \mnras}} \bibinfo{volume}{528} (\bibinfo{number}{2}): \bibinfo{pages}{3249--3259}. \bibinfo{doi}{\doi{10.1093/mnras/stae086}}.
\eprint{2310.13695}.

\bibtype{Article}%
\bibitem[{Alves} et al.(2024)]{Alves}
\bibinfo{author}{{Alves} LMB}, \bibinfo{author}{{Sullivan} AG}, \bibinfo{author}{{Yang} Y}, \bibinfo{author}{{Gayathri} V}, \bibinfo{author}{{M{\'a}rka} Z}, \bibinfo{author}{{M{\'a}rka} S} and  \bibinfo{author}{{Bartos} I} (\bibinfo{year}{2024}), \bibinfo{month}{Jul.}
\bibinfo{title}{{Determining the Hubble constant with AGN-assisted black hole mergers}}.
\bibinfo{journal}{{\em \mnras}} \bibinfo{volume}{531} (\bibinfo{number}{3}): \bibinfo{pages}{3679--3683}. \bibinfo{doi}{\doi{10.1093/mnras/stae1360}}.
\eprint{2009.13739}.

\bibtype{Article}%
\bibitem[{Amati} et al.(2021)]{2021ExA....52..183A}
\bibinfo{author}{{Amati} L}, \bibinfo{author}{{O'Brien} PT}, \bibinfo{author}{{G{\"o}tz} D}, \bibinfo{author}{{Bozzo} E}, \bibinfo{author}{{Santangelo} A}, \bibinfo{author}{{Tanvir} N}, \bibinfo{author}{{Frontera} F}, \bibinfo{author}{{Mereghetti} S}, \bibinfo{author}{{Osborne} JP}, \bibinfo{author}{{Blain} A}, \bibinfo{author}{{Basa} S}, \bibinfo{author}{{Branchesi} M}, \bibinfo{author}{{Burderi} L}, \bibinfo{author}{{Caballero-Garc{\'\i}a} M}, \bibinfo{author}{{Castro-Tirado} AJ}, \bibinfo{author}{{Christensen} L}, \bibinfo{author}{{Ciolfi} R}, \bibinfo{author}{{De Rosa} A}, \bibinfo{author}{{Doroshenko} V}, \bibinfo{author}{{Ferrara} A}, \bibinfo{author}{{Ghirlanda} G}, \bibinfo{author}{{Hanlon} L}, \bibinfo{author}{{Heddermann} P}, \bibinfo{author}{{Hutchinson} I}, \bibinfo{author}{{Labanti} C}, \bibinfo{author}{{Le Floch} E}, \bibinfo{author}{{Lerman} H}, \bibinfo{author}{{Paltani} S}, \bibinfo{author}{{Reglero} V}, \bibinfo{author}{{Rezzolla} L}, \bibinfo{author}{{Rosati} P},
  \bibinfo{author}{{Salvaterra} R}, \bibinfo{author}{{Stratta} G}, \bibinfo{author}{{Tenzer} C} and  \bibinfo{author}{{Theseus Consortium}} (\bibinfo{year}{2021}), \bibinfo{month}{Dec.}
\bibinfo{title}{{The THESEUS space mission: science goals, requirements and mission concept}}.
\bibinfo{journal}{{\em Experimental Astronomy}} \bibinfo{volume}{52} (\bibinfo{number}{3}): \bibinfo{pages}{183--218}. \bibinfo{doi}{\doi{10.1007/s10686-021-09807-8}}.
\eprint{2104.09531}.

\bibtype{Article}%
\bibitem[{Arcavi} et al.(2017)]{Arcavi_2017}
\bibinfo{author}{{Arcavi} I}, \bibinfo{author}{{Hosseinzadeh} G}, \bibinfo{author}{{Howell} DA}, \bibinfo{author}{{McCully} C}, \bibinfo{author}{{Poznanski} D}, \bibinfo{author}{{Kasen} D}, \bibinfo{author}{{Barnes} J}, \bibinfo{author}{{Zaltzman} M}, \bibinfo{author}{{Vasylyev} S}, \bibinfo{author}{{Maoz} D} and  \bibinfo{author}{{Valenti} S} (\bibinfo{year}{2017}), \bibinfo{month}{Nov.}
\bibinfo{title}{{Optical emission from a kilonova following a gravitational-wave-detected neutron-star merger}}.
\bibinfo{journal}{{\em \nat}} \bibinfo{volume}{551} (\bibinfo{number}{7678}): \bibinfo{pages}{64--66}. \bibinfo{doi}{\doi{10.1038/nature24291}}.
\eprint{1710.05843}.

\bibtype{Article}%
\bibitem[Balaudo et al.(2023)]{Balaudo:2022znx}
\bibinfo{author}{Balaudo A}, \bibinfo{author}{Garoffolo A}, \bibinfo{author}{Martinelli M}, \bibinfo{author}{Mukherjee S} and  \bibinfo{author}{Silvestri A} (\bibinfo{year}{2023}).
\bibinfo{title}{{Prospects of testing late-time cosmology with weak lensing of gravitational waves and galaxy surveys}}.
\bibinfo{journal}{{\em JCAP}} \bibinfo{volume}{06}: \bibinfo{pages}{050}. \bibinfo{doi}{\doi{10.1088/1475-7516/2023/06/050}}.
\eprint{2210.06398}.

\bibtype{Article}%
\bibitem[{Ballard} et al.(2023)]{ballard}
\bibinfo{author}{{Ballard} W}, \bibinfo{author}{{Palmese} A}, \bibinfo{author}{{Hernandez} IM}, \bibinfo{author}{{BenZvi} S}, \bibinfo{author}{{Moon} J}, \bibinfo{author}{{Ross} AJ}, \bibinfo{author}{{Rossi} G}, \bibinfo{author}{{Aguilar} J}, \bibinfo{author}{{Ahlen} S}, \bibinfo{author}{{Blum} R}, \bibinfo{author}{{Brooks} D}, \bibinfo{author}{{Claybaugh} T}, \bibinfo{author}{{de la Macorra} A}, \bibinfo{author}{{Dey} A}, \bibinfo{author}{{Doel} P}, \bibinfo{author}{{Forero-Romero} JE}, \bibinfo{author}{{Gontcho A Gontcho} S}, \bibinfo{author}{{Honscheid} K}, \bibinfo{author}{{Kremin} A}, \bibinfo{author}{{Manera} M}, \bibinfo{author}{{Meisner} A}, \bibinfo{author}{{Miquel} R}, \bibinfo{author}{{Moustakas} J}, \bibinfo{author}{{Prada} F}, \bibinfo{author}{{Sanchez} E}, \bibinfo{author}{{Tarl{\'e}} G}, \bibinfo{author}{{Zhou} Z} and  \bibinfo{author}{{DESI Collaboration}} (\bibinfo{year}{2023}), \bibinfo{month}{Nov.}
\bibinfo{title}{{A Dark Siren Measurement of the Hubble Constant with the LIGO/Virgo Gravitational Wave Event GW190412 and DESI Galaxies}}.
\bibinfo{journal}{{\em Research Notes of the American Astronomical Society}} \bibinfo{volume}{7} (\bibinfo{number}{11}), \bibinfo{eid}{250}. \bibinfo{doi}{\doi{10.3847/2515-5172/ad0eda}}.
\eprint{2311.13062}.

\bibtype{Article}%
\bibitem[{Bartos} et al.(2017)]{BartosRapid}
\bibinfo{author}{{Bartos} I}, \bibinfo{author}{{Kocsis} B}, \bibinfo{author}{{Haiman} Z} and  \bibinfo{author}{{M{\'a}rka} S} (\bibinfo{year}{2017}), \bibinfo{month}{Feb.}
\bibinfo{title}{{Rapid and Bright Stellar-mass Binary Black Hole Mergers in Active Galactic Nuclei}}.
\bibinfo{journal}{{\em \apj}} \bibinfo{volume}{835} (\bibinfo{number}{2}), \bibinfo{eid}{165}. \bibinfo{doi}{\doi{10.3847/1538-4357/835/2/165}}.
\eprint{1602.03831}.

\bibtype{Article}%
\bibitem[Bera et al.(2020)]{Bera:2020jhx}
\bibinfo{author}{Bera S}, \bibinfo{author}{Rana D}, \bibinfo{author}{More S} and  \bibinfo{author}{Bose S} (\bibinfo{year}{2020}).
\bibinfo{title}{{Incompleteness Matters Not: Inference of $H_0$ from Binary Black Hole\textendash{}Galaxy Cross-correlations}}.
\bibinfo{journal}{{\em Astrophys. J.}} \bibinfo{volume}{902} (\bibinfo{number}{1}): \bibinfo{pages}{79}. \bibinfo{doi}{\doi{10.3847/1538-4357/abb4e0}}.
\eprint{2007.04271}.

\bibtype{Article}%
\bibitem[{Blanchard} et al.(2017)]{Blanchard}
\bibinfo{author}{{Blanchard} PK} and  et al. (\bibinfo{year}{2017}), \bibinfo{month}{Oct.}
\bibinfo{title}{{The Electromagnetic Counterpart of the Binary Neutron Star Merger LIGO/Virgo GW170817. VII. Properties of the Host Galaxy and Constraints on the Merger Timescale}}.
\bibinfo{journal}{{\em \apjl}} \bibinfo{volume}{848} (\bibinfo{number}{2}), \bibinfo{eid}{L22}. \bibinfo{doi}{\doi{10.3847/2041-8213/aa9055}}.
\eprint{1710.05458}.

\bibtype{Article}%
\bibitem[{Bom} and {Palmese}(2024)]{Bom24}
\bibinfo{author}{{Bom} CR} and  \bibinfo{author}{{Palmese} A} (\bibinfo{year}{2024}), \bibinfo{month}{Oct.}
\bibinfo{title}{{Standard siren cosmology with gravitational waves from binary black hole mergers in active galactic nuclei}}.
\bibinfo{journal}{{\em \prd}} \bibinfo{volume}{110} (\bibinfo{number}{8}), \bibinfo{eid}{083005}. \bibinfo{doi}{\doi{10.1103/PhysRevD.110.083005}}.
\eprint{2307.01330}.

\bibtype{Article}%
\bibitem[{Bom} et al.(2024)]{Bom24_darksiren}
\bibinfo{author}{{Bom} CR}, \bibinfo{author}{{Alfradique} V}, \bibinfo{author}{{Palmese} A}, \bibinfo{author}{{Teixeira} G}, \bibinfo{author}{{Santana-Silva} L}, \bibinfo{author}{{Santos} A} and  \bibinfo{author}{{Darc} P} (\bibinfo{year}{2024}), \bibinfo{month}{Nov.}
\bibinfo{title}{{A dark standard siren measurement of the Hubble constant following LIGO/Virgo/KAGRA O4a and previous runs}}.
\bibinfo{journal}{{\em \mnras}} \bibinfo{volume}{535} (\bibinfo{number}{1}): \bibinfo{pages}{961--975}. \bibinfo{doi}{\doi{10.1093/mnras/stae2390}}.
\eprint{2404.16092}.

\bibtype{Article}%
\bibitem[Borghi et al.(2024)]{Borghi:2023opd}
\bibinfo{author}{Borghi N}, \bibinfo{author}{Mancarella M}, \bibinfo{author}{Moresco M}, \bibinfo{author}{Tagliazucchi M}, \bibinfo{author}{Iacovelli F}, \bibinfo{author}{Cimatti A} and  \bibinfo{author}{Maggiore M} (\bibinfo{year}{2024}).
\bibinfo{title}{{Cosmology and Astrophysics with Standard Sirens and Galaxy Catalogs in View of Future Gravitational Wave Observations}}.
\bibinfo{journal}{{\em Astrophys. J.}} \bibinfo{volume}{964} (\bibinfo{number}{2}): \bibinfo{pages}{191}. \bibinfo{doi}{\doi{10.3847/1538-4357/ad20eb}}.
\eprint{2312.05302}.

\bibtype{Article}%
\bibitem[{Borhanian} et al.(2020)]{Borhanian}
\bibinfo{author}{{Borhanian} S}, \bibinfo{author}{{Dhani} A}, \bibinfo{author}{{Gupta} A}, \bibinfo{author}{{Arun} KG} and  \bibinfo{author}{{Sathyaprakash} BS} (\bibinfo{year}{2020}), \bibinfo{month}{Jul.}
\bibinfo{title}{{Dark Sirens to Resolve the Hubble-Lema{\^\i}tre Tension}}.
\bibinfo{journal}{{\em arXiv e-prints}} , \bibinfo{eid}{arXiv:2007.02883}\bibinfo{doi}{\doi{10.48550/arXiv.2007.02883}}.
\eprint{2007.02883}.

\bibtype{Article}%
\bibitem[{Cabrera} et al.(2024)]{Cabrera}
\bibinfo{author}{{Cabrera} T}, \bibinfo{author}{{Palmese} A}, \bibinfo{author}{{Hu} L}, \bibinfo{author}{{O'Connor} B}, \bibinfo{author}{{Ford} KES}, \bibinfo{author}{{McKernan} B}, \bibinfo{author}{{Andreoni} I}, \bibinfo{author}{{Ahumada} T}, \bibinfo{author}{{Amsellem} A}, \bibinfo{author}{{Busmann} M}, \bibinfo{author}{{Clark} P}, \bibinfo{author}{{Coughlin} MW}, \bibinfo{author}{{Dadiani} E}, \bibinfo{author}{{Diaz} V}, \bibinfo{author}{{Graham} MJ}, \bibinfo{author}{{Gruen} D}, \bibinfo{author}{{Kunnumkai} K}, \bibinfo{author}{{Postiglione} J}, \bibinfo{author}{{Sommer} JS} and  \bibinfo{author}{{Valdes} F} (\bibinfo{year}{2024}), \bibinfo{month}{Jul.}
\bibinfo{title}{{Searching for electromagnetic emission in an AGN from the gravitational wave binary black hole merger candidate S230922g}}.
\bibinfo{journal}{{\em \prd}} , \bibinfo{eid}{arXiv:2407.10698}\bibinfo{doi}{\doi{10.48550/arXiv.2407.10698}}.
\eprint{2407.10698}.

\bibtype{Article}%
\bibitem[Calder\'on~Bustillo et al.(2021)]{CalderonBustillo:2020kcg}
\bibinfo{author}{Calder\'on~Bustillo J}, \bibinfo{author}{Leong SHW}, \bibinfo{author}{Dietrich T} and  \bibinfo{author}{Lasky PD} (\bibinfo{year}{2021}).
\bibinfo{title}{{Mapping the Universe Expansion: Enabling Percent-level Measurements of the Hubble Constant with a Single Binary Neutron-star Merger Detection}}.
\bibinfo{journal}{{\em Astrophys. J. Lett.}} \bibinfo{volume}{912} (\bibinfo{number}{1}): \bibinfo{pages}{L10}. \bibinfo{doi}{\doi{10.3847/2041-8213/abf502}}.
\eprint{2006.11525}.

\bibtype{Article}%
\bibitem[Califano et al.(2023{\natexlab{a}})]{Califano:2022cmo}
\bibinfo{author}{Califano M}, \bibinfo{author}{de~Martino I}, \bibinfo{author}{Vernieri D} and  \bibinfo{author}{Capozziello S} (\bibinfo{year}{2023}{\natexlab{a}}).
\bibinfo{title}{{Constraining $\Lambda$CDM cosmological parameters with Einstein Telescope mock data}}.
\bibinfo{journal}{{\em Mon. Not. Roy. Astron. Soc.}} \bibinfo{volume}{518}: \bibinfo{pages}{3372--3385}. \bibinfo{doi}{\doi{10.1093/mnras/stac3230}}.
\eprint{2205.11221}.

\bibtype{Article}%
\bibitem[Califano et al.(2023{\natexlab{b}})]{Califano:2022syd}
\bibinfo{author}{Califano M}, \bibinfo{author}{de~Martino I}, \bibinfo{author}{Vernieri D} and  \bibinfo{author}{Capozziello S} (\bibinfo{year}{2023}{\natexlab{b}}).
\bibinfo{title}{{Exploiting the Einstein Telescope to solve the Hubble tension}}.
\bibinfo{journal}{{\em Phys. Rev. D}} \bibinfo{volume}{107} (\bibinfo{number}{12}): \bibinfo{pages}{123519}. \bibinfo{doi}{\doi{10.1103/PhysRevD.107.123519}}.
\eprint{2208.13999}.

\bibtype{Article}%
\bibitem[Callister(2024)]{Callister:2024cdx}
\bibinfo{author}{Callister TA} (\bibinfo{year}{2024}), \bibinfo{month}{10}.
\bibinfo{title}{{Observed Gravitational-Wave Populations}} \eprint{2410.19145}.

\bibtype{Article}%
\bibitem[Chatterjee et al.(2021)]{Chatterjee:2021xrm}
\bibinfo{author}{Chatterjee D}, \bibinfo{author}{Hegade K~R A}, \bibinfo{author}{Holder G}, \bibinfo{author}{Holz DE}, \bibinfo{author}{Perkins S}, \bibinfo{author}{Yagi K} and  \bibinfo{author}{Yunes N} (\bibinfo{year}{2021}).
\bibinfo{title}{{Cosmology with Love: Measuring the Hubble constant using neutron star universal relations}}.
\bibinfo{journal}{{\em Phys. Rev. D}} \bibinfo{volume}{104} (\bibinfo{number}{8}): \bibinfo{pages}{083528}. \bibinfo{doi}{\doi{10.1103/PhysRevD.104.083528}}.
\eprint{2106.06589}.

\bibtype{Article}%
\bibitem[Chen(2020)]{Chen:2020dyt}
\bibinfo{author}{Chen HY} (\bibinfo{year}{2020}).
\bibinfo{title}{{Systematic Uncertainty of Standard Sirens from the Viewing Angle of Binary Neutron Star Inspirals}}.
\bibinfo{journal}{{\em Phys. Rev. Lett.}} \bibinfo{volume}{125} (\bibinfo{number}{20}): \bibinfo{pages}{201301}. \bibinfo{doi}{\doi{10.1103/PhysRevLett.125.201301}}.
\eprint{2006.02779}.

\bibtype{Article}%
\bibitem[{Chen} et al.(2018)]{chen17}
\bibinfo{author}{{Chen} HY}, \bibinfo{author}{{Fishbach} M} and  \bibinfo{author}{{Holz} DE} (\bibinfo{year}{2018}), \bibinfo{month}{Oct}.
\bibinfo{title}{{A two per cent Hubble constant measurement from standard sirens within five years}}.
\bibinfo{journal}{{\em \nat}} \bibinfo{volume}{562}: \bibinfo{pages}{545--547}. \bibinfo{doi}{\doi{10.1038/s41586-018-0606-0}}.
\eprint{1712.06531}.

\bibtype{Article}%
\bibitem[Chen et al.(2024{\natexlab{a}})]{Chen:2024gdn}
\bibinfo{author}{Chen HY}, \bibinfo{author}{Ezquiaga JM} and  \bibinfo{author}{Gupta I} (\bibinfo{year}{2024}{\natexlab{a}}).
\bibinfo{title}{{Cosmography with next-generation gravitational wave detectors}}.
\bibinfo{journal}{{\em Class. Quant. Grav.}} \bibinfo{volume}{41} (\bibinfo{number}{12}): \bibinfo{pages}{125004}. \bibinfo{doi}{\doi{10.1088/1361-6382/ad424f}}.
\eprint{2402.03120}.

\bibtype{Article}%
\bibitem[Chen et al.(2024{\natexlab{b}})]{Chen:2023dgw}
\bibinfo{author}{Chen HY}, \bibinfo{author}{Talbot C} and  \bibinfo{author}{Chase EA} (\bibinfo{year}{2024}{\natexlab{b}}).
\bibinfo{title}{{Mitigating the Counterpart Selection Effect for Standard Sirens}}.
\bibinfo{journal}{{\em Phys. Rev. Lett.}} \bibinfo{volume}{132} (\bibinfo{number}{19}): \bibinfo{pages}{191003}. \bibinfo{doi}{\doi{10.1103/PhysRevLett.132.191003}}.
\eprint{2307.10402}.

\bibtype{Article}%
\bibitem[{Cigarr{\'a}n D{\'\i}az} and {Mukherjee}(2022)]{diaz22}
\bibinfo{author}{{Cigarr{\'a}n D{\'\i}az} C} and  \bibinfo{author}{{Mukherjee} S} (\bibinfo{year}{2022}), \bibinfo{month}{Apr.}
\bibinfo{title}{{Mapping the cosmic expansion history from LIGO-Virgo-KAGRA in synergy with DESI and SPHEREx}}.
\bibinfo{journal}{{\em \mnras}} \bibinfo{volume}{511} (\bibinfo{number}{2}): \bibinfo{pages}{2782--2795}. \bibinfo{doi}{\doi{10.1093/mnras/stac208}}.
\eprint{2107.12787}.

\bibtype{Article}%
\bibitem[Colombo et al.(2024)]{Colombo:2023une}
\bibinfo{author}{Colombo A} and  et al. (\bibinfo{year}{2024}).
\bibinfo{title}{{Multi-messenger prospects for black hole - neutron star mergers in the O4 and O5 runs}}.
\bibinfo{journal}{{\em Astron. Astrophys.}} \bibinfo{volume}{686}: \bibinfo{pages}{A265}. \bibinfo{doi}{\doi{10.1051/0004-6361/202348384}}.
\eprint{2310.16894}.

\bibtype{Article}%
\bibitem[{Coulter} et al.(2017)]{Coulter_2017}
\bibinfo{author}{{Coulter} DA}, \bibinfo{author}{{Foley} RJ}, \bibinfo{author}{{Kilpatrick} CD}, \bibinfo{author}{{Drout} MR}, \bibinfo{author}{{Piro} AL}, \bibinfo{author}{{Shappee} BJ}, \bibinfo{author}{{Siebert} MR}, \bibinfo{author}{{Simon} JD}, \bibinfo{author}{{Ulloa} N}, \bibinfo{author}{{Kasen} D}, \bibinfo{author}{{Madore} BF}, \bibinfo{author}{{Murguia-Berthier} A}, \bibinfo{author}{{Pan} YC}, \bibinfo{author}{{Prochaska} JX}, \bibinfo{author}{{Ramirez-Ruiz} E}, \bibinfo{author}{{Rest} A} and  \bibinfo{author}{{Rojas-Bravo} C} (\bibinfo{year}{2017}), \bibinfo{month}{Dec.}
\bibinfo{title}{{Swope Supernova Survey 2017a (SSS17a), the optical counterpart to a gravitational wave source}}.
\bibinfo{journal}{{\em Science}} \bibinfo{volume}{358} (\bibinfo{number}{6370}): \bibinfo{pages}{1556--1558}. \bibinfo{doi}{\doi{10.1126/science.aap9811}}.
\eprint{1710.05452}.

\bibtype{Article}%
\bibitem[Cozzumbo et al.(2024)]{Cozzumbo:2024vxw}
\bibinfo{author}{Cozzumbo A}, \bibinfo{author}{Dupletsa U}, \bibinfo{author}{Calder\'on R}, \bibinfo{author}{Murgia R}, \bibinfo{author}{Oganesyan G} and  \bibinfo{author}{Branchesi M} (\bibinfo{year}{2024}), \bibinfo{month}{11}.
\bibinfo{title}{{Model-independent cosmology with joint observations of gravitational waves and $\gamma$-ray bursts}} \eprint{2411.02490}.

\bibtype{Article}%
\bibitem[{D{\'a}lya} et al.(2018)]{2018MNRAS.479.2374D}
\bibinfo{author}{{D{\'a}lya} G}, \bibinfo{author}{{Galg{\'o}czi} G}, \bibinfo{author}{{Dobos} L}, \bibinfo{author}{{Frei} Z}, \bibinfo{author}{{Henƒg} IS}, \bibinfo{author}{{Macas} R}, \bibinfo{author}{{Messenger} C}, \bibinfo{author}{{Raffai} P} and  \bibinfo{author}{{de Souza} RS} (\bibinfo{year}{2018}), \bibinfo{month}{Sep.}
\bibinfo{title}{{GLADE: A galaxy catalogue for multimessenger searches in the advanced gravitational-wave detector era}}.
\bibinfo{journal}{{\em \mnras}} \bibinfo{volume}{479} (\bibinfo{number}{2}): \bibinfo{pages}{2374--2381}. \bibinfo{doi}{\doi{10.1093/mnras/sty1703}}.
\eprint{1804.05709}.

\bibtype{Article}%
\bibitem[{Dark Energy Survey Collaboration}(2016)]{DESoverview}
\bibinfo{author}{{Dark Energy Survey Collaboration}} (\bibinfo{year}{2016}), \bibinfo{month}{Aug.}
\bibinfo{title}{{The Dark Energy Survey: more than dark energy - an overview}}.
\bibinfo{journal}{{\em \mnras}} \bibinfo{volume}{460} (\bibinfo{number}{2}): \bibinfo{pages}{1270--1299}. \bibinfo{doi}{\doi{10.1093/mnras/stw641}}.
\eprint{1601.00329}.

\bibtype{Article}%
\bibitem[Dehghani et al.(2024)]{Dehghani:2024wsh}
\bibinfo{author}{Dehghani A}, \bibinfo{author}{Kim JL}, \bibinfo{author}{Hosseini DS}, \bibinfo{author}{Krolewski A}, \bibinfo{author}{Mukherjee S} and  \bibinfo{author}{Geshnizjani G} (\bibinfo{year}{2024}), \bibinfo{month}{11}.
\bibinfo{title}{{The Gravitational Wave Bias Parameter from Angular Power Spectra: Bridging Between Galaxies and Binary Black Holes}} \eprint{2411.11965}.

\bibtype{Article}%
\bibitem[Del~Pozzo et al.(2017)]{DelPozzo:2015bna}
\bibinfo{author}{Del~Pozzo W}, \bibinfo{author}{Li TGF} and  \bibinfo{author}{Messenger C} (\bibinfo{year}{2017}).
\bibinfo{title}{{Cosmological inference using only gravitational wave observations of binary neutron stars}}.
\bibinfo{journal}{{\em Phys. Rev. D}} \bibinfo{volume}{95} (\bibinfo{number}{4}): \bibinfo{pages}{043502}. \bibinfo{doi}{\doi{10.1103/PhysRevD.95.043502}}.
\eprint{1506.06590}.

\bibtype{Article}%
\bibitem[{Dhawan} et al.(2020)]{2020ApJ...888...67D}
\bibinfo{author}{{Dhawan} S}, \bibinfo{author}{{Bulla} M}, \bibinfo{author}{{Goobar} A}, \bibinfo{author}{{Sagu{\'e}s Carracedo} A} and  \bibinfo{author}{{Setzer} CN} (\bibinfo{year}{2020}), \bibinfo{month}{Jan.}
\bibinfo{title}{{Constraining the Observer Angle of the Kilonova AT2017gfo Associated with GW170817: Implications for the Hubble Constant}}.
\bibinfo{journal}{{\em \apj}} \bibinfo{volume}{888} (\bibinfo{number}{2}), \bibinfo{eid}{67}. \bibinfo{doi}{\doi{10.3847/1538-4357/ab5799}}.
\eprint{1909.13810}.

\bibtype{Article}%
\bibitem[Dálya et al.(2022)]{Dalya_2022}
\bibinfo{author}{Dálya G}, \bibinfo{author}{Díaz R}, \bibinfo{author}{Bouchet FR}, \bibinfo{author}{Frei Z}, \bibinfo{author}{Jasche J}, \bibinfo{author}{Lavaux G}, \bibinfo{author}{Macas R}, \bibinfo{author}{Mukherjee S}, \bibinfo{author}{Pálfi M}, \bibinfo{author}{de Souza RS}, \bibinfo{author}{Wandelt BD}, \bibinfo{author}{Bilicki M} and  \bibinfo{author}{Raffai P} (\bibinfo{year}{2022}), \bibinfo{month}{May}.
\bibinfo{title}{Glade+: an extended galaxy catalogue for multimessenger searches with advanced gravitational-wave detectors}.
\bibinfo{journal}{{\em Monthly Notices of the Royal Astronomical Society}} \bibinfo{volume}{514} (\bibinfo{number}{1}): \bibinfo{pages}{1403–1411}.
ISSN \bibinfo{issn}{1365-2966}. \bibinfo{doi}{\doi{10.1093/mnras/stac1443}}.
\bibinfo{url}{\url{http://dx.doi.org/10.1093/mnras/stac1443}}.

\bibtype{Article}%
\bibitem[Ezquiaga(2021)]{Ezquiaga:2021ayr}
\bibinfo{author}{Ezquiaga JM} (\bibinfo{year}{2021}).
\bibinfo{title}{{Hearing gravity from the cosmos: GWTC-2 probes general relativity at cosmological scales}}.
\bibinfo{journal}{{\em Phys. Lett. B}} \bibinfo{volume}{822}: \bibinfo{pages}{136665}. \bibinfo{doi}{\doi{10.1016/j.physletb.2021.136665}}.
\eprint{2104.05139}.

\bibtype{Article}%
\bibitem[Ezquiaga and Holz(2021)]{Ezquiaga:2020tns}
\bibinfo{author}{Ezquiaga JM} and  \bibinfo{author}{Holz DE} (\bibinfo{year}{2021}).
\bibinfo{title}{{Jumping the Gap: Searching for LIGO\textquoteright{}s Biggest Black Holes}}.
\bibinfo{journal}{{\em Astrophys. J. Lett.}} \bibinfo{volume}{909} (\bibinfo{number}{2}): \bibinfo{pages}{L23}. \bibinfo{doi}{\doi{10.3847/2041-8213/abe638}}.
\eprint{2006.02211}.

\bibtype{Article}%
\bibitem[{Ezquiaga} and {Holz}(2022)]{spectral_sirens}
\bibinfo{author}{{Ezquiaga} JM} and  \bibinfo{author}{{Holz} DE} (\bibinfo{year}{2022}), \bibinfo{month}{Aug.}
\bibinfo{title}{{Spectral Sirens: Cosmology from the Full Mass Distribution of Compact Binaries}}.
\bibinfo{journal}{{\em \prl}} \bibinfo{volume}{129} (\bibinfo{number}{6}), \bibinfo{eid}{061102}. \bibinfo{doi}{\doi{10.1103/PhysRevLett.129.061102}}.
\eprint{2202.08240}.

\bibtype{Article}%
\bibitem[{Farah} et al.(2025)]{2025ApJ...978..153F}
\bibinfo{author}{{Farah} AM}, \bibinfo{author}{{Callister} TA}, \bibinfo{author}{{Ezquiaga} JM}, \bibinfo{author}{{Zevin} M} and  \bibinfo{author}{{Holz} DE} (\bibinfo{year}{2025}), \bibinfo{month}{Jan.}
\bibinfo{title}{{No Need to Know: Toward Astrophysics-free Gravitational-wave Cosmology}}.
\bibinfo{journal}{{\em ApJ}} \bibinfo{volume}{978} (\bibinfo{number}{2}), \bibinfo{eid}{153}. \bibinfo{doi}{\doi{10.3847/1538-4357/ad9253}}.
\eprint{2404.02210}.

\bibtype{Article}%
\bibitem[Farr et al.(2019)]{Farr_2019}
\bibinfo{author}{Farr WM}, \bibinfo{author}{Fishbach M}, \bibinfo{author}{Ye J} and  \bibinfo{author}{Holz DE} (\bibinfo{year}{2019}), \bibinfo{month}{oct}.
\bibinfo{title}{A future percent-level measurement of the hubble expansion at redshift 0.8 with advanced ligo}.
\bibinfo{journal}{{\em \apjl}} \bibinfo{volume}{883} (\bibinfo{number}{2}): \bibinfo{pages}{L42}. \bibinfo{doi}{\doi{10.3847/2041-8213/ab4284}}.

\bibtype{Article}%
\bibitem[{Feeney} et al.(2019)]{2019PhRvL.122f1105F}
\bibinfo{author}{{Feeney} SM}, \bibinfo{author}{{Peiris} HV}, \bibinfo{author}{{Williamson} AR}, \bibinfo{author}{{Nissanke} SM}, \bibinfo{author}{{Mortlock} DJ}, \bibinfo{author}{{Alsing} J} and  \bibinfo{author}{{Scolnic} D} (\bibinfo{year}{2019}), \bibinfo{month}{Feb.}
\bibinfo{title}{{Prospects for Resolving the Hubble Constant Tension with Standard Sirens}}.
\bibinfo{journal}{{\em \prl}} \bibinfo{volume}{122} (\bibinfo{number}{6}), \bibinfo{eid}{061105}. \bibinfo{doi}{\doi{10.1103/PhysRevLett.122.061105}}.
\eprint{1802.03404}.

\bibtype{Article}%
\bibitem[Ferri et al.(2024)]{Ferri:2024amc}
\bibinfo{author}{Ferri Ja}, \bibinfo{author}{Tashiro IL}, \bibinfo{author}{Abramo LR}, \bibinfo{author}{Matos I}, \bibinfo{author}{Quartin M} and  \bibinfo{author}{Sturani R} (\bibinfo{year}{2024}), \bibinfo{month}{11}.
\bibinfo{title}{{A robust cosmic standard ruler from the cross-correlations of galaxies and dark sirens}} \eprint{2412.00202}.

\bibtype{Article}%
\bibitem[Finke et al.(2021)]{Finke_2021}
\bibinfo{author}{Finke A}, \bibinfo{author}{Foffa S}, \bibinfo{author}{Iacovelli F}, \bibinfo{author}{Maggiore M} and  \bibinfo{author}{Mancarella M} (\bibinfo{year}{2021}), \bibinfo{month}{Aug.}
\bibinfo{title}{Cosmology with ligo/virgo dark sirens: Hubble parameter and modified gravitational wave propagation}.
\bibinfo{journal}{{\em Journal of Cosmology and Astroparticle Physics}} \bibinfo{volume}{2021} (\bibinfo{number}{08}): \bibinfo{pages}{026}.
ISSN \bibinfo{issn}{1475-7516}. \bibinfo{doi}{\doi{10.1088/1475-7516/2021/08/026}}.
\bibinfo{url}{\url{http://dx.doi.org/10.1088/1475-7516/2021/08/026}}.

\bibtype{Article}%
\bibitem[{Gair} et al.(2022)]{Hitchhiker}
\bibinfo{author}{{Gair} JR}, \bibinfo{author}{{Ghosh} A}, \bibinfo{author}{{Gray} R}, \bibinfo{author}{{Holz} DE}, \bibinfo{author}{{Mastrogiovanni} S}, \bibinfo{author}{{Mukherjee} S}, \bibinfo{author}{{Palmese} A}, \bibinfo{author}{{Tamanini} N} and  et al. (\bibinfo{year}{2022}), \bibinfo{month}{Dec.}
\bibinfo{title}{{The Hitchhiker's guide to the galaxy catalog approach for gravitational wave cosmology}}.
\bibinfo{journal}{{\em arXiv e-prints}} , \bibinfo{eid}{arXiv:2212.08694}\bibinfo{doi}{\doi{10.48550/arXiv.2212.08694}}.
\eprint{2212.08694}.

\bibtype{Article}%
\bibitem[Ghosh et al.(2023)]{Ghosh:2023ksl}
\bibinfo{author}{Ghosh T}, \bibinfo{author}{More S}, \bibinfo{author}{Bera S} and  \bibinfo{author}{Bose S} (\bibinfo{year}{2023}), \bibinfo{month}{12}.
\bibinfo{title}{{Bayesian framework to infer the Hubble constant from cross-correlation of individual gravitational wave events with galaxies}} \eprint{2312.16305}.

\bibtype{Article}%
\bibitem[{Gianfagna} et al.(2024)]{Gianfragna}
\bibinfo{author}{{Gianfagna} G}, \bibinfo{author}{{Piro} L}, \bibinfo{author}{{Pannarale} F}, \bibinfo{author}{{Van Eerten} H}, \bibinfo{author}{{Ricci} F} and  \bibinfo{author}{{Ryan} G} (\bibinfo{year}{2024}), \bibinfo{month}{Feb.}
\bibinfo{title}{{Potential biases and prospects for the Hubble constant estimation via electromagnetic and gravitational-wave joint analyses}}.
\bibinfo{journal}{{\em \mnras}} \bibinfo{volume}{528} (\bibinfo{number}{2}): \bibinfo{pages}{2600--2613}. \bibinfo{doi}{\doi{10.1093/mnras/stae198}}.
\eprint{2309.17073}.

\bibtype{Article}%
\bibitem[Graff et al.(2015)]{Graff:2015bba}
\bibinfo{author}{Graff PB}, \bibinfo{author}{Buonanno A} and  \bibinfo{author}{Sathyaprakash BS} (\bibinfo{year}{2015}).
\bibinfo{title}{{Missing Link: Bayesian detection and measurement of intermediate-mass black-hole binaries}}.
\bibinfo{journal}{{\em Phys. Rev. D}} \bibinfo{volume}{92} (\bibinfo{number}{2}): \bibinfo{pages}{022002}. \bibinfo{doi}{\doi{10.1103/PhysRevD.92.022002}}.
\eprint{1504.04766}.

\bibtype{Article}%
\bibitem[{Graham} et al.(2020)]{graham20}
\bibinfo{author}{{Graham} MJ}, \bibinfo{author}{{Ford} KES}, \bibinfo{author}{{McKernan} B}, \bibinfo{author}{{Ross} NP}, \bibinfo{author}{{Stern} D}, \bibinfo{author}{{Burdge} K}, \bibinfo{author}{{Coughlin} M}, \bibinfo{author}{{Djorgovski} SG}, \bibinfo{author}{{Drake} AJ}, \bibinfo{author}{{Duev} D}, \bibinfo{author}{{Kasliwal} M}, \bibinfo{author}{{Mahabal} AA}, \bibinfo{author}{{van Velzen} S}, \bibinfo{author}{{Belecki} J}, \bibinfo{author}{{Bellm} EC}, \bibinfo{author}{{Burruss} R}, \bibinfo{author}{{Cenko} SB}, \bibinfo{author}{{Cunningham} V}, \bibinfo{author}{{Helou} G}, \bibinfo{author}{{Kulkarni} SR}, \bibinfo{author}{{Masci} FJ}, \bibinfo{author}{{Prince} T}, \bibinfo{author}{{Reiley} D}, \bibinfo{author}{{Rodriguez} H}, \bibinfo{author}{{Rusholme} B}, \bibinfo{author}{{Smith} RM} and  \bibinfo{author}{{Soumagnac} MT} (\bibinfo{year}{2020}), \bibinfo{month}{Jun.}
\bibinfo{title}{{Candidate Electromagnetic Counterpart to the Binary Black Hole Merger Gravitational-Wave Event S190521g$^{*}$}}.
\bibinfo{journal}{{\em \prl}} \bibinfo{volume}{124} (\bibinfo{number}{25}), \bibinfo{eid}{251102}. \bibinfo{doi}{\doi{10.1103/PhysRevLett.124.251102}}.
\eprint{2006.14122}.

\bibtype{Article}%
\bibitem[{Graham} et al.(2023)]{graham23}
\bibinfo{author}{{Graham} MJ}, \bibinfo{author}{{McKernan} B}, \bibinfo{author}{{Ford} KES}, \bibinfo{author}{{Stern} D}, \bibinfo{author}{{Djorgovski} SG}, \bibinfo{author}{{Coughlin} M}, \bibinfo{author}{{Burdge} KB}, \bibinfo{author}{{Bellm} EC}, \bibinfo{author}{{Helou} G}, \bibinfo{author}{{Mahabal} AA}, \bibinfo{author}{{Masci} FJ}, \bibinfo{author}{{Purdum} J}, \bibinfo{author}{{Rosnet} P} and  \bibinfo{author}{{Rusholme} B} (\bibinfo{year}{2023}), \bibinfo{month}{Jan.}
\bibinfo{title}{{A Light in the Dark: Searching for Electromagnetic Counterparts to Black Hole-Black Hole Mergers in LIGO/Virgo O3 with the Zwicky Transient Facility}}.
\bibinfo{journal}{{\em \apj}} \bibinfo{volume}{942} (\bibinfo{number}{2}), \bibinfo{eid}{99}. \bibinfo{doi}{\doi{10.3847/1538-4357/aca480}}.
\eprint{2209.13004}.

\bibtype{Article}%
\bibitem[{Gray} et al.(2023)]{2023JCAP...12..023G}
\bibinfo{author}{{Gray} R}, \bibinfo{author}{{Beirnaert} F}, \bibinfo{author}{{Karathanasis} C}, \bibinfo{author}{{Revenu} B}, \bibinfo{author}{{Turski} C}, \bibinfo{author}{{Chen} A}, \bibinfo{author}{{Baker} T}, \bibinfo{author}{{Vallejo} S}, \bibinfo{author}{{Romano} AE}, \bibinfo{author}{{Ghosh} T}, \bibinfo{author}{{Ghosh} A}, \bibinfo{author}{{Leyde} K}, \bibinfo{author}{{Mastrogiovanni} S} and  \bibinfo{author}{{More} S} (\bibinfo{year}{2023}), \bibinfo{month}{Dec.}
\bibinfo{title}{{Joint cosmological and gravitational-wave population inference using dark sirens and galaxy catalogues}}.
\bibinfo{journal}{{\em \jcap}} \bibinfo{volume}{2023} (\bibinfo{number}{12}), \bibinfo{eid}{023}. \bibinfo{doi}{\doi{10.1088/1475-7516/2023/12/023}}.
\eprint{2308.02281}.

\bibtype{Article}%
\bibitem[Guidorzi et al.(2017)]{Guidorzi:2017ogy}
\bibinfo{author}{Guidorzi C} and  et al. (\bibinfo{year}{2017}).
\bibinfo{title}{{Improved Constraints on $H_0$ from a Combined Analysis of Gravitational-wave and Electromagnetic Emission from GW170817}}.
\bibinfo{journal}{{\em Astrophys. J. Lett.}} \bibinfo{volume}{851} (\bibinfo{number}{2}): \bibinfo{pages}{L36}. \bibinfo{doi}{\doi{10.3847/2041-8213/aaa009}}.
\eprint{1710.06426}.

\bibtype{Article}%
\bibitem[Hanselman et al.(2024)]{hanselman2024}
\bibinfo{author}{Hanselman AG}, \bibinfo{author}{Vijaykumar A}, \bibinfo{author}{Fishbach M} and  \bibinfo{author}{Holz DE} (\bibinfo{year}{2024}).
\bibinfo{title}{Gravitational-wave dark siren cosmology systematics from galaxy weighting} \eprint{2405.14818}, \bibinfo{url}{\url{https://arxiv.org/abs/2405.14818}}.

\bibtype{Article}%
\bibitem[{Heinzel} et al.(2021)]{2021MNRAS.502.3057H}
\bibinfo{author}{{Heinzel} J}, \bibinfo{author}{{Coughlin} MW}, \bibinfo{author}{{Dietrich} T}, \bibinfo{author}{{Bulla} M}, \bibinfo{author}{{Antier} S}, \bibinfo{author}{{Christensen} N}, \bibinfo{author}{{Coulter} DA}, \bibinfo{author}{{Foley} RJ}, \bibinfo{author}{{Issa} L} and  \bibinfo{author}{{Khetan} N} (\bibinfo{year}{2021}), \bibinfo{month}{Apr.}
\bibinfo{title}{{Comparing inclination-dependent analyses of kilonova transients}}.
\bibinfo{journal}{{\em \mnras}} \bibinfo{volume}{502} (\bibinfo{number}{2}): \bibinfo{pages}{3057--3065}. \bibinfo{doi}{\doi{10.1093/mnras/stab221}}.
\eprint{2010.10746}.

\bibtype{Article}%
\bibitem[{Holz} and {Hughes}(2005)]{2005ApJ...629...15H}
\bibinfo{author}{{Holz} DE} and  \bibinfo{author}{{Hughes} SA} (\bibinfo{year}{2005}), \bibinfo{month}{Aug.}
\bibinfo{title}{{Using Gravitational-Wave Standard Sirens}}.
\bibinfo{journal}{{\em \apj}} \bibinfo{volume}{629} (\bibinfo{number}{1}): \bibinfo{pages}{15--22}. \bibinfo{doi}{\doi{10.1086/431341}}.
\eprint{astro-ph/0504616}.

\bibtype{Article}%
\bibitem[{Hotokezaka} et al.(2019)]{Hotokezaka}
\bibinfo{author}{{Hotokezaka} K}, \bibinfo{author}{{Nakar} E}, \bibinfo{author}{{Gottlieb} O}, \bibinfo{author}{{Nissanke} S}, \bibinfo{author}{{Masuda} K}, \bibinfo{author}{{Hallinan} G}, \bibinfo{author}{{Mooley} KP} and  \bibinfo{author}{{Deller} AT} (\bibinfo{year}{2019}), \bibinfo{month}{Jul.}
\bibinfo{title}{{A Hubble constant measurement from superluminal motion of the jet in GW170817}}.
\bibinfo{journal}{{\em Nature Astronomy}} \bibinfo{volume}{3}: \bibinfo{pages}{940--944}. \bibinfo{doi}{\doi{10.1038/s41550-019-0820-1}}.
\eprint{1806.10596}.

\bibtype{Article}%
\bibitem[{Howlett} and {Davis}(2020)]{Howlett_2020}
\bibinfo{author}{{Howlett} C} and  \bibinfo{author}{{Davis} TM} (\bibinfo{year}{2020}), \bibinfo{month}{Mar.}
\bibinfo{title}{{Standard siren speeds: improving velocities in gravitational-wave measurements of H$_{0}$}}.
\bibinfo{journal}{{\em \mnras}} \bibinfo{volume}{492} (\bibinfo{number}{3}): \bibinfo{pages}{3803--3815}. \bibinfo{doi}{\doi{10.1093/mnras/staa049}}.
\eprint{1909.00587}.

\bibtype{Article}%
\bibitem[Huang et al.(2022)]{Huang:2022rdg}
\bibinfo{author}{Huang Y}, \bibinfo{author}{Chen HY}, \bibinfo{author}{Haster CJ}, \bibinfo{author}{Sun L}, \bibinfo{author}{Vitale S} and  \bibinfo{author}{Kissel J} (\bibinfo{year}{2022}), \bibinfo{month}{4}.
\bibinfo{title}{{Impact of calibration uncertainties on Hubble constant measurements from gravitational-wave sources}} \eprint{2204.03614}.

\bibtype{Article}%
\bibitem[Karathanasis et al.(2023)]{Karathanasis:2022rtr}
\bibinfo{author}{Karathanasis C}, \bibinfo{author}{Mukherjee S} and  \bibinfo{author}{Mastrogiovanni S} (\bibinfo{year}{2023}).
\bibinfo{title}{{Binary black holes population and cosmology in new lights: signature of PISN mass and formation channel in GWTC-3}}.
\bibinfo{journal}{{\em Mon. Not. Roy. Astron. Soc.}} \bibinfo{volume}{523} (\bibinfo{number}{3}): \bibinfo{pages}{4539--4555}. \bibinfo{doi}{\doi{10.1093/mnras/stad1373}}.
\eprint{2204.13495}.

\bibtype{Misc}%
\bibitem[Kaur et al.(2024)]{kaur2024}
\bibinfo{author}{Kaur R}, \bibinfo{author}{O'Connor B}, \bibinfo{author}{Palmese A} and  \bibinfo{author}{Kunnumkai K} (\bibinfo{year}{2024}).
\bibinfo{title}{Detecting prompt and afterglow jet emission of gravitational wave events from ligo/virgo/kagra and next generation detectors}.
\eprint{2410.10579}, \bibinfo{url}{\url{https://arxiv.org/abs/2410.10579}}.

\bibtype{Article}%
\bibitem[{Kiendrebeogo} et al.(2023)]{Kiendrebeogo}
\bibinfo{author}{{Kiendrebeogo} RW}, \bibinfo{author}{{Farah} AM}, \bibinfo{author}{{Foley} EM}, \bibinfo{author}{{Gray} A}, \bibinfo{author}{{Kunert} N}, \bibinfo{author}{{Puecher} A}, \bibinfo{author}{{Toivonen} A}, \bibinfo{author}{{VandenBerg} RO}, \bibinfo{author}{{Anand} S}, \bibinfo{author}{{Ahumada} T}, \bibinfo{author}{{Karambelkar} V}, \bibinfo{author}{{Coughlin} MW}, \bibinfo{author}{{Dietrich} T}, \bibinfo{author}{{Kam} SZ}, \bibinfo{author}{{Pang} PTH}, \bibinfo{author}{{Singer} LP} and  \bibinfo{author}{{Sravan} N} (\bibinfo{year}{2023}), \bibinfo{month}{Dec.}
\bibinfo{title}{{Updated Observing Scenarios and Multimessenger Implications for the International Gravitational-wave Networks O4 and O5}}.
\bibinfo{journal}{{\em \apj}} \bibinfo{volume}{958} (\bibinfo{number}{2}), \bibinfo{eid}{158}. \bibinfo{doi}{\doi{10.3847/1538-4357/acfcb1}}.
\eprint{2306.09234}.

\bibtype{Article}%
\bibitem[{Kunert} et al.(2024)]{kunert}
\bibinfo{author}{{Kunert} N}, \bibinfo{author}{{Gair} J}, \bibinfo{author}{{Pang} PTH} and  \bibinfo{author}{{Dietrich} T} (\bibinfo{year}{2024}), \bibinfo{month}{Aug.}
\bibinfo{title}{{Impact of gravitational waveform model systematics on the measurement of the Hubble constant}}.
\bibinfo{journal}{{\em \prd}} \bibinfo{volume}{110} (\bibinfo{number}{4}), \bibinfo{eid}{043520}. \bibinfo{doi}{\doi{10.1103/PhysRevD.110.043520}}.
\eprint{2405.18158}.

\bibtype{Article}%
\bibitem[{Kunnumkai} et al.(2024)]{Kunnumkai_2024_NSBH}
\bibinfo{author}{{Kunnumkai} K}, \bibinfo{author}{{Palmese} A}, \bibinfo{author}{{Bulla} M}, \bibinfo{author}{{Dietrich} T}, \bibinfo{author}{{Farah} AM} and  \bibinfo{author}{{Pang} PTH} (\bibinfo{year}{2024}), \bibinfo{month}{Sep.}
\bibinfo{title}{{Kilonova emission from GW230529 and mass gap neutron star-black hole mergers}}.
\bibinfo{journal}{{\em arXiv e-prints}} , \bibinfo{eid}{arXiv:2409.10651}\bibinfo{doi}{\doi{10.48550/arXiv.2409.10651}}.
\eprint{2409.10651}.

\bibtype{Misc}%
\bibitem[Kunnumkai et al.(2024)]{kunnumkai2024detecting}
\bibinfo{author}{Kunnumkai K}, \bibinfo{author}{Palmese A}, \bibinfo{author}{Farah AM}, \bibinfo{author}{Bulla M}, \bibinfo{author}{Dietrich T}, \bibinfo{author}{Pang PTH}, \bibinfo{author}{Anand S}, \bibinfo{author}{Andreoni I}, \bibinfo{author}{Cabrera T} and  \bibinfo{author}{Connor BO} (\bibinfo{year}{2024}).
\bibinfo{title}{Detecting electromagnetic counterparts to ligo/virgo/kagra gravitational wave events with decam: Neutron star mergers}.
\eprint{2411.13673}, \bibinfo{url}{\url{https://arxiv.org/abs/2411.13673}}.

\bibtype{Article}%
\bibitem[Leyde et al.(2022)]{Leyde:2022fsc}
\bibinfo{author}{Leyde K}, \bibinfo{author}{Mastrogiovanni S}, \bibinfo{author}{Steer DA}, \bibinfo{author}{Chassande-Mottin E} and  \bibinfo{author}{Karathanasis C} (\bibinfo{year}{2022}).
\bibinfo{title}{{Current and future constraints on cosmology and modified gravitational wave friction from binary black holes}}.
\bibinfo{journal}{{\em JCAP}} \bibinfo{volume}{09}: \bibinfo{pages}{012}. \bibinfo{doi}{\doi{10.1088/1475-7516/2022/09/012}}.
\eprint{2202.00025}.

\bibtype{Article}%
\bibitem[Li and Paczynski(1998)]{Li:1998bw}
\bibinfo{author}{Li LX} and  \bibinfo{author}{Paczynski B} (\bibinfo{year}{1998}).
\bibinfo{title}{{Transient events from neutron star mergers}}.
\bibinfo{journal}{{\em Astrophys. J. Lett.}} \bibinfo{volume}{507}: \bibinfo{pages}{L59}. \bibinfo{doi}{\doi{10.1086/311680}}.
\eprint{astro-ph/9807272}.

\bibtype{Article}%
\bibitem[Libanore et al.(2021)]{Libanore:2020fim}
\bibinfo{author}{Libanore S}, \bibinfo{author}{Artale MC}, \bibinfo{author}{Karagiannis D}, \bibinfo{author}{Liguori M}, \bibinfo{author}{Bartolo N}, \bibinfo{author}{Bouffanais Y}, \bibinfo{author}{Giacobbo N}, \bibinfo{author}{Mapelli M} and  \bibinfo{author}{Matarrese S} (\bibinfo{year}{2021}).
\bibinfo{title}{{Gravitational Wave mergers as tracers of Large Scale Structures}}.
\bibinfo{journal}{{\em JCAP}} \bibinfo{volume}{02}: \bibinfo{pages}{035}. \bibinfo{doi}{\doi{10.1088/1475-7516/2021/02/035}}.
\eprint{2007.06905}.

\bibtype{Article}%
\bibitem[Libanore et al.(2022)]{Libanore:2021jqv}
\bibinfo{author}{Libanore S}, \bibinfo{author}{Artale MC}, \bibinfo{author}{Karagiannis D}, \bibinfo{author}{Liguori M}, \bibinfo{author}{Bartolo N}, \bibinfo{author}{Bouffanais Y}, \bibinfo{author}{Mapelli M} and  \bibinfo{author}{Matarrese S} (\bibinfo{year}{2022}).
\bibinfo{title}{{Clustering of Gravitational Wave and Supernovae events: a multitracer analysis in Luminosity Distance Space}}.
\bibinfo{journal}{{\em JCAP}} \bibinfo{volume}{02} (\bibinfo{number}{02}): \bibinfo{pages}{003}. \bibinfo{doi}{\doi{10.1088/1475-7516/2022/02/003}}.
\eprint{2109.10857}.

\bibtype{Article}%
\bibitem[{LIGO Scientific Collaboration} et al.(2017)]{MMApaper}
\bibinfo{author}{{LIGO Scientific Collaboration}}, \bibinfo{author}{{Virgo Collaboration}}, \bibinfo{author}{{GBM} F}, \bibinfo{author}{{INTEGRAL}}, \bibinfo{author}{{IceCube Collaboration}}, \bibinfo{author}{{AstroSat Cadmium Zinc Telluride Imager Team}}, \bibinfo{author}{{IPN Collaboration}}, \bibinfo{author}{{The Insight-Hxmt Collaboration}}, \bibinfo{author}{{ANTARES Collaboration}}, \bibinfo{author}{{The Swift Collaboration}}, \bibinfo{author}{{AGILE Team}}, \bibinfo{author}{{The 1M2H Team}}, \bibinfo{author}{{The Dark Energy Camera GW-EM Collaboration}}, \bibinfo{author}{{the DES Collaboration}}, \bibinfo{author}{{The DLT40 Collaboration}}, \bibinfo{author}{{GRAWITA}}, \bibinfo{author}{{:}}, \bibinfo{author}{{GRAvitational Wave Inaf TeAm}}, \bibinfo{author}{{The Fermi Large Area Telescope Collaboration}}, \bibinfo{author}{{ATCA}}, \bibinfo{author}{{:}}, \bibinfo{author}{{Telescope Compact Array} A}, \bibinfo{author}{{ASKAP}}, \bibinfo{author}{{:}}, \bibinfo{author}{{SKA Pathfinder} A},
  \bibinfo{author}{{Las Cumbres Observatory Group}}, \bibinfo{author}{{OzGrav}}, \bibinfo{author}{{DWF}}, \bibinfo{author}{{AST3}}, \bibinfo{author}{{CAASTRO Collaborations}}, \bibinfo{author}{{The VINROUGE Collaboration}}, \bibinfo{author}{{MASTER Collaboration}}, \bibinfo{author}{{J-GEM}}, \bibinfo{author}{{GROWTH}}, \bibinfo{author}{{JAGWAR}}, \bibinfo{author}{{NRAO} C}, \bibinfo{author}{{TTU-NRAO}}, \bibinfo{author}{{NuSTAR Collaborations}}, \bibinfo{author}{{Pan-STARRS}}, \bibinfo{author}{{The MAXI Team}}, \bibinfo{author}{{Consortium} T}, \bibinfo{author}{{KU Collaboration}}, \bibinfo{author}{{Optical Telescope} N}, \bibinfo{author}{{ePESSTO}}, \bibinfo{author}{{GROND}}, \bibinfo{author}{{Tech University} T}, \bibinfo{author}{{SALT Group}}, \bibinfo{author}{{TOROS}}, \bibinfo{author}{{:}}, \bibinfo{author}{{Transient Robotic Observatory of the South Collaboration}}, \bibinfo{author}{{The BOOTES Collaboration}}, \bibinfo{author}{{MWA}}, \bibinfo{author}{{:}}, \bibinfo{author}{{Widefield Array} M},
  \bibinfo{author}{{The CALET Collaboration}}, \bibinfo{author}{{IKI-GW Follow-up Collaboration}}, \bibinfo{author}{{H.~E.~S.~S.~Collaboration}}, \bibinfo{author}{{LOFAR Collaboration}}, \bibinfo{author}{{LWA}}, \bibinfo{author}{{:}}, \bibinfo{author}{{Wavelength Array} L}, \bibinfo{author}{{HAWC Collaboration}}, \bibinfo{author}{{The Pierre Auger Collaboration}}, \bibinfo{author}{{ALMA Collaboration}}, \bibinfo{author}{{Euro VLBI Team}}, \bibinfo{author}{{Pi of the Sky Collaboration}}, \bibinfo{author}{{The Chandra Team at McGill University}}, \bibinfo{author}{{DFN}}, \bibinfo{author}{{:}}, \bibinfo{author}{{Fireball Network} D}, \bibinfo{author}{{ATLAS}}, \bibinfo{author}{{Time Resolution Universe Survey} H}, \bibinfo{author}{{RIMAS}}, \bibinfo{author}{{RATIR}} and  \bibinfo{author}{{South Africa/MeerKAT} S} (\bibinfo{year}{2017}), \bibinfo{month}{Oct.}
\bibinfo{title}{{Multi-messenger Observations of a Binary Neutron Star Merger}}.
\bibinfo{journal}{{\em \apjl}} \bibinfo{volume}{848} (\bibinfo{number}{2}), \bibinfo{eid}{L12}. \bibinfo{doi}{\doi{10.3847/2041-8213/aa91c9}}.
\eprint{1710.05833}.

\bibtype{Article}%
\bibitem[{Lipunov} et al.(2017)]{Lipunov}
\bibinfo{author}{{Lipunov} VM}, \bibinfo{author}{{Gorbovskoy} E}, \bibinfo{author}{{Kornilov} VG}, \bibinfo{author}{{. Tyurina} N}, \bibinfo{author}{{Balanutsa} P}, \bibinfo{author}{{Kuznetsov} A}, \bibinfo{author}{{Vlasenko} D}, \bibinfo{author}{{Kuvshinov} D}, \bibinfo{author}{{Gorbunov} I}, \bibinfo{author}{{Buckley} DAH}, \bibinfo{author}{{Krylov} AV}, \bibinfo{author}{{Podesta} R}, \bibinfo{author}{{Lopez} C}, \bibinfo{author}{{Podesta} F}, \bibinfo{author}{{Levato} H}, \bibinfo{author}{{Saffe} C}, \bibinfo{author}{{Mallamachi} C}, \bibinfo{author}{{Potter} S}, \bibinfo{author}{{Budnev} NM}, \bibinfo{author}{{Gress} O}, \bibinfo{author}{{Ishmuhametova} Y}, \bibinfo{author}{{Vladimirov} V}, \bibinfo{author}{{Zimnukhov} D}, \bibinfo{author}{{Yurkov} V}, \bibinfo{author}{{Sergienko} Y}, \bibinfo{author}{{Gabovich} A}, \bibinfo{author}{{Rebolo} R}, \bibinfo{author}{{Serra-Ricart} M}, \bibinfo{author}{{Israelyan} G}, \bibinfo{author}{{Chazov} V}, \bibinfo{author}{{Wang} X}, \bibinfo{author}{{Tlatov} A} and
  \bibinfo{author}{{Panchenko} MI} (\bibinfo{year}{2017}), \bibinfo{month}{Nov.}
\bibinfo{title}{{MASTER Optical Detection of the First LIGO/Virgo Neutron Star Binary Merger GW170817}}.
\bibinfo{journal}{{\em \apjl}} \bibinfo{volume}{850} (\bibinfo{number}{1}), \bibinfo{eid}{L1}. \bibinfo{doi}{\doi{10.3847/2041-8213/aa92c0}}.
\eprint{1710.05461}.

\bibtype{Article}%
\bibitem[Maga\~na Hernandez and Ray(2024)]{MaganaHernandez:2024uty}
\bibinfo{author}{Maga\~na Hernandez I} and  \bibinfo{author}{Ray A} (\bibinfo{year}{2024}), \bibinfo{month}{4}.
\bibinfo{title}{{Beyond Gaps and Bumps: Spectral Siren Cosmology with Non-Parametric Population Models}} \eprint{2404.02522}.

\bibtype{Article}%
\bibitem[Magana~Hernandez(2023)]{MaganaHernandez:2021zyc}
\bibinfo{author}{Magana~Hernandez I} (\bibinfo{year}{2023}).
\bibinfo{title}{{Constraining the number of spacetime dimensions from GWTC-3 binary black hole mergers}}.
\bibinfo{journal}{{\em Phys. Rev. D}} \bibinfo{volume}{107} (\bibinfo{number}{8}): \bibinfo{pages}{084033}. \bibinfo{doi}{\doi{10.1103/PhysRevD.107.084033}}.
\eprint{2112.07650}.

\bibtype{Book}%
\bibitem[Maggiore(2007)]{Maggiore:1900zz}
\bibinfo{author}{Maggiore M} (\bibinfo{year}{2007}).
\bibinfo{title}{{Gravitational Waves. Vol. 1: Theory and Experiments}}, \bibinfo{series}{Oxford Master Series in Physics}, \bibinfo{publisher}{Oxford University Press}.
\bibinfo{comment}{ISBN} \bibinfo{isbn}{978-0-19-857074-5, 978-0-19-852074-0}.

\bibtype{Book}%
\bibitem[Maggiore(2018)]{Maggiore:2018sht}
\bibinfo{author}{Maggiore M} (\bibinfo{year}{2018}), \bibinfo{month}{3}.
\bibinfo{title}{{Gravitational Waves. Vol. 2: Astrophysics and Cosmology}}, \bibinfo{publisher}{Oxford University Press}.
\bibinfo{comment}{ISBN} \bibinfo{isbn}{978-0-19-857089-9}.

\bibtype{Article}%
\bibitem[Mali and Essick(2024)]{Mali:2024wpq}
\bibinfo{author}{Mali U} and  \bibinfo{author}{Essick R} (\bibinfo{year}{2024}), \bibinfo{month}{10}.
\bibinfo{title}{{Striking a Chord with Spectral Sirens: multiple features in the compact binary population correlate with $H_0$}} \eprint{2410.07416}.

\bibtype{Article}%
\bibitem[{Mancarella} et al.(2022)]{Mancarella_2022}
\bibinfo{author}{{Mancarella} M}, \bibinfo{author}{{Genoud-Prachex} E} and  \bibinfo{author}{{Maggiore} M} (\bibinfo{year}{2022}), \bibinfo{month}{Mar.}
\bibinfo{title}{{Cosmology and modified gravitational wave propagation from binary black hole population models}}.
\bibinfo{journal}{{\em \prd}} \bibinfo{volume}{105} (\bibinfo{number}{6}), \bibinfo{eid}{064030}. \bibinfo{doi}{\doi{10.1103/PhysRevD.105.064030}}.
\eprint{2112.05728}.

\bibtype{Article}%
\bibitem[Mancarella et al.(2024)]{Mancarella:2024qle}
\bibinfo{author}{Mancarella M}, \bibinfo{author}{Iacovelli F}, \bibinfo{author}{Foffa S}, \bibinfo{author}{Muttoni N} and  \bibinfo{author}{Maggiore M} (\bibinfo{year}{2024}).
\bibinfo{title}{{Accurate Standard Siren Cosmology with Joint Gravitational-Wave and \ensuremath{\gamma}-Ray Burst Observations}}.
\bibinfo{journal}{{\em Phys. Rev. Lett.}} \bibinfo{volume}{133} (\bibinfo{number}{26}): \bibinfo{pages}{261001}. \bibinfo{doi}{\doi{10.1103/PhysRevLett.133.261001}}.
\eprint{2405.02286}.

\bibtype{Article}%
\bibitem[Mandel et al.(2019)]{Mandel_2019}
\bibinfo{author}{Mandel I}, \bibinfo{author}{Farr WM} and  \bibinfo{author}{Gair JR} (\bibinfo{year}{2019}), \bibinfo{month}{Mar.}
\bibinfo{title}{Extracting distribution parameters from multiple uncertain observations with selection biases}.
\bibinfo{journal}{{\em Monthly Notices of the Royal Astronomical Society}} \bibinfo{volume}{486} (\bibinfo{number}{1}): \bibinfo{pages}{1086–1093}.
ISSN \bibinfo{issn}{1365-2966}. \bibinfo{doi}{\doi{10.1093/mnras/stz896}}.
\bibinfo{url}{\url{http://dx.doi.org/10.1093/mnras/stz896}}.

\bibtype{incollection}%
\bibitem[{Mapelli}(2021)]{2021hgwa.bookE..16M}
\bibinfo{author}{{Mapelli} M} (\bibinfo{year}{2021}), \bibinfo{title}{{Formation Channels of Single and Binary Stellar-Mass Black Holes}}, \bibinfo{editor}{{Bambi} C}, \bibinfo{editor}{{Katsanevas} S} and  \bibinfo{editor}{{Kokkotas} KD}, (Eds.), \bibinfo{booktitle}{Handbook of Gravitational Wave Astronomy}, pp.~\bibinfo{pages}{16}.

\bibtype{Article}%
\bibitem[{Margutti} et al.(2018)]{Margutti_2018}
\bibinfo{author}{{Margutti} R}, \bibinfo{author}{{Alexander} KD}, \bibinfo{author}{{Xie} X}, \bibinfo{author}{{Sironi} L}, \bibinfo{author}{{Metzger} BD}, \bibinfo{author}{{Kathirgamaraju} A}, \bibinfo{author}{{Fong} W}, \bibinfo{author}{{Blanchard} PK}, \bibinfo{author}{{Berger} E}, \bibinfo{author}{{MacFadyen} A}, \bibinfo{author}{{Giannios} D}, \bibinfo{author}{{Guidorzi} C}, \bibinfo{author}{{Hajela} A}, \bibinfo{author}{{Chornock} R}, \bibinfo{author}{{Cowperthwaite} PS}, \bibinfo{author}{{Eftekhari} T}, \bibinfo{author}{{Nicholl} M}, \bibinfo{author}{{Villar} VA}, \bibinfo{author}{{Williams} PKG} and  \bibinfo{author}{{Zrake} J} (\bibinfo{year}{2018}), \bibinfo{month}{Mar.}
\bibinfo{title}{{The Binary Neutron Star Event LIGO/Virgo GW170817 160 Days after Merger: Synchrotron Emission across the Electromagnetic Spectrum}}.
\bibinfo{journal}{{\em \apjl}} \bibinfo{volume}{856} (\bibinfo{number}{1}), \bibinfo{eid}{L18}. \bibinfo{doi}{\doi{10.3847/2041-8213/aab2ad}}.
\eprint{1801.03531}.

\bibtype{Article}%
\bibitem[Mastrogiovanni et al.(2020)]{Mastrogiovanni:2020gua}
\bibinfo{author}{Mastrogiovanni S}, \bibinfo{author}{Steer D} and  \bibinfo{author}{Barsuglia M} (\bibinfo{year}{2020}).
\bibinfo{title}{{Probing modified gravity theories and cosmology using gravitational-waves and associated electromagnetic counterparts}}.
\bibinfo{journal}{{\em Phys. Rev. D}} \bibinfo{volume}{102} (\bibinfo{number}{4}): \bibinfo{pages}{044009}. \bibinfo{doi}{\doi{10.1103/PhysRevD.102.044009}}.
\eprint{2004.01632}.

\bibtype{Article}%
\bibitem[{Mastrogiovanni} et al.(2021)]{2021PhRvD.104f2009M}
\bibinfo{author}{{Mastrogiovanni} S}, \bibinfo{author}{{Leyde} K}, \bibinfo{author}{{Karathanasis} C}, \bibinfo{author}{{Chassande-Mottin} E}, \bibinfo{author}{{Steer} DA}, \bibinfo{author}{{Gair} J}, \bibinfo{author}{{Ghosh} A}, \bibinfo{author}{{Gray} R}, \bibinfo{author}{{Mukherjee} S} and  \bibinfo{author}{{Rinaldi} S} (\bibinfo{year}{2021}), \bibinfo{month}{Sep.}
\bibinfo{title}{{On the importance of source population models for gravitational-wave cosmology}}.
\bibinfo{journal}{{\em \prd}} \bibinfo{volume}{104} (\bibinfo{number}{6}), \bibinfo{eid}{062009}. \bibinfo{doi}{\doi{10.1103/PhysRevD.104.062009}}.
\eprint{2103.14663}.

\bibtype{Article}%
\bibitem[Mastrogiovanni et al.(2021)]{Mastrogiovanni:2021wsd}
\bibinfo{author}{Mastrogiovanni S}, \bibinfo{author}{Leyde K}, \bibinfo{author}{Karathanasis C}, \bibinfo{author}{Chassande-Mottin E}, \bibinfo{author}{Steer DA}, \bibinfo{author}{Gair J}, \bibinfo{author}{Ghosh A}, \bibinfo{author}{Gray R}, \bibinfo{author}{Mukherjee S} and  \bibinfo{author}{Rinaldi S} (\bibinfo{year}{2021}).
\bibinfo{title}{{On the importance of source population models for gravitational-wave cosmology}}.
\bibinfo{journal}{{\em Phys. Rev. D}} \bibinfo{volume}{104} (\bibinfo{number}{6}): \bibinfo{pages}{062009}. \bibinfo{doi}{\doi{10.1103/PhysRevD.104.062009}}.
\eprint{2103.14663}.

\bibtype{Article}%
\bibitem[{Mastrogiovanni} et al.(2023)]{Mastrogiovanni2023}
\bibinfo{author}{{Mastrogiovanni} S}, \bibinfo{author}{{Laghi} D}, \bibinfo{author}{{Gray} R}, \bibinfo{author}{{Santoro} GC}, \bibinfo{author}{{Ghosh} A}, \bibinfo{author}{{Karathanasis} C}, \bibinfo{author}{{Leyde} K}, \bibinfo{author}{{Steer} DA}, \bibinfo{author}{{Perri{\`e}s} S} and  \bibinfo{author}{{Pierra} G} (\bibinfo{year}{2023}), \bibinfo{month}{Aug.}
\bibinfo{title}{{Joint population and cosmological properties inference with gravitational waves standard sirens and galaxy surveys}}.
\bibinfo{journal}{{\em \prd}} \bibinfo{volume}{108} (\bibinfo{number}{4}), \bibinfo{eid}{042002}. \bibinfo{doi}{\doi{10.1103/PhysRevD.108.042002}}.
\eprint{2305.10488}.

\bibtype{Article}%
\bibitem[Mastrogiovanni et al.(2024)]{Mastrogiovanni:2024mqc}
\bibinfo{author}{Mastrogiovanni S}, \bibinfo{author}{Karathanasis C}, \bibinfo{author}{Gair J}, \bibinfo{author}{Ashton G}, \bibinfo{author}{Rinaldi S}, \bibinfo{author}{Huang HY} and  \bibinfo{author}{D\'alya G} (\bibinfo{year}{2024}).
\bibinfo{title}{{Cosmology with Gravitational Waves: A Review}}.
\bibinfo{journal}{{\em Annalen Phys.}} \bibinfo{volume}{536} (\bibinfo{number}{2}): \bibinfo{pages}{2200180}. \bibinfo{doi}{\doi{10.1002/andp.202200180}}.

\bibtype{Article}%
\bibitem[{Mastrogiovanni} et al.(2024)]{2024A&A...682A.167M}
\bibinfo{author}{{Mastrogiovanni} S}, \bibinfo{author}{{Pierra} G}, \bibinfo{author}{{Perri{\`e}s} S}, \bibinfo{author}{{Laghi} D}, \bibinfo{author}{{Santoro} GC}, \bibinfo{author}{{Ghosh} A}, \bibinfo{author}{{Gray} R}, \bibinfo{author}{{Karathanasis} C} and  \bibinfo{author}{{Leyde} K} (\bibinfo{year}{2024}), \bibinfo{month}{Feb.}
\bibinfo{title}{{ICAROGW: A python package for inference of astrophysical population properties of noisy, heterogeneous, and incomplete observations}}.
\bibinfo{journal}{{\em \aap}} \bibinfo{volume}{682}, \bibinfo{eid}{A167}. \bibinfo{doi}{\doi{10.1051/0004-6361/202347007}}.

\bibtype{Article}%
\bibitem[{McKernan} et al.(2012)]{McKernan12}
\bibinfo{author}{{McKernan} B}, \bibinfo{author}{{Ford} KES}, \bibinfo{author}{{Lyra} W} and  \bibinfo{author}{{Perets} HB} (\bibinfo{year}{2012}), \bibinfo{month}{Sep.}
\bibinfo{title}{{Intermediate mass black holes in AGN discs - I. Production and growth}}.
\bibinfo{journal}{{\em \mnras}} \bibinfo{volume}{425} (\bibinfo{number}{1}): \bibinfo{pages}{460--469}. \bibinfo{doi}{\doi{10.1111/j.1365-2966.2012.21486.x}}.
\eprint{1206.2309}.

\bibtype{Article}%
\bibitem[{McKernan} et al.(2019)]{mckernanRam}
\bibinfo{author}{{McKernan} B}, \bibinfo{author}{{Ford} KES}, \bibinfo{author}{{Bartos} I}, \bibinfo{author}{{Graham} MJ}, \bibinfo{author}{{Lyra} W}, \bibinfo{author}{{Marka} S}, \bibinfo{author}{{Marka} Z}, \bibinfo{author}{{Ross} NP}, \bibinfo{author}{{Stern} D} and  \bibinfo{author}{{Yang} Y} (\bibinfo{year}{2019}), \bibinfo{month}{Oct.}
\bibinfo{title}{{Ram-pressure Stripping of a Kicked Hill Sphere: Prompt Electromagnetic Emission from the Merger of Stellar Mass Black Holes in an AGN Accretion Disk}}.
\bibinfo{journal}{{\em \apjl}} \bibinfo{volume}{884} (\bibinfo{number}{2}), \bibinfo{eid}{L50}. \bibinfo{doi}{\doi{10.3847/2041-8213/ab4886}}.
\eprint{1907.03746}.

\bibtype{Article}%
\bibitem[Messenger and Read(2012)]{Messenger:2011gi}
\bibinfo{author}{Messenger C} and  \bibinfo{author}{Read J} (\bibinfo{year}{2012}).
\bibinfo{title}{{Measuring a cosmological distance-redshift relationship using only gravitational wave observations of binary neutron star coalescences}}.
\bibinfo{journal}{{\em Phys. Rev. Lett.}} \bibinfo{volume}{108}: \bibinfo{pages}{091101}. \bibinfo{doi}{\doi{10.1103/PhysRevLett.108.091101}}.
\eprint{1107.5725}.

\bibtype{Article}%
\bibitem[Metzger et al.(2010)]{Metzger_2010}
\bibinfo{author}{Metzger BD}, \bibinfo{author}{Martínez-Pinedo G}, \bibinfo{author}{Darbha S}, \bibinfo{author}{Quataert E}, \bibinfo{author}{Arcones A}, \bibinfo{author}{Kasen D}, \bibinfo{author}{Thomas R}, \bibinfo{author}{Nugent P}, \bibinfo{author}{Panov IV} and  \bibinfo{author}{Zinner NT} (\bibinfo{year}{2010}), \bibinfo{month}{Jun.}
\bibinfo{title}{Electromagnetic counterparts of compact object mergers powered by the radioactive decay of r-process nuclei: Transients from compact object mergers}.
\bibinfo{journal}{{\em Monthly Notices of the Royal Astronomical Society}} \bibinfo{volume}{406} (\bibinfo{number}{4}): \bibinfo{pages}{2650–2662}.
ISSN \bibinfo{issn}{0035-8711}. \bibinfo{doi}{\doi{10.1111/j.1365-2966.2010.16864.x}}.
\bibinfo{url}{\url{http://dx.doi.org/10.1111/j.1365-2966.2010.16864.x}}.

\bibtype{Article}%
\bibitem[{Mooley} et al.(2018)]{mooley}
\bibinfo{author}{{Mooley} KP}, \bibinfo{author}{{Deller} AT}, \bibinfo{author}{{Gottlieb} O}, \bibinfo{author}{{Nakar} E}, \bibinfo{author}{{Hallinan} G}, \bibinfo{author}{{Bourke} S}, \bibinfo{author}{{Frail} DA}, \bibinfo{author}{{Horesh} A}, \bibinfo{author}{{Corsi} A} and  \bibinfo{author}{{Hotokezaka} K} (\bibinfo{year}{2018}), \bibinfo{month}{Sep.}
\bibinfo{title}{{Superluminal motion of a relativistic jet in the neutron-star merger GW170817}}.
\bibinfo{journal}{{\em \nat}} \bibinfo{volume}{561} (\bibinfo{number}{7723}): \bibinfo{pages}{355--359}. \bibinfo{doi}{\doi{10.1038/s41586-018-0486-3}}.
\eprint{1806.09693}.

\bibtype{Article}%
\bibitem[Mooley et al.(2022)]{Mooley_2022}
\bibinfo{author}{Mooley KP}, \bibinfo{author}{Anderson J} and  \bibinfo{author}{Lu W} (\bibinfo{year}{2022}), \bibinfo{month}{oct}.
\bibinfo{title}{Optical superluminal motion measurement in the neutron-star merger {GW}170817}.
\bibinfo{journal}{{\em Nature}} \bibinfo{volume}{610} (\bibinfo{number}{7931}): \bibinfo{pages}{273--276}. \bibinfo{doi}{\doi{10.1038/s41586-022-05145-7}}.
\bibinfo{url}{\url{https://doi.org/10.1038\%2Fs41586-022-05145-7}}.

\bibtype{Article}%
\bibitem[Mukherjee(2022)]{Mukherjee:2021rtw}
\bibinfo{author}{Mukherjee S} (\bibinfo{year}{2022}).
\bibinfo{title}{{The redshift dependence of black hole mass distribution: is it reliable for standard sirens cosmology?}}
\bibinfo{journal}{{\em Mon. Not. Roy. Astron. Soc.}} \bibinfo{volume}{515} (\bibinfo{number}{4}): \bibinfo{pages}{5495--5505}. \bibinfo{doi}{\doi{10.1093/mnras/stac2152}}.
\eprint{2112.10256}.

\bibtype{Article}%
\bibitem[Mukherjee et al.(2021{\natexlab{a}})]{Mukherjee2020}
\bibinfo{author}{Mukherjee S}, \bibinfo{author}{Lavaux G}, \bibinfo{author}{Bouchet FR}, \bibinfo{author}{Jasche J}, \bibinfo{author}{Wandelt BD}, \bibinfo{author}{Nissanke SM}, \bibinfo{author}{Leclercq F} and  \bibinfo{author}{Hotokezaka K} (\bibinfo{year}{2021}{\natexlab{a}}).
\bibinfo{title}{{Velocity correction for Hubble constant measurements from standard sirens}}.
\bibinfo{journal}{{\em Astron. Astrophys.}} \bibinfo{volume}{646}: \bibinfo{pages}{A65}. \bibinfo{doi}{\doi{10.1051/0004-6361/201936724}}.
\eprint{1909.08627}.

\bibtype{Article}%
\bibitem[Mukherjee et al.(2021{\natexlab{b}})]{Mukherjee:2020hyn}
\bibinfo{author}{Mukherjee S}, \bibinfo{author}{Wandelt BD}, \bibinfo{author}{Nissanke SM} and  \bibinfo{author}{Silvestri A} (\bibinfo{year}{2021}{\natexlab{b}}).
\bibinfo{title}{{Accurate precision Cosmology with redshift unknown gravitational wave sources}}.
\bibinfo{journal}{{\em Phys. Rev. D}} \bibinfo{volume}{103} (\bibinfo{number}{4}): \bibinfo{pages}{043520}. \bibinfo{doi}{\doi{10.1103/PhysRevD.103.043520}}.
\eprint{2007.02943}.

\bibtype{Article}%
\bibitem[Mukherjee et al.(2021{\natexlab{c}})]{Mukherjee:2020mha}
\bibinfo{author}{Mukherjee S}, \bibinfo{author}{Wandelt BD} and  \bibinfo{author}{Silk J} (\bibinfo{year}{2021}{\natexlab{c}}).
\bibinfo{title}{{Testing the general theory of relativity using gravitational wave propagation from dark standard sirens}}.
\bibinfo{journal}{{\em Mon. Not. Roy. Astron. Soc.}} \bibinfo{volume}{502} (\bibinfo{number}{1}): \bibinfo{pages}{1136--1144}. \bibinfo{doi}{\doi{10.1093/mnras/stab001}}.
\eprint{2012.15316}.

\bibtype{Article}%
\bibitem[Mukherjee et al.(2024)]{Mukherjee:2022afz}
\bibinfo{author}{Mukherjee S}, \bibinfo{author}{Krolewski A}, \bibinfo{author}{Wandelt BD} and  \bibinfo{author}{Silk J} (\bibinfo{year}{2024}).
\bibinfo{title}{{Cross-correlating dark sirens and galaxies: constraints on $H_0$ from GWTC-3 of LIGO-Virgo-KAGRA}}.
\bibinfo{journal}{{\em Astrophys. J.}} \bibinfo{volume}{975} (\bibinfo{number}{2}): \bibinfo{pages}{189}. \bibinfo{doi}{\doi{10.3847/1538-4357/ad7d90}}.
\eprint{2203.03643}.

\bibtype{Misc}%
\bibitem[Müller et al.(2024)]{muller2024}
\bibinfo{author}{Müller M}, \bibinfo{author}{Mukherjee S} and  \bibinfo{author}{Ryan G} (\bibinfo{year}{2024}).
\bibinfo{title}{Be careful in multi-messenger inference of the hubble constant: A path forward for robust inference}.
\eprint{2406.11965}, \bibinfo{url}{\url{https://arxiv.org/abs/2406.11965}}.

\bibtype{Article}%
\bibitem[Namikawa et al.(2016{\natexlab{a}})]{Namikawa:2015prh}
\bibinfo{author}{Namikawa T}, \bibinfo{author}{Nishizawa A} and  \bibinfo{author}{Taruya A} (\bibinfo{year}{2016}{\natexlab{a}}).
\bibinfo{title}{{Anisotropies of gravitational-wave standard sirens as a new cosmological probe without redshift information}}.
\bibinfo{journal}{{\em Phys. Rev. Lett.}} \bibinfo{volume}{116} (\bibinfo{number}{12}): \bibinfo{pages}{121302}. \bibinfo{doi}{\doi{10.1103/PhysRevLett.116.121302}}.
\eprint{1511.04638}.

\bibtype{Article}%
\bibitem[Namikawa et al.(2016{\natexlab{b}})]{Namikawa:2016edr}
\bibinfo{author}{Namikawa T}, \bibinfo{author}{Nishizawa A} and  \bibinfo{author}{Taruya A} (\bibinfo{year}{2016}{\natexlab{b}}).
\bibinfo{title}{{Detecting Black-Hole Binary Clustering via the Second-Generation Gravitational-Wave Detectors}}.
\bibinfo{journal}{{\em Phys. Rev. D}} \bibinfo{volume}{94} (\bibinfo{number}{2}): \bibinfo{pages}{024013}. \bibinfo{doi}{\doi{10.1103/PhysRevD.94.024013}}.
\eprint{1603.08072}.

\bibtype{Article}%
\bibitem[{Ng} et al.(2024)]{2024arXiv241023541N}
\bibinfo{author}{{Ng} TCK}, \bibinfo{author}{{Rinaldi} S} and  \bibinfo{author}{{Hannuksela} OA} (\bibinfo{year}{2024}), \bibinfo{month}{Oct.}
\bibinfo{title}{{Inferring cosmology from gravitational waves using non-parametric detector-frame mass distribution}}.
\bibinfo{journal}{{\em arXiv}} , \bibinfo{eid}{arXiv:2410.23541}\bibinfo{doi}{\doi{10.48550/arXiv.2410.23541}}.
\eprint{2410.23541}.

\bibtype{Article}%
\bibitem[Nicolaou et al.(2020)]{Nicolaou_2020}
\bibinfo{author}{Nicolaou C}, \bibinfo{author}{Lahav O}, \bibinfo{author}{Lemos P}, \bibinfo{author}{Hartley W} and  \bibinfo{author}{Braden J} (\bibinfo{year}{2020}), \bibinfo{month}{Apr.}
\bibinfo{title}{The impact of peculiar velocities on the estimation of the hubble constant from gravitational wave standard sirens}.
\bibinfo{journal}{{\em Monthly Notices of the Royal Astronomical Society}} \bibinfo{volume}{495} (\bibinfo{number}{1}): \bibinfo{pages}{90–97}.
ISSN \bibinfo{issn}{1365-2966}. \bibinfo{doi}{\doi{10.1093/mnras/staa1120}}.
\bibinfo{url}{\url{http://dx.doi.org/10.1093/mnras/staa1120}}.

\bibtype{Article}%
\bibitem[Nimonkar and Mukherjee(2023)]{Nimonkar:2023pyt}
\bibinfo{author}{Nimonkar H} and  \bibinfo{author}{Mukherjee S} (\bibinfo{year}{2023}).
\bibinfo{title}{{Dependence of peculiar velocity on the host properties of the gravitational wave sources and its impact on the measurement of Hubble constant}}.
\bibinfo{journal}{{\em Mon. Not. Roy. Astron. Soc.}} \bibinfo{volume}{527} (\bibinfo{number}{2}): \bibinfo{pages}{2152--2164}. \bibinfo{doi}{\doi{10.1093/mnras/stad3256}}.
\eprint{2307.05688}.

\bibtype{Article}%
\bibitem[Oguri(2016)]{Oguri:2016dgk}
\bibinfo{author}{Oguri M} (\bibinfo{year}{2016}).
\bibinfo{title}{{Measuring the distance-redshift relation with the cross-correlation of gravitational wave standard sirens and galaxies}}.
\bibinfo{journal}{{\em Phys. Rev. D}} \bibinfo{volume}{93} (\bibinfo{number}{8}): \bibinfo{pages}{083511}. \bibinfo{doi}{\doi{10.1103/PhysRevD.93.083511}}.
\eprint{1603.02356}.

\bibtype{Article}%
\bibitem[{Palmese} et al.(2017)]{Palmese_2017}
\bibinfo{author}{{Palmese} A} and  et al. (\bibinfo{year}{2017}), \bibinfo{month}{Nov.}
\bibinfo{title}{{Evidence for Dynamically Driven Formation of the GW170817 Neutron Star Binary in NGC 4993}}.
\bibinfo{journal}{{\em \apjl}} \bibinfo{volume}{849}, \bibinfo{eid}{L34}. \bibinfo{doi}{\doi{10.3847/2041-8213/aa9660}}.
\eprint{1710.06748}.

\bibtype{Article}%
\bibitem[{Palmese} et al.(2020)]{palmese20_sts}
\bibinfo{author}{{Palmese} A}, \bibinfo{author}{{deVicente} J}, \bibinfo{author}{{Pereira} MES}, \bibinfo{author}{{Annis} J}, \bibinfo{author}{{Hartley} W}, \bibinfo{author}{{Herner} K}, \bibinfo{author}{{Soares-Santos} M}, \bibinfo{author}{{Crocce} M}, \bibinfo{author}{{Huterer} D}, \bibinfo{author}{{Maga{\~n}a Hernandez} I} and  et al. (\bibinfo{year}{2020}), \bibinfo{month}{Sep.}
\bibinfo{title}{{A Statistical Standard Siren Measurement of the Hubble Constant from the LIGO/Virgo Gravitational Wave Compact Object Merger GW190814 and Dark Energy Survey Galaxies}}.
\bibinfo{journal}{{\em \apjl}} \bibinfo{volume}{900} (\bibinfo{number}{2}), \bibinfo{eid}{L33}. \bibinfo{doi}{\doi{10.3847/2041-8213/abaeff}}.
\eprint{2006.14961}.

\bibtype{Article}%
\bibitem[Palmese et al.(2021)]{Palmese_AGN}
\bibinfo{author}{Palmese A}, \bibinfo{author}{Fishbach M}, \bibinfo{author}{Burke CJ}, \bibinfo{author}{Annis J} and  \bibinfo{author}{Liu X} (\bibinfo{year}{2021}), \bibinfo{month}{Jun}.
\bibinfo{title}{Do ligo/virgo black hole mergers produce agn flares? the case of gw190521 and prospects for reaching a confident association}.
\bibinfo{journal}{{\em The Astrophysical Journal Letters}} \bibinfo{volume}{914} (\bibinfo{number}{2}): \bibinfo{pages}{L34}.
ISSN \bibinfo{issn}{2041-8213}. \bibinfo{doi}{\doi{10.3847/2041-8213/ac0883}}.
\bibinfo{url}{\url{http://dx.doi.org/10.3847/2041-8213/ac0883}}.

\bibtype{Article}%
\bibitem[{Palmese} et al.(2023)]{darksiren_DESI}
\bibinfo{author}{{Palmese} A}, \bibinfo{author}{{Bom} CR}, \bibinfo{author}{{Mucesh} S} and  \bibinfo{author}{{Hartley} WG} (\bibinfo{year}{2023}), \bibinfo{month}{Jan.}
\bibinfo{title}{A standard siren measurement of the hubble constant using gravitational-wave events from the first three ligo/virgo observing runs and the desi legacy survey}.
\bibinfo{journal}{{\em The Astrophysical Journal}} \bibinfo{volume}{943} (\bibinfo{number}{1}): \bibinfo{pages}{56}.
ISSN \bibinfo{issn}{1538-4357}. \bibinfo{doi}{\doi{10.3847/1538-4357/aca6e3}}.
\bibinfo{url}{\url{http://dx.doi.org/10.3847/1538-4357/aca6e3}}.

\bibtype{Article}%
\bibitem[{Palmese} et al.(2024)]{Palmese_2024}
\bibinfo{author}{{Palmese} A}, \bibinfo{author}{{Kaur} R}, \bibinfo{author}{{Hajela} A}, \bibinfo{author}{{Margutti} R}, \bibinfo{author}{{McDowell} A} and  \bibinfo{author}{{MacFadyen} A} (\bibinfo{year}{2024}), \bibinfo{month}{Mar.}
\bibinfo{title}{{Standard siren measurement of the Hubble constant using GW170817 and the latest observations of the electromagnetic counterpart afterglow}}.
\bibinfo{journal}{{\em \prd}} \bibinfo{volume}{109} (\bibinfo{number}{6}), \bibinfo{eid}{063508}. \bibinfo{doi}{\doi{10.1103/PhysRevD.109.063508}}.
\eprint{2305.19914}.

\bibtype{Article}%
\bibitem[{Perna} et al.(2024)]{Perna2024}
\bibinfo{author}{{Perna} G}, \bibinfo{author}{{Mastrogiovanni} S} and  \bibinfo{author}{{Ricciardone} A} (\bibinfo{year}{2024}), \bibinfo{month}{May}.
\bibinfo{title}{{Investigating the impact of galaxies' compact binary hosting probability for gravitational-wave cosmology}}.
\bibinfo{journal}{{\em arXiv e-prints}} , \bibinfo{eid}{arXiv:2405.07904}\bibinfo{doi}{\doi{10.48550/arXiv.2405.07904}}.
\eprint{2405.07904}.

\bibtype{Article}%
\bibitem[Peron et al.(2024)]{Peron:2023zae}
\bibinfo{author}{Peron M}, \bibinfo{author}{Ravenni A}, \bibinfo{author}{Libanore S}, \bibinfo{author}{Liguori M} and  \bibinfo{author}{Artale MC} (\bibinfo{year}{2024}).
\bibinfo{title}{{Clustering of binary black hole mergers: a detailed analysis of the eagle~+~mobse simulation}}.
\bibinfo{journal}{{\em Mon. Not. Roy. Astron. Soc.}} \bibinfo{volume}{530} (\bibinfo{number}{1}): \bibinfo{pages}{1129--1143}. \bibinfo{doi}{\doi{10.1093/mnras/stae893}}.
\eprint{2305.18003}.

\bibtype{Article}%
\bibitem[Pierra et al.(2024)]{Pierra:2023deu}
\bibinfo{author}{Pierra G}, \bibinfo{author}{Mastrogiovanni S}, \bibinfo{author}{Perri\`es S} and  \bibinfo{author}{Mapelli M} (\bibinfo{year}{2024}).
\bibinfo{title}{{Study of systematics on the cosmological inference of the Hubble constant from gravitational wave standard sirens}}.
\bibinfo{journal}{{\em Phys. Rev. D}} \bibinfo{volume}{109} (\bibinfo{number}{8}): \bibinfo{pages}{083504}. \bibinfo{doi}{\doi{10.1103/PhysRevD.109.083504}}.
\eprint{2312.11627}.

\bibtype{Article}%
\bibitem[{Planck Collaboration}(2020)]{planckcolab}
\bibinfo{author}{{Planck Collaboration}} (\bibinfo{year}{2020}), \bibinfo{month}{Sep.}
\bibinfo{title}{{Planck 2018 results. VI. Cosmological parameters}}.
\bibinfo{journal}{{\em \aap}} \bibinfo{volume}{641}, \bibinfo{eid}{A6}. \bibinfo{doi}{\doi{10.1051/0004-6361/201833910}}.
\eprint{1807.06209}.

\bibtype{Article}%
\bibitem[{Riess} et al.(2022)]{Riess2022}
\bibinfo{author}{{Riess} AG}, \bibinfo{author}{{Yuan} W}, \bibinfo{author}{{Macri} LM}, \bibinfo{author}{{Scolnic} D}, \bibinfo{author}{{Brout} D}, \bibinfo{author}{{Casertano} S}, \bibinfo{author}{{Jones} DO}, \bibinfo{author}{{Murakami} Y}, \bibinfo{author}{{Anand} GS}, \bibinfo{author}{{Breuval} L}, \bibinfo{author}{{Brink} TG}, \bibinfo{author}{{Filippenko} AV}, \bibinfo{author}{{Hoffmann} S}, \bibinfo{author}{{Jha} SW}, \bibinfo{author}{{D'arcy Kenworthy} W}, \bibinfo{author}{{Mackenty} J}, \bibinfo{author}{{Stahl} BE} and  \bibinfo{author}{{Zheng} W} (\bibinfo{year}{2022}), \bibinfo{month}{Jul.}
\bibinfo{title}{{A Comprehensive Measurement of the Local Value of the Hubble Constant with 1 km s$^{-1}$ Mpc$^{-1}$ Uncertainty from the Hubble Space Telescope and the SH0ES Team}}.
\bibinfo{journal}{{\em \apjl}} \bibinfo{volume}{934} (\bibinfo{number}{1}), \bibinfo{eid}{L7}. \bibinfo{doi}{\doi{10.3847/2041-8213/ac5c5b}}.
\eprint{2112.04510}.

\bibtype{Article}%
\bibitem[Salvarese and Chen(2024)]{Salvarese:2024jpq}
\bibinfo{author}{Salvarese A} and  \bibinfo{author}{Chen HY} (\bibinfo{year}{2024}).
\bibinfo{title}{{Mitigating the Binary Viewing Angle Bias for Standard Sirens}}.
\bibinfo{journal}{{\em Astrophys. J. Lett.}} \bibinfo{volume}{974} (\bibinfo{number}{1}): \bibinfo{pages}{L16}. \bibinfo{doi}{\doi{10.3847/2041-8213/ad7bbc}}.
\eprint{2406.11126}.

\bibtype{Article}%
\bibitem[Scelfo et al.(2020)]{Scelfo:2020jyw}
\bibinfo{author}{Scelfo G}, \bibinfo{author}{Boco L}, \bibinfo{author}{Lapi A} and  \bibinfo{author}{Viel M} (\bibinfo{year}{2020}).
\bibinfo{title}{{Exploring galaxies-gravitational waves cross-correlations as an astrophysical probe}}.
\bibinfo{journal}{{\em JCAP}} \bibinfo{volume}{10}: \bibinfo{pages}{045}. \bibinfo{doi}{\doi{10.1088/1475-7516/2020/10/045}}.
\eprint{2007.08534}.

\bibtype{Article}%
\bibitem[Scelfo et al.(2022)]{Scelfo:2021fqe}
\bibinfo{author}{Scelfo G}, \bibinfo{author}{Spinelli M}, \bibinfo{author}{Raccanelli A}, \bibinfo{author}{Boco L}, \bibinfo{author}{Lapi A} and  \bibinfo{author}{Viel M} (\bibinfo{year}{2022}).
\bibinfo{title}{{Gravitational waves \texttimes{} HI intensity mapping: cosmological and astrophysical applications}}.
\bibinfo{journal}{{\em JCAP}} \bibinfo{volume}{01} (\bibinfo{number}{01}): \bibinfo{pages}{004}. \bibinfo{doi}{\doi{10.1088/1475-7516/2022/01/004}}.
\eprint{2106.09786}.

\bibtype{Article}%
\bibitem[{Schutz}(1986)]{schutz}
\bibinfo{author}{{Schutz} BF} (\bibinfo{year}{1986}), \bibinfo{month}{Sep.}
\bibinfo{title}{{Determining the Hubble constant from gravitational wave observations}}.
\bibinfo{journal}{{\em \nat}} \bibinfo{volume}{323}: \bibinfo{pages}{310}. \bibinfo{doi}{\doi{10.1038/323310a0}}.

\bibtype{Article}%
\bibitem[{Soares-Santos} et al.(2017)]{Soaressantos_2017}
\bibinfo{author}{{Soares-Santos} M} and  et al. (\bibinfo{year}{2017}), \bibinfo{month}{Oct.}
\bibinfo{title}{{The Electromagnetic Counterpart of the Binary Neutron Star Merger LIGO/Virgo GW170817. I. Discovery of the Optical Counterpart Using the Dark Energy Camera}}.
\bibinfo{journal}{{\em \apjl}} \bibinfo{volume}{848} (\bibinfo{number}{2}), \bibinfo{eid}{L16}. \bibinfo{doi}{\doi{10.3847/2041-8213/aa9059}}.
\eprint{1710.05459}.

\bibtype{Article}%
\bibitem[{Soares-Santos} et al.(2019)]{DES:2019}
\bibinfo{author}{{Soares-Santos} M}, \bibinfo{author}{{Palmese} A} and  et al. (\bibinfo{year}{2019}), \bibinfo{month}{May}.
\bibinfo{title}{{First Measurement of the Hubble Constant from a Dark Standard Siren using the Dark Energy Survey Galaxies and the LIGO/Virgo Binary-Black-hole Merger GW170814}}.
\bibinfo{journal}{{\em \apjl}} \bibinfo{volume}{876}, \bibinfo{eid}{L7}. \bibinfo{doi}{\doi{10.3847/2041-8213/ab14f1}}.
\eprint{1901.01540}.

\bibtype{Article}%
\bibitem[{Tagawa} et al.(2023)]{tagawa23}
\bibinfo{author}{{Tagawa} H}, \bibinfo{author}{{Kimura} SS}, \bibinfo{author}{{Haiman} Z}, \bibinfo{author}{{Perna} R} and  \bibinfo{author}{{Bartos} I} (\bibinfo{year}{2023}), \bibinfo{month}{Mar.}
\bibinfo{title}{{Observable signatures of stellar-mass black holes in active galactic nuclei}}.
\bibinfo{journal}{{\em arXiv e-prints}} , \bibinfo{eid}{arXiv:2303.02172}\bibinfo{doi}{\doi{10.48550/arXiv.2303.02172}}.
\eprint{2303.02172}.

\bibtype{Article}%
\bibitem[{Talbot} and {Thrane}(2018)]{2018ApJ...856..173T}
\bibinfo{author}{{Talbot} C} and  \bibinfo{author}{{Thrane} E} (\bibinfo{year}{2018}), \bibinfo{month}{Apr.}
\bibinfo{title}{{Measuring the Binary Black Hole Mass Spectrum with an Astrophysically Motivated Parameterization}}.
\bibinfo{journal}{{\em \apj}} \bibinfo{volume}{856} (\bibinfo{number}{2}), \bibinfo{eid}{173}. \bibinfo{doi}{\doi{10.3847/1538-4357/aab34c}}.
\eprint{1801.02699}.

\bibtype{Article}%
\bibitem[{Tanvir} et al.(2017)]{Tanvir_2017}
\bibinfo{author}{{Tanvir} NR} and  et al. (\bibinfo{year}{2017}), \bibinfo{month}{Oct.}
\bibinfo{title}{{The Emergence of a Lanthanide-rich Kilonova Following the Merger of Two Neutron Stars}}.
\bibinfo{journal}{{\em \apjl}} \bibinfo{volume}{848} (\bibinfo{number}{2}), \bibinfo{eid}{L27}. \bibinfo{doi}{\doi{10.3847/2041-8213/aa90b6}}.
\eprint{1710.05455}.

\bibtype{Article}%
\bibitem[Taylor et al.(2012)]{Taylor2012}
\bibinfo{author}{Taylor SR}, \bibinfo{author}{Gair JR} and  \bibinfo{author}{Mandel I} (\bibinfo{year}{2012}), \bibinfo{month}{Jan}.
\bibinfo{title}{Cosmology using advanced gravitational-wave detectors alone}.
\bibinfo{journal}{{\em Phys. Rev. D}} \bibinfo{volume}{85}: \bibinfo{pages}{023535}. \bibinfo{doi}{\doi{10.1103/PhysRevD.85.023535}}.
\bibinfo{url}{\url{https://link.aps.org/doi/10.1103/PhysRevD.85.023535}}.

\bibtype{Article}%
\bibitem[Tiwari(2018)]{Tiwari:2017ndi}
\bibinfo{author}{Tiwari V} (\bibinfo{year}{2018}).
\bibinfo{title}{{Estimation of the Sensitive Volume for Gravitational-wave Source Populations Using Weighted Monte Carlo Integration}}.
\bibinfo{journal}{{\em Class. Quant. Grav.}} \bibinfo{volume}{35} (\bibinfo{number}{14}): \bibinfo{pages}{145009}. \bibinfo{doi}{\doi{10.1088/1361-6382/aac89d}}.
\eprint{1712.00482}.

\bibtype{Article}%
\bibitem[{Valenti} et al.(2017)]{Valenti_2017}
\bibinfo{author}{{Valenti} S}, \bibinfo{author}{{Sand} DJ}, \bibinfo{author}{{Yang} S}, \bibinfo{author}{{Cappellaro} E}, \bibinfo{author}{{Tartaglia} L}, \bibinfo{author}{{Corsi} A}, \bibinfo{author}{{Jha} SW}, \bibinfo{author}{{Reichart} DE}, \bibinfo{author}{{Haislip} J} and  \bibinfo{author}{{Kouprianov} V} (\bibinfo{year}{2017}), \bibinfo{month}{Oct.}
\bibinfo{title}{{The Discovery of the Electromagnetic Counterpart of GW170817: Kilonova AT 2017gfo/DLT17ck}}.
\bibinfo{journal}{{\em \apjl}} \bibinfo{volume}{848} (\bibinfo{number}{2}), \bibinfo{eid}{L24}. \bibinfo{doi}{\doi{10.3847/2041-8213/aa8edf}}.
\eprint{1710.05854}.

\bibtype{Article}%
\bibitem[{Vitale} and {Chen}(2018)]{2018PhRvL.121b1303V}
\bibinfo{author}{{Vitale} S} and  \bibinfo{author}{{Chen} HY} (\bibinfo{year}{2018}), \bibinfo{month}{Jul.}
\bibinfo{title}{{Measuring the Hubble Constant with Neutron Star Black Hole Mergers}}.
\bibinfo{journal}{{\em \prl}} \bibinfo{volume}{121} (\bibinfo{number}{2}), \bibinfo{eid}{021303}. \bibinfo{doi}{\doi{10.1103/PhysRevLett.121.021303}}.
\eprint{1804.07337}.

\bibtype{incollection}%
\bibitem[{Vitale} et al.(2022)]{2022hgwa.bookE..45V}
\bibinfo{author}{{Vitale} S}, \bibinfo{author}{{Gerosa} D}, \bibinfo{author}{{Farr} WM} and  \bibinfo{author}{{Taylor} SR} (\bibinfo{year}{2022}), \bibinfo{title}{{Inferring the Properties of a Population of Compact Binaries in Presence of Selection Effects}}, \bibinfo{editor}{{Bambi} C}, \bibinfo{editor}{{Katsanevas} S} and  \bibinfo{editor}{{Kokkotas} KD}, (Eds.), \bibinfo{booktitle}{Handbook of Gravitational Wave Astronomy}, pp.~\bibinfo{pages}{45}.

\bibtype{Article}%
\bibitem[Zazzera et al.(2024)]{Zazzera:2024agl}
\bibinfo{author}{Zazzera S}, \bibinfo{author}{Fonseca J}, \bibinfo{author}{Baker T} and  \bibinfo{author}{Clarkson C} (\bibinfo{year}{2024}), \bibinfo{month}{12}.
\bibinfo{title}{{Gravitational waves and galaxies cross-correlations: a forecast on GW biases for future detectors}} \eprint{2412.01678}.

\bibtype{Article}%
\bibitem[Zhang(2018)]{Zhang:2018nea}
\bibinfo{author}{Zhang P} (\bibinfo{year}{2018}), \bibinfo{month}{10}.
\bibinfo{title}{{The large scale structure in the 3D luminosity-distance space and its cosmological applications}} \eprint{1810.11915}.

\end{thebibliography*}

\end{document}